\documentclass[a4paper,11pt]{article}
\pdfoutput=1
\usepackage{notation}

\allowdisplaybreaks

\preprint{UWThPh2024-4\hspace{1em}}

\title{
Higgs branch RG-flows via Decay and Fission
}

\author[a]{Antoine Bourget}
\author[b]{, Marcus Sperling}
\author[c]{and Zhenghao Zhong}
\affiliation[a]{Université Paris-Saclay, CNRS, CEA, Institut de physique théorique,\\ 91191, Gif-sur-Yvette, France}
\affiliation[b]{Fakultät für Physik, Universität Wien,\\
Boltzmanngasse 5, 1090 Wien, Austria
}
\affiliation[c]{Mathematical Institute, University of Oxford,\\
Andrew Wiles Building, Woodstock Road, Oxford, OX2 6GG, UK}
\emailAdd{antoine.bourget@ipht.fr}
\emailAdd{marcus.sperling@univie.ac.at}
\emailAdd{zhenghao.zhong@maths.ox.ac.uk}

\abstract{Magnetic quivers have been an instrumental technique for advancing our understanding of Higgs branches of supersymmetric theories with 8 supercharges. In this work, we present the \emph{decay and fission} algorithm for unitary magnetic quivers. It enables the derivation of the complete phase (Hasse) diagram and is characterised by the following key attributes: First and foremost, the algorithm is inherently simple; just relying on convex linear algebra. Second, any magnetic quiver can only undergo \emph{decay} or \emph{fission} processes; these reflect the possible Higgs branch RG-flows (Higgsings), and the quivers thereby generated are the magnetic quivers of the new RG fixed points. Third, the geometry of the \emph{decay} or \emph{fission} transition (i.e.\ the transverse slice) is simply read off. As a consequence, the algorithm does not rely on a complete list of minimal transitions, but rather outputs the transverse slice geometry automatically. As a proof of concept, its efficacy is showcased across various scenarios, encompassing SCFTs from dimensions 3 to 6, instanton moduli spaces, and little string theories.}

\begin{document} 
\maketitle

\section{Introduction}

The Higgs mechanism is a well-known concept in quantum field theories (QFTs) wherein a scalar field acquires a vacuum expectation value (VEV) that subsequently breaks the gauge symmetry \cite{englert1964broken,higgs1964broken,guralnik1964global,kibble1967symmetry}. For example, the electroweak theory has a $\surm(2)\times \urm(1)$ gauge symmetry, and when the scalar field (Higgs boson) acquires a non-zero VEV, the symmetry is broken to $\urm(1)$. Such a Higgsing also constitutes a phase transition in the theory. Supersymmetric QFTs with 8 supercharges in space-time dimension 3, 4, 5 and 6 generically possess a continuous space of vacua known as the Higgs branch, denoted $\Higgs$. As the Higgs branch is parameterised by many scalar fields, the theory can be Higgsed in multiple ways. This rich structure can then be encoded in a phase diagram \cite{Bourget:2019aer}.

Phase diagrams are important as they show how various theories are related to each other through means such as mass deformations, tuning of gauge couplings, Coulomb branch deformations, etc. In this paper, we derive phase diagrams that encodes how different theories are related via the (partial) Higgs mechanism, or to be more precise, Higgsing along the Higgs branch\footnote{In the literature, this has also been called \emph{Higgs branch deformations} or \emph{Higgs branch RG-flows}.}. 
While performing a partial Higgs mechanism might not pose significant challenges for theories with known Lagrangian descriptions, the realm of superconformal field theories (SCFTs) frequently lacks such descriptions, particularly in space-time dimensions 4, 5, and 6.
Thus, to investigate characteristics of SCFTs, such as their Higgs branches, new methods had to be devised.
One particularly powerful technique that allows one to study the Higgs branches of gauge theories and SCFTs, regardless of their space-time dimension, and even little string theories is the \emph{magnetic quiver} \cite{Cabrera:2018jxt,Cabrera:2019izd,Bourget:2019rtl,Cabrera:2019dob, Bourget:2020gzi,Bourget:2020asf,Bourget:2020xdz,Closset:2020scj,Akhond:2020vhc,vanBeest:2020kou,Giacomelli:2020gee,Bourget:2020mez,VanBeest:2020kxw,Closset:2020afy,Akhond:2021knl,Bourget:2021csg,vanBeest:2021xyt,Sperling:2021fcf,Nawata:2021nse,Akhond:2022jts,Giacomelli:2022drw,Hanany:2022itc,Fazzi:2022hal,Bourget:2022tmw,Fazzi:2022yca,Nawata:2023rdx,Bourget:2023cgs,DelZotto:2023nrb,Lawrie:2023uiu,Mansi:2023faa,Fazzi:2023ulb}. 

Consider a theory $\Tcal$ in space-time dimension $d=3,4,5,6$ with 8 supercharges: the magnetic quiver is a $3d$ $\Ncal=4$ (generalised) quiver gauge theory whose Coulomb branch is, by construction, the same as the Higgs branch of $\Tcal$. If the Higgs branch is a union of several hyper-K\"ahler cones, then there exist several magnetic quivers, one for each cone \cite{Ferlito:2016grh,Bourget:2023cgs}. Therefore, studying the magnetic quiver is an indirect path of studying the Higgs branch of $\Tcal$. 
We pursue this indirect route due to the array of recently developed techniques tailored for the study of the Coulomb branch of $3d$ $\Ncal=4$ theories, starting with \cite{Cremonesi:2013lqa,Cremonesi:2014xha,Cabrera:2018ann}.
This transforms a historically challenging subject into a realm of familiarity and ease.

In \cite{Bourget:2019aer}, an algorithm, which we henceforth refer to as the \emph{quiver subtraction} algorithm, was introduced. The algorithm provides the stratification of the Higgs branch of $\Tcal$, encoding it in a phase diagram called a (Higgs branch) \emph{Hasse diagram}. This phase diagram encodes much of the Higgs branch including the nature of the transverse space between Higgsed theories. However, starting from $\Tcal$, this algorithm is unable to determine the set of possible theories $\Tcal_i$ that $\Tcal$ can be Higgsed to. 

While being a powerful approach that has found wide-spread application \cite{Cabrera:2019dob,Bourget:2020gzi,VanBeest:2020kxw,vanBeest:2020kou,Bourget:2020mez,Closset:2020scj,Eckhard:2020jyr,Martone:2021ixp,Closset:2021lwy,Santilli:2021rlf,Arias-Tamargo:2021ppf,Bourget:2021csg,Giacomelli:2022drw,Hanany:2022itc,Fazzi:2022hal,Fazzi:2022yca,Mu:2023uws,Lawrie:2023uiu,Mansi:2023faa,Fazzi:2023ulb}, the quiver subtraction algorithm is plagued by a few shortcomings: firstly, the algorithm requires a list of all possible elementary slices. This list is mathematically complete for the class of nilpotent orbits closures \cite{Kraft1980,Kraft1982,fu2017generic}, but more general symplectic singularities may include other minimal slices. Such a new isolated symplectic singularity has recently been described in \cite{bellamy2023new}. Secondly, it requires the knowledge of all possible quiver realisation of the elementary slices. For example, the (closure of the) minimal nilpotent orbit of $E_6$ has up to now four known magnetic quiver realisations: a unitary affine Dynkin quiver, a unitary twisted affine Dynkin quiver, an orthosymplectic quiver \cite{Bourget:2020gzi,Bourget:2020xdz}, and a folded orthosymplectic quiver \cite{Bourget:2021xex}.
Thirdly, quiver subtraction is only partially understood in the case of repeated identical transitions --- leading to \emph{decorated quivers} \cite{Bourget:2022ehw}, which are, for example, relevant for moduli spaces involving multiple instantons. At present, it is unclear how to define the Coulomb branch of the decorated quiver; however, the quiver subtraction algorithm supplemented by decorations passed numerous consistency checks.

\begin{figure}[t]
    \centering
    \includegraphics[page=2]{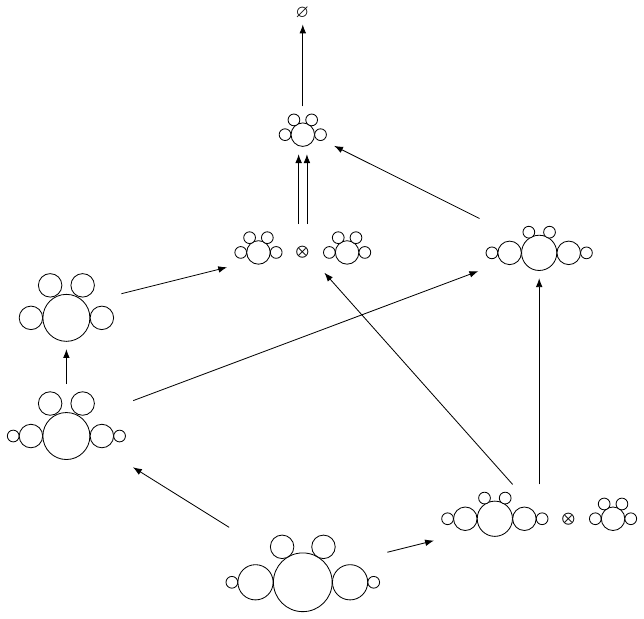}
    \caption{Cartoon of decay and fission of quivers: The circle diameters symbolise the ranks. Upon a decay, one or several circles decrease in size; while a fission leads to the splitting into two.}
    \label{fig:intro_cartoon}
\end{figure}

This paper serves to fill this gap by expanding and further developing the \emph{decay and fission} algorithm \cite{Bourget:2023dkj} for magnetic quivers, whereby the diagram is generated by recursively Higgsing, i.e.\ starting from the bottom leaves. In a nutshell, the main results of this paper are: 
\begin{itemize}
    \item The Higgs branch phase diagram is shown to coincide with the Hasse diagram of simple objects in convex linear algebra. It does not require any a priori knowledge of the list of possible elementary transitions. 
    \item Elementary slices, which correspond to elementary Higgsing phase transitions, are associated to one of two fundamental processes that magnetic quivers can undergo: 
    \begin{itemize}
        \item \emph{Decay}, where the magnetic quiver reduces to one with smaller rank. There are infinitely many decay types, which correspond geometrically to infinitely many elementary slices. 
        \item \emph{Fission}, where the magnetic quiver splits into two parts, while preserving the total rank. There are only two fission types, and correspondingly two possible elementary slices.    
    \end{itemize}
    Figure \ref{fig:intro_cartoon} displays a cartoon version of these transitions.
    \item The decay transitions above can be implemented practically: upon inspection of balanced subquivers and over-balanced nodes, decay transitions can be read off.
\end{itemize}

The \emph{decay and fission} algorithm not only reproduces the results of the standard quiver subtraction, but also allows to deduce the daughter theories $\Tcal_i$ and granddaughter theories $\Tcal_i'$ etc. obtained from partial Higgsing. 
Of course, the information within the magnetic quiver does not comprehensively encode all aspects of $\Tcal$. To be more accurate, the decay and fission algorithm enables one to deduce the Higgs branches for $\Tcal_i$, $\Tcal_i'$, and so forth.
Understanding these Higgs branches provides extensive information which almost always enables one to identify the theories after Higgsing through existing literature. If not, the absence of literature suggests a potentially new theory, where the algorithm predicts its Higgs branch structure.

The true strength of this algorithm lies in its inherent simplicity.
In contrast to standard quiver subtraction, which needs to introduce additional gauge groups with successive subtractions, the decay and fission algorithm never requires new nodes and simplifies the magnetic quivers with each step. 
This characteristic not only streamlines the process, but also facilitates the creation of a computer algorithm. 
Furthermore, when addressing complicated quivers, the full phase or Hasse diagram can often be exceedingly complex, rendering it less practical for analysis. The decay and fission algorithm, serving as a partial Higgsing technique, eliminates the need to construct the entire phase diagram for valuable insights. For instance, one can halt the procedure as soon as one arrives at theories that one is already familiar with. 

\paragraph{Comparison to other methods of Higgsing.}
For supersymmetric theories with 8 supercharges, there are many different approaches to Higgsing. For instance, the Higgsing of class $\mathcal{S}$ theories via closing of partial punctures \cite{Chacaltana:2012zy}. However, the short-comings of this and most other Higgsing algorithms include: 
\begin{enumerate}[a)]
   \item the Higgsing may not be minimal,
   \item the exact nature (e.g.\ a Kleinian singularity) of the transverse space is not known,
   \item it does not contain some of the more unconventional Higgsings (e.g.\ transitions that causes the magnetic quiver to fission).
\end{enumerate}
The decay and fission algorithm addresses all these short-comings and still remains a very simple and straightforward algorithm that can be applied to any SUSY theory with 8 supercharges that has a known unitary magnetic quiver.

\paragraph{Organisation of the paper.}
In this paper, we demonstrate that the infinite richness of the (generalised) Higgs mechanism in theories with 8 supercharges follows from a unique simple construction, provided the Higgs branch admits a unitary magnetic quiver realisation.
The goal of Section \ref{sec:decay_fission} is to spell out this construction and to provide useful tips to implement it in practice. Thereafter, Section \ref{sec:examples} provides a substantial list of examples from a large landscape of physical theories in various space-time dimensions. In Section \ref{sec:conclusion}, we conclude and discuss open challenges.

\paragraph{Notation.}
This paper focuses on two diagrammatic techniques: magnetic quivers and Hasse diagrams. Throughout, the following conventions are used:
\begin{itemize}
   \item \ul{Magnetic quivers:} The unframed quiver graph is composed of nodes and edges. Nodes $\bigcirc$ denote unitary gauge nodes (a 3d $\Ncal=4$ vector multiplet) with the rank indicated next to it. Simply laced edges $\bigcirc \text{---} \bigcirc$ denote bifundamental hypermultiplets. Non-simply laced egdes are understood as in \cite{Cremonesi:2014xha}. \\
   The balance for a unitary gauge group $\urm(k)$ is given by $b=\sum_ix_in_i-2k$ where $n_i$ is the rank of the adjacent gauge/flavor nodes that are connected to $\urm(k)$ with a multiplicity $x_i$ edge.
   We also use black nodes to denote \emph{overbalanced} nodes ($b>0$) and white nodes for \emph{balanced} nodes ($b=0$). The magnetic quivers are always presented as unframed (i.e.\ without flavour nodes), meaning that it is implied that an $\mathrm{U}(1)_{\mathrm{diag}}$ is always decoupled. 

   \item \ul{Hasse diagrams:} This graph, again composed of vertices and edges, encodes the partially ordered set of symplectic leaves that constitute the finite stratification of a symplectic singularity. 
   The symplectic leaves $\Lcal_i$ are the vertices $\bullet$ of the diagram. Between any pair $(\Lcal_i,\Lcal_j)$ of partially order leaves $\Lcal_i < \Lcal_j$, meaning $\Lcal_i \subset \overline{\Lcal}_j$, there exists a transverse slice $\Scal_{i,j}$. Pictorially, the slice is indicated as line between the two adjacent vertices.
Importantly, for conical symplectic singularities, there is a unique lowest leaf $\Lcal_0$ which consists of a single point. The slice $\Scal_{0,i} = \overline{\Lcal}_i$ is then the closure of the non-trivial leaf $\Lcal_i$.

   In this work, the Hasse diagram is oriented with $\Lcal_0$ at the bottom.  In terms of the physics of RG-flows, the vertex at the bottom denotes the ``mother theory'' and the other vertices are ``daughter/granddaughter'' theories that can be reached from the mother theory via the Higgs mechanism. We note that in some physics literature the mother theory is placed at the top instead; thus, leading to an inverted picture.
\end{itemize}

\clearpage

\FloatBarrier

\section{Decay and fission}
\label{sec:decay_fission}

All the results of this paper rely on one simple idea, which was presented in essence in the letter \cite{Bourget:2023dkj}, and in much more detail, and with important additions, in this section. 
To help readers  get familiar with this idea, we provide an intuitive description in Section~\ref{sec:informal}. Combined with the practical realisation in Section~\ref{sec:algorithm}, most of the examples in Section~\ref{sec:examples} can then be worked out. Section~\ref{sec:formal} gives a precise (but maybe somewhat indigest) formulation of the fission and decay algorithm. It can be skipped at first reading, but it is necessary to grasp fully certain peculiarities discussed in Section~\ref{sec:examples}. Finally, an implementation in \texttt{Mathematica} is provided at the address
\begin{center}
    \href{https://www.antoinebourget.org/attachments/files/FissionDecay.nb}{https://www.antoinebourget.org/attachments/files/FissionDecay.nb}
\end{center}

\subsection{Intuitive statement}
\label{sec:informal}

\paragraph{An example with decay only. }
Consider a good unitary quiver $\MQ$, seen as a magnetic quiver for the Higgs branch of a given theory. For instance, 
\begin{align}
    \raisebox{-.5\height}{
\includegraphics[page=1]{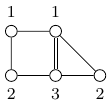}
} \label{eq:example1}
\end{align}
We say the quiver is \emph{good} if each node has non-negative balance. 

\begin{figure}
    \centering
    \begin{tabular}{ccc}
\raisebox{-.5\height}{\includegraphics[page=3]{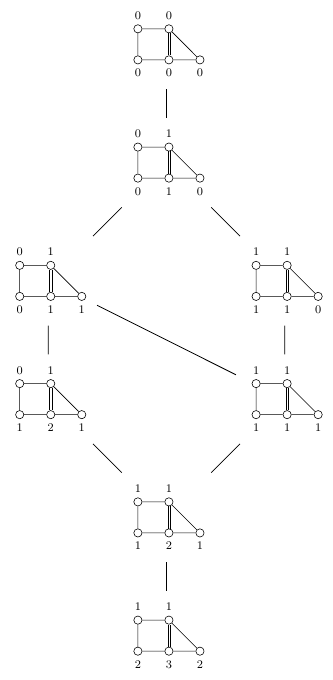}} & \hspace{1cm} & \raisebox{-.5\height}{\includegraphics[page=8]{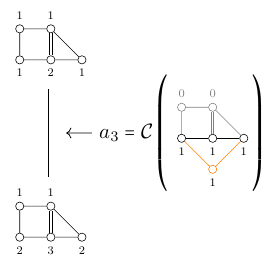}}
    \end{tabular}
    \caption{Hasse diagram obtained from decays of the quiver \eqref{eq:example1} (left), and example of the subsequent computation of the geometry of an elementary transverse slice (right). }
    \label{fig:hasseExample}
\end{figure}

We then compile the list of all good quivers with the same shape,\footnote{As explained in Section \ref{sec:formal}, one needs to remove from this list the quivers that correspond to moduli spaces of instantons.} and with ranks on each node smaller or equal to those in $\MQ$  --- such quivers are said to be smaller or equal to $\MQ$, and this defines a partial order $\leq$. In the example \eqref{eq:example1}, we find exactly 8 such quivers, namely: 
\begin{align}
    \raisebox{-.5\height}{
\includegraphics[page=2]{figures_fission_decay.pdf}
}
\label{eq:8quivers}
\end{align}
Note that if we add up any two non-zero quivers in this list, we end up with a quiver which is not smaller or equal to $\MQ$. This means that the quiver cannot fission (the next example contains fissions). Therefore, it can only \textbf{decay}, and the decay products are precisely the 8 quivers above. The partial order between them can be summarized using a Hasse diagram,  obtained by comparing in all possible ways the 8 quivers in \eqref{eq:8quivers}, which gives 25 pairs out of the $\binom{8}{2} = 28$ possible pairs of distinct quivers, and deleting a pair $\MQ_1 < \MQ_2$ if there exists $\MQ_3$ such that $\MQ_1 < \MQ_3 < \MQ_2$. The diagram is shown on the left of Figure \ref{fig:hasseExample}.

We claim this is the Coulomb branch diagram for $\MQ$ in \eqref{eq:example1}. The nature of the transverse slices can be read out from the difference of the two quivers at the ends of each edge. More precisely, align the two quivers, subtract the smaller one from the larger one, and rebalance;\footnote{Again, the precise way of rebalancing is given in Section \ref{sec:formal}. In general, it can involve a non-simply laced edge. } then the Coulomb branch of the rebalanced quiver is the transverse slice. This is illustrated on the right of Figure \ref{fig:hasseExample}. 

\begin{figure}
    \centering
\includegraphics[page=9]{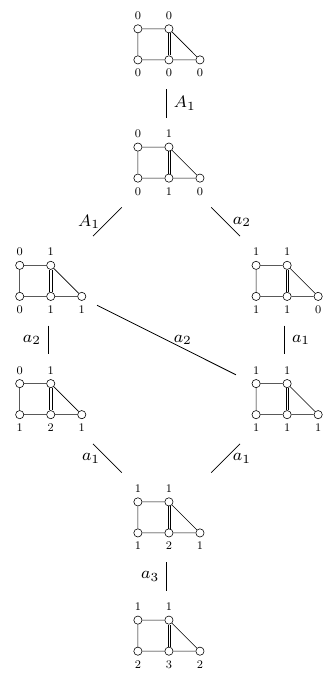}
    \caption{Diagrams of decays from Figure \ref{fig:hasseExample} with the geometry of elementary transverse slices. }
    \label{fig:hasseExampleComplete}
\end{figure}

We insist on the fact that the Hasse diagram is obtained first, and the elementary transitions are \emph{computed} in a second step. This is the main difference with the quiver subtraction algorithm, which is an iterative process in which elementary transitions are an \emph{input}. Repeating a computation similar to the right of Figure \ref{fig:hasseExample} for all elementary transitions, one obtains the final diagram of Figure \ref{fig:hasseExampleComplete}. 

\begin{figure}
    \centering
    \begin{tabular}{ccc}
\raisebox{-.5\height}{\includegraphics[page=7]{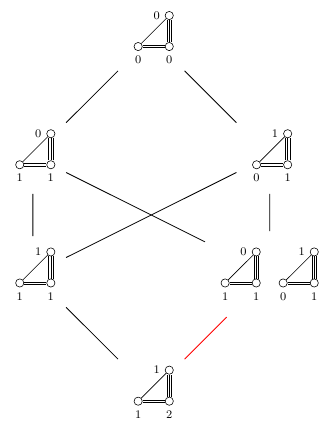}
} & \hspace{1cm} & \raisebox{-.5\height}{\includegraphics[page=10]{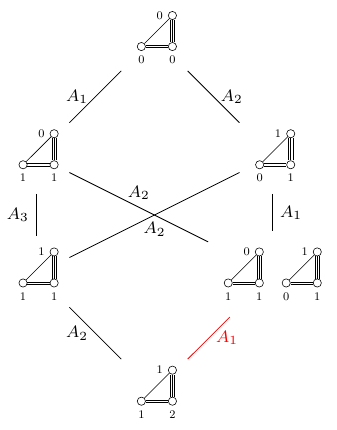}}
    \end{tabular}
    \caption{Hasse diagram obtained from decays and fissions of the quiver \eqref{eq:example2} (left), and the same diagram after the geometry of the transverse slices has been computed (right). The red line denotes the only fission transition, all other transitions are decays. }
    \label{fig:hasseExample2}
\end{figure}

\paragraph{An example with fission. }
It can also happen that the sum of two or more non-zero good quivers that are $\leq \MQ$ is also $\leq \MQ$. When this is the case, the quiver (or one of its decay products) can \textbf{fission} successively into smaller parts, and each part keeps decaying and fissioning, following the same rules. For instance, consider the quiver 
\begin{align}
    \raisebox{-.5\height}{
\includegraphics[page=4]{figures_fission_decay.pdf}
}
\label{eq:example2}
\end{align}
There are exactly 5 good lower quivers: 
\begin{align}
    \raisebox{-.5\height}{
\includegraphics[page=5]{figures_fission_decay.pdf}
}
\label{eq:example2V}
\end{align}
and among them, there is a pair of quivers whose sum gives the original quiver. This means the latter can fission: 
\begin{align}
    \raisebox{-.5\height}{
\includegraphics[page=6]{figures_fission_decay.pdf}
}
\label{eq:example2pair}
\end{align}
The Coulomb branch Hasse diagram of \eqref{eq:example2} is therefore given in the left part of Figure \ref{fig:hasseExample2}. 
The nature of the transverse slice corresponding to fission is obtained by comparing the greatest common divisors of the ranks of the fission products (for each of the 8 quivers in \eqref{eq:8quivers}, this gcd is equal to 1).  When they are equal, the transverse slice has an $A_1$ geometry (this is the case here). When they are different, the geometry is the non-normal slice $m$ \cite{vinberg1972class,fu2017generic}. In the present case, the red transition in Figure \ref{fig:hasseExample2} corresponds to $A_1$, while all other transitions are computed using the rebalancing as in Figure \ref{fig:hasseExample}. We finally get the result shown in the right part of Figure \ref{fig:hasseExample2}.

Note that prominent examples for fission arises from $k$-instanton moduli spaces with $k>1$ \cite{Cremonesi:2014xha}, $\Scal$-fold theories \cite{Bourget:2020mez}, magnetic quivers of rank 2 4d SCFTs \cite{Bourget:2021csg}, 6d $\Ncal=(1,0)$ (higher rank) E-string theories \cite{Cabrera:2019izd}, etc. Even a simple class $\mathcal{S}$ theory such as $T_6$ possesses this feature. Section \ref{sec:examples} contains a selection of illustrative examples.

\subsection{Formal statement}
\label{sec:formal}

The previous subsection gave an overview of the algorithm. We now provide all the technical details, beginning with introducing the appropriate formalism. 

Let $\MQ$ be unitary quiver. It is convenient to encode a quiver using a matrix, which encodes the underlying graph, and a rank vector, which specifies the ranks of the unitary groups at each node. This leads us to the following definitions.

\paragraph{Definitions. } 
Let $n$ be a positive integer. We define a partial order on $\mathbb{Z}^n$ as follows. We say that $K \in \mathbb{Z}^n$ is \emph{non-negative}, written $K \geq 0$, if all its entries are non-negative, i.e.\ if $K \in \mathbb{N}^n$. Then, given two elements $K_1 , K_2 \in \mathbb{Z}^n$, we say that $K_1 \leq K_2$ if $K_2 - K_1 \geq 0$. 

A quiver $\MQ$ with $n$ nodes is a pair $(A,K)$, where $A$ is an $n \times n$ matrix with coefficients in $\mathbb{Z}$ and $K \in \mathbb{N}^n$ is the rank vector, such that
\begin{itemize}
    \item $A$ has diagonal coefficients $A_{i,i} = -2 + 2g$ where $g \in \mathbb{N}$.
    \item The off-diagonal coefficients $A_{i,j}$ of $A$ are non-negative integers with either $A_{i,j} = A_{j,i} = 0$ or $A_{i,j}   A_{j,i} \neq 0$, and in this last case, either $A_{i,j} | A_{j,i}$ or $A_{j,i} | A_{i,j}$. 
    \item The \emph{balance vector} $B \coloneqq AK$ has non-negative entries $B_i \geq 0$. 
\end{itemize}
When $A_{i,j}=A_{j,i}$, we draw $A_{i,j}$ links between nodes $i$ and $j$. When $0 < A_{i,j} | A_{j,i}$, we draw $A_{i,j}$ arrows, each of them $\frac{A_{j,i}}{A_{i,j}}$-laced, from node $j$ to node $i$. The balance $b$ of the $i$-th node is the $i$-th entry of the balance vector, $b = B_i$. In some drawings, when no confusion is possible, only the nodes with $K_i >0$ are represented, with the value $K_i$ indicated next to the node. 

We say that a quiver is \emph{reducible} if it contains a $\urm(1)$ node on the long side such that, when this node is deleted, the quiver breaks into several connected components. We assume the quiver under consideration is \emph{not reducible}.\footnote{When a quiver is reducible, the corresponding theory is a product, and all its features factorise. Therefore, we focus on irreducible quivers, and it is always implied that whenever a reducible quiver is reached in the algorithm, it should be reduced and each irreducible part should be studied separately. }

\paragraph{Leaves. } 
Given a quiver, defined by $(A,K)$, let $K' \in \mathbb{Z}^n$ such that $0 \leq K' \leq K$ and $AK' \geq 0$. We define the property 
\begin{equation}
    \mathcal{P}_1 (K') = \left[ (K')^t A (K') \geq 0 \textrm{  and  } (K')^t   (K') = 1 \right]
\end{equation}
This property $\mathcal{P}_1$ simply detects the subquivers of the form U(1) with one or more adjoints.
We also define the property $\mathcal{P}_2 (K')$ as follows. First, define for $1 \leq i \leq n$ the vector $\delta (i) \in \mathbb{N}^n$ by $\delta (i)_j = \delta_{i,j}$. Then $  \mathcal{P}_2 (K')$ is true if there exists a \emph{long node} index $1 \leq i \leq n$ such that $K'_i>0$ and, defining $K'' = K' - \delta(i)$, for all $j$, $(AK'')_j \times (K'')_j=0$ (no sum on $j$) and $\mathrm{gcd}(K'') = (K'')^t A \delta(i)$. This detects the subquivers that correspond to instanton moduli spaces.\footnote{We can unpack condition $\mathcal{P}_2 (K')$ as follows. The idea is that a quiver corresponds to moduli space of instantons if there is a $\urm(1)$ node which is connected to a subquiver by one simply laced edge, and that subquiver is by itself fully balanced after deletion of the $\urm(1)$ node. }

Consider now the finite set 
\begin{equation}
    \mathcal{V} = \{  K' \; | \;  K' \in \mathbb{N}^n \, , \quad  K' \leq K \, , A K' \geq 0 \, , \quad \textrm{not} \,  \mathcal{P}_1 (K')  \, , \quad  \textrm{not} \,   \mathcal{P}_2 (K')  \}  
    \label{eq:decay_fission_products}
\end{equation} 
and the set
\begin{equation}
    \mathcal{V}_0 = \{ K' \in \mathcal{V} \; | \; K' \neq 0 \} \, . 
\end{equation}
The elements of $\mathcal{V}$ are the possible decay and fission products. We now just have to assemble them in all possible ways. We define, $\mathcal{L}_0 = \{ \emptyset \}$ and for $m  \in \mathbb{N}_{>0}$, 
\begin{equation}
    \mathcal{L}_m = \{ \multiset{ K'_1 , \dots , K'_m  }  \quad| \quad   \forall\; 1 \leq j \leq m \, , \, K'_j \in \mathcal{V}_0 \textrm{ and } K'_1 + \dots + K'_m \leq K \} \, . 
\end{equation}  
Elements of $\mathcal{L}_m$ are the multisets\footnote{A multiset can be seen as a set where elements can have multiplicities, i.e.\ appear more than one time, or equivalently, as a list up to permutation.  We denote multisets with the symbols $\multiset{\cdots}$. For instance, $\multiset{5,5,2} = \multiset{5,2,5} \neq \multiset{5,2}$. The multiplicity of 5 in  $\multiset{5,5,2}$ is 2, and the multiplicity of 2 is 1.   } of $m$ vectors of $\mathcal{V}_0$ whose sum is $\leq K$ (repetitions are allowed).\footnote{For our example \eqref{eq:example2}, $\mathcal{V}$ consists of the five vectors listed in \eqref{eq:example2V}, $\mathcal{V}_0 \simeq \mathcal{L}_1$ consists of the four vectors on the right of \eqref{eq:example2V}, and $\mathcal{L}_2$ has only one element, namely the pair that corresponds to \eqref{eq:example2pair}. All $\mathcal{L}_{m >2}$ are empty. } Finally, we write 
\begin{equation}
    \mathcal{L} = \bigcup\limits_{m \in \mathbb{N}} \mathcal{L}_m \, . 
    \label{eq:set_of_leaves}
\end{equation}
This is the set of vertices of the Hasse diagram, which correspond to the symplectic leaves of the 3d Coulomb branch of the initial quiver $(A,K)$. Note that $\mathcal{L}_m = \emptyset$ for $m >> 1$, so the above union is finite. For an element $\ell \in \mathcal{L}$, there is a unique $m \in \mathbb{N}$ such that $\ell \in \mathcal{L}_m$, we call it the \emph{length} of $\ell$, denoted $\mathrm{length}(\ell)$. We also denote by $\Sigma \ell \in \mathbb{N}^n$ the sum of the elements of $\ell$. 

\paragraph{Partial order. }
We now have to define a partial order on $\mathcal{L}$. Let $\ell_1 , \ell_2 \in \mathcal{L}$. We write $\ell_1 \rightsquigarrow \ell_2$ if 
\begin{equation}
    | \mathrm{length}(\ell_1) - \mathrm{length}(\ell_2) | \leq 1 \textrm{ and } |\ell_1 \cap \ell_2| =  \mathrm{length}(\ell_1) - 1  \textrm{ and } \Sigma \ell_1 \geq \Sigma \ell_2 \, . 
    \label{eq:partialOrder}
\end{equation}
The relation $\rightsquigarrow$ is reflexive and antisymmetric, but not transitive in general. Let us denote by $\succcurlyeq$ its transitive closure, i.e.\ $\ell_1 \succcurlyeq \ell_2$ if there exist a chain $\ell_1 \rightsquigarrow \dots \rightsquigarrow \ell_2$. This is a partial order relation. We claim that $(\mathcal{L} , \succcurlyeq)$ coincides with the poset of symplectic leaves in the 3d Coulomb branch of the quiver. This concludes our construction of the Hasse diagram.

\paragraph{Elementary transitions.}
The last ingredient we need is the geometry of the transverse slice between two adjacent leaves in the partial order, which is a minimal degeneration. Let $\ell_1 \succcurlyeq \ell_2$ be two \emph{adjacent} leaves, i.e.\ such that $\ell_1 \neq \ell_2$ and there is no leaf $\ell_3 \neq \ell_1 , \ell_2$ satisfying $\ell_1 \succcurlyeq \ell_3 \succcurlyeq \ell_2$. Since they are adjacent, they satisfy \eqref{eq:partialOrder}. There are three possibilities, depending on the value of $\mathrm{length}(\ell_1) - \mathrm{length}(\ell_2)$: 
\begin{itemize}
    \item If $\mathrm{length}(\ell_1) - \mathrm{length}(\ell_2) = +1$, then $\ell_2 \subset \ell_1$. The transition corresponds to the Coulomb branch of the unique element in $\ell_1 \backslash \ell_2$; let $\mu$ be its multiplicity in the multiset $\ell_1$. This is a \emph{terminal decay}: one quiver disappears entirely. The geometry of the transition is simply given by a union of $\mu$ copies of the Coulomb branch of the vanishing quiver.  
    \item If $\mathrm{length}(\ell_1) - \mathrm{length}(\ell_2) = 0$, there is a unique $K_1 \in \ell_1 \backslash \ell_2$ and a unique $K_2 \in \ell_2 \backslash \ell_1$. Let $\mu$ be the multiplicity of $K_1$ in the multiset $\ell_1$. The transition is a non terminal decay from $K_1$ to $K_2$. Let $\MQ$ be the quiver obtained from $K_1 - K_2$ by rebalancing using one additional $\urm(1)$ node with an adjoint for each connected component $K_2^{\alpha}$ of $K_2$: the rebalancing is done using $\mathrm{gcd}(K_2^{\alpha})$-laced edges, pointing towards the new $\urm(1)$ node. Then the geometry of the transition is the union of $\mu$ copies of the Coulomb branch of $\MQ$.\footnote{Several examples of such non-trivial rebalancings are shown in Section \ref{sec:LST}. }  
    \item If $\mathrm{length}(\ell_1) - \mathrm{length}(\ell_2) = -1$, there is a unique $K_1 \in \ell_1 \backslash \ell_2$ and exactly two vectors $K_2 , K'_2 \in \ell_2 \backslash \ell_1$. This corresponds to \emph{fission}. Let $\mu$ be the multiplicity of $K_1$ in the multiset $\ell_1$. If the quiver contains 0 or 1 loop, the geometry of the transition is $\mu \cdot A_1$ if $\mathrm{gcd}(K_2) = \mathrm{gcd}(K'_2)$, and $\mu \cdot  m$ if $\mathrm{gcd}(K_2) \neq \mathrm{gcd}(K'_2)$.  For two loops or higher, a generalization is needed and left for future work.\footnote{A reasonable guess, based on the first line of Table \ref{tab:smallQuivers}, is that $\mu \cdot A_1$ should be extended to $\mu \cdot D_{g+1}$ for a genus $g \geq 2$ quiver. } 
\end{itemize}

\paragraph{Derivation.} 
We do not have a formal proof that the algorithm presented here is correct. Rather, we have used a physical, almost experimental approach, and this should be kept in mind. We have combined insights from various perspectives in order to infer general rules, that we have tested in as many cases as possible, finding in all cases perfect agreement. Specifically, the algorithm was built from 
\begin{itemize}
    \item Comparison with the quiver subtraction algorithm (including decorations, if needed);
    \item Physical intuition coming from 3d mirror symmetry and brane physics, as reviewed in Section \ref{sec:3d} -- for simple classes of quivers, this constitutes a proof of the algorithm; 
    \item Agreement with the Higgs mechanism when a weakly coupled description is available, as exemplified in Section \ref{sec:CompleteExample}. Our results are  also compatible with partial results in class $\mathcal{S}$ theories where a subset of Higgsings can be done via partial closing of punctures.  Fissions are precisely the Higgsings of class $\mathcal{S}$ theories that are not given by partially closing  punctures;
    \item Agreement with results in the mathematical literature, in particular for symmetric products and affine Grassmannian slices. 
\end{itemize}

\subsection{The decay algorithm in practice}
\label{sec:algorithm}

We now have described how to compute the Hasse diagram of the Coulomb branch of any good unitary quiver. 
The algorithm is very general, and requires no input other than the initial quiver. However, it can be difficult to implement without the help of a computer, as the list of good subquivers to consider to build $\mathcal{V}$ (see \eqref{eq:decay_fission_products}) can be very large. Geometrically, it boils down to finding integral points in a convex cone in $\mathbb{R}^n$. 
In this subsection we describe a convenient shortcut that applies to a class of simple quivers, namely those for which only decay can occur, but never fission. This allows to compute the diagrams efficiently by hand for these simple quivers. 

Consider a quiver with unitary gauge nodes only, potentially non-simply laced.
Locate all the gauge nodes that have balance $b=0$. These balanced connected subquivers take the form of a union of finite Dynkin diagrams.\footnote{One can prove, by arguments similar to \cite{Nekrasov:2012xe}, that balanced (sub-)quivers with unitary gauge groups need to take the shape of a finite Dynkin diagram.} Assuming first that the quiver cannot fission into two good quivers, the decay transitions can in practice be implemented by the following subtractions\protect\daggerfootnote{In some talks given by ZZ, this was previously called the ``Inverted Quiver Subtraction'' algorithm.}:   
\begin{enumerate}[(1)]
	\item \label{Rule:1} \textbf{A-type Kleinian singularity.}
	For an overbalanced $\urm(k)$ node ($b>0$), decay simply turns $\urm(k) \rightarrow \urm(k-1)$ and the transition is an $A_{b+1}$ Kleinian singularity. This subtraction is only allowed if the quiver does not have any bad/ugly nodes ($b<0$) after this subtraction. 
	\item \label{Rule:2} \textbf{Closure of minimal $\mathfrak{g}$ orbit (one instanton).}
	From a balanced, connected Dynkin-type subquiver, subtract the respective weighted finite Dynkin diagram of the Lie algebra $\mathfrak{g}$, see Table \ref{tab:Dynkin}.  The transverse space of this transition is then the one-$\mathfrak{g}$ instanton moduli space, provided there is no enhancement.
	
	\item \label{Rule:3} \textbf{$h_{n,k}$ and $\overline{h}_{n,k }$ singularities.} Suppose there exists a linear chain of $n-1$ balanced nodes on the short side of a $k$-laced edge which is connected to a node on the long side with balance $b$.
	  \begin{enumerate}
		 \item \label{rule:NonSim1}  For $b=k-2$, subtract the $n$ node quiver $(1)-(1)-\ldots-(1)\xLeftarrow{k} (1)$ with a $k$-laced edge, such that the transverse slice is the $h_{n,k}$ singularity, see Table \ref{tab:non-sim_quivers}. 
    \item \label{rule:NonSim2} For $b=k-1$, subtract the $n$ node quiver $(1)-(1)-\ldots-(1)\xLeftarrow{k}(1)$ with a $k$-laced edge such that the transverse slice is the $\overline{h}_{n,k}$ singularity, see Table \ref{tab:non-sim_quivers}. 
    \end{enumerate}
\end{enumerate}
Note that not all the transverse slices obtained from decay that appear in this paper are of the types in  (\ref{Rule:1})-(\ref{Rule:3}). However, the algorithm is still sensitive to these other slices. If one performs the algorithm (\ref{Rule:1})-(\ref{Rule:3}) and ends up with an ugly quiver, which contains a decoupled free hypermultiplet(s), the free hypermultiplet(s) enhance the transverse slice; for instance, into one of the more exotic slices in Table \ref{tab:smallQuivers}.

The rules above provide a full list of decay products. One can then work out the possible fissions from there, by combining these decay products in all possible ways compatible with the original quiver.

\begin{table}[]
    \centering
    \begin{subtable}[t]{0.7\textwidth}
    \begin{tabular}{cc|cc}
    \toprule
    \multicolumn{2}{c|}{classical algebras} & 
        \multicolumn{2}{c}{exceptional algebras} \\ \midrule
        $A_n$ & \raisebox{-.5\height}{\includegraphics[page=1]{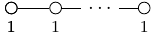}} & $E_6$ & 
        \raisebox{-.5\height}{\includegraphics[page=5]{figures_inverted_sub_algo.pdf}}\\
         $B_n$ & \raisebox{-.5\height}{\includegraphics[page=2]{figures_inverted_sub_algo.pdf}} &  $E_7$ & 
        \raisebox{-.5\height}{\includegraphics[page=6]{figures_inverted_sub_algo.pdf}} \\
          $C_n$ & \raisebox{-.5\height}{\includegraphics[page=3]{figures_inverted_sub_algo.pdf}} &  $E_8$ & 
        \raisebox{-.5\height}{\includegraphics[page=7]{figures_inverted_sub_algo.pdf}} \\
           $D_n$ & \raisebox{-.5\height}{\includegraphics[page=4]{figures_inverted_sub_algo.pdf}} &  $F_4$ & 
        \raisebox{-.5\height}{\includegraphics[page=8]{figures_inverted_sub_algo.pdf}} \\
         & &  $G_2$ & 
        \raisebox{-.5\height}{\includegraphics[page=9]{figures_inverted_sub_algo.pdf}} \\ \bottomrule
    \end{tabular}
    \caption{}
    \label{tab:Dynkin}
\end{subtable}
\begin{subtable}[t]{0.28\textwidth}
    \centering
    \begin{tabular}{c}
    \toprule
       $h_{n,k}$ and $\overline{h}_{n,k}$  \\\midrule
         \raisebox{-.5\height}{\includegraphics[page=15]{figures_inverted_sub_algo.pdf}}\\
         \bottomrule
    \end{tabular}
    \caption{}
    \label{tab:non-sim_quivers}
    \end{subtable}
    \caption{\subref{tab:Dynkin} Dynkin diagrams of classical and exceptional Lie algebras. The numbers at the nodes are the dual Coxeter labels.
   \subref{tab:non-sim_quivers}: For both cases the non-simply laced quiver contains $n$ $\urm(1)$ nodes, from which the first $n-1$ are balanced and on the short side of the $k$-laced edge. The long $\urm(1)$ node has balance $b=k-2$ and $k-1$ for $h_{n,l}$ and $\overline{h}_{n,k}$, respectively.
    }
    \label{tab:rules}
\end{table}

\FloatBarrier

\subsection{A complete example}
\label{sec:CompleteExample}

\begin{sidewaysfigure}
    \centering
    \begin{subfigure}[t]{0.37\textwidth}
    \centering
  \includegraphics[page=2]{figures_inverted_sub_splitting.pdf}
  \caption{Decay and fission}
    \label{fig:splitting_inverted}
    \end{subfigure}
    \hfill
    \begin{subfigure}[t]{0.37\textwidth}
    \centering
  \includegraphics[page=3]{figures_inverted_sub_splitting.pdf}
  \caption{Quiver subtraction}
    \label{fig:splitting_standard}
    \end{subfigure}
    \hfill
      \begin{subfigure}[t]{0.22\textwidth}
    \centering
  \includegraphics[page=1]{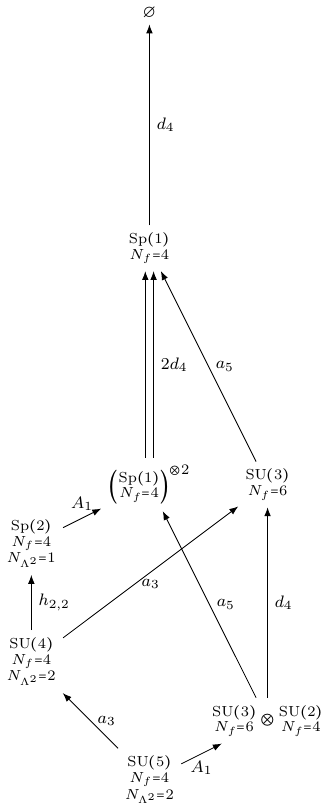}
  \caption{Partial Higgsing}
  \label{fig:splitting_Higgs}
    \end{subfigure}
    \caption{Contrasting decay and fission, quiver subtraction, and partial Higgs mechanism in the case of an $\surm(5)$ theory and its magnetic quiver. Here, $2d_4 = d_4 \cup d_4$ denotes a union, and in (\subref{fig:splitting_standard}) green denotes ``decoration''.}
    \label{fig:splitting_ex}
\end{sidewaysfigure}

Now, it is time to show-cast the algorithm on an example that contains fission and decay, allows for a partial Higgsing interpretation, and can be compared to standard quiver subtraction (with all its subtleties).

As an example, consider the (electric) $\surm(5)$ gauge theory\footnote{For concreteness, consider this as a 3d $\Ncal=4$ theory.} with $N_f=4$ hypermultiplets in the fundamental representation and $N_{\Lambda^2}=2$ hypermultiplets in the traceless 2nd rank anti-symmetric representation. Its magnetic quiver is given by\footnote{In spirit, this example is a truncated version of the magnetic quiver for the little string theory with an effective description given by an $\surmL(5)$ gauge algebra supported on a curve of self-intersection $0$, see \cite{Mansi:2023faa}. In this example, the number of flavours is reduced to 4, which is well-defined as 3d $\Ncal=4$ theory. } 
\begin{align}
\raisebox{-.5\height}{
    \includegraphics[page=4]{figures_inverted_sub_splitting.pdf}}
    \label{eq:complete_ex_MQ}
\end{align}
As a first step, the decay and fission algorithm is applied to \eqref{eq:complete_ex_MQ}, which outputs the Hasse diagram in Figure \ref{fig:splitting_inverted}; which was displayed in the introductory cartoon of Figure \ref{fig:intro_cartoon}. 
The next step is to run the standard quiver subtraction algorithm on \eqref{eq:complete_ex_MQ}, which then results in the Hasse diagram in Figure \ref{fig:splitting_standard}.
As claimed above, the two Hasse diagrams agree: i.e.\ the same number of leaves and the same minimal transverse slices in between them. Moreover, the drastic difference in magnetic quivers obtained by the two algorithms becomes apparent. Fission and decay keeps the shape and reduces ranks at each step. In contrast, quiver subtraction forces us to include additional nodes due to re-balancing, requires decoration (here in green), or requires non-simply laced edges.  

The advantage of this example is that one can analyse branching rules for the $\surm(5)$ gauge theory and directly follow the partial Higgs mechanism. Let us focus one the two transitions that the $\surm(5)$ theory can undergo. There exists a partial Higgs mechanism
\begin{align}
\begin{pmatrix}
\surm(5) \\
N_f=4 \\
N_{\Lambda^2}=2
\end{pmatrix} 
\longrightarrow 
\begin{pmatrix}
\surm(4) \\
N_f=4 \\
N_{\Lambda^2}=2
\end{pmatrix} 
\end{align}
which is straightforward in terms of branching rules $\surm(5)\to\surm(4)$:
\begin{subequations}
 \begin{align}
[1,0,0,0] &\to [1,0,0] + [0,0,0] \\
[0,1,0,0] &\to [0,1,0] + [1,0,0] \\
[1,0,0,1] &\to [1,0,1] +   [1,0,0] + [0,0,1] + [0,0,0]
 \end{align}
\end{subequations}
where irreps are labelled by their Dynkin labels. This Higgsing is a decay in terms of the magnetic quiver \eqref{eq:complete_ex_MQ} and shown as the $a_3$ transition of \eqref{eq:complete_ex_MQ} in Figure \ref{fig:splitting_inverted}. The ``decay product'' is, in fact, the magnetic quiver for the $\surm(4)$ gauge theory.

Additionally, there exists a partial Higgs mechanism of the form
\begin{align}
\begin{pmatrix}
\surm(5) \\
N_f=4 \\
N_{\Lambda^2}=2
\end{pmatrix} 
\longrightarrow 
\begin{pmatrix}
\surm(3) \\
N_f=6 
\end{pmatrix} 
\otimes
\begin{pmatrix}
\surm(2) \\
N_f=4 
\end{pmatrix} 
\end{align}
which can be verified by inspecting branching rules $\surm(5) \to \surm(3) \times \surm(2)$:
 \begin{subequations}
 \begin{align}
[1,0,0,0] &\to [1]\otimes [0,0] + [0]\otimes [1,0] \\
[0,1,0,0] &\to [0]\otimes [0,0] + [0]\otimes [0,1] + [1] \otimes [1,0] \\
[1,0,0,1] &\to [0]\otimes[0,0] +   [2]\otimes [0,0] + [0]\otimes[1,1] + [1]\otimes [1,0] + [1] \otimes [0,1] \;.
 \end{align}
\end{subequations}
The reason why the $\surm(5)$ theory splits into a direct product is that the irreps $ [1]\otimes [1,0] + [1] \otimes [0,1]$, charged under both gauge group factors, precisely cancel between the decomposition of the adjoint and the 2nd anti-symmetric (and its complex conjugate). From the magnetic quiver perspective, this Higgsing is the $A_1$ fission of \eqref{eq:complete_ex_MQ} in Figure \ref{fig:splitting_inverted}. The two ``fission fragments'' are indeed the magnetic quivers for the $\surm(3)$ and $\surm(2)$ gauge theory, respectively.

Continuing this branching analysis, yields the Higgs branch Hasse diagram in Figure \ref{fig:splitting_Higgs}. Comparing to Figure \ref{fig:splitting_inverted} demonstrates that all quivers obtained via the decay and fission algorithm are the magnetic quivers for the (electric) theories obtained from the partial Higgs mechanism.
In the next section, a host of examples is detailed. For most of them, one does not have a weakly coupled description such that Higgs branch RG-flows are substantially more involved than vanilla partial Higgs mechanism. The decay and fission algorithm then offers a unique approach to analyse the possible Higgs branch RG-flows in a systematic way.

\subsection{Classification of isolated singularities}
\label{sec:classification}

The algorithm presented in Section~\ref{sec:algorithm} can be used to identify quivers whose Coulomb branch is an \emph{isolated} conical symplectic singularity (ICSS). In practice, this amounts to classifying pairs $(A,K)$ such that the set of leaves \eqref{eq:set_of_leaves} contains exactly two elements (the trivial leaf, i.e.\ the singularity, and the non-trivial leaf). This is a well-defined mathematical question, which can presumably be addressed abstractly, thereby providing a full classification. We postpone this for future work, but we present here a first step in that direction, by using a brute force approach that is applicable to quivers with a small number of nodes. Specifically, we focus on quivers with $n = 1, 2, 3$ nodes.

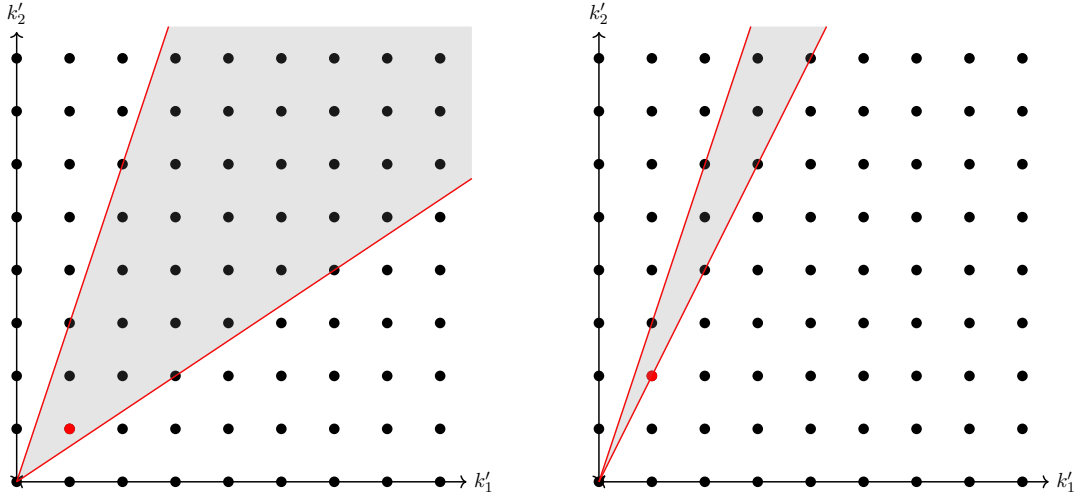
\begin{figure}[t]
\begin{center}
\scalebox{.7}{
\begin{tikzpicture}
\node at (0,0) {\begin{tikzpicture}
  \draw [<->,thick] (0,0) -- (8.5,0) node [right] {$k_1'$};
  \draw [<->,thick] (0,0) -- (0,8.5) node [above] {$k_2'$};

  \foreach \x in {0,...,8}{
    \foreach \y in {0,...,8}{
      \fill (\x,\y) circle (.1);
    }
  }

\fill[red] (1,2) circle (.1);

  \draw [red, thick] (0,0) -- (4.3,8.6);  
  \draw [red, thick] (0,0) -- (2.87,8.6);  

  \fill[fill=gray, fill opacity=0.2] (0,0) -- (4.3,8.6) -- (2.87,8.6) -- cycle;
\end{tikzpicture}};
\node at (-11,0) {\begin{tikzpicture}
  \draw [<->,thick] (0,0) -- (8.5,0) node [right] {$k_1'$};
  \draw [<->,thick] (0,0) -- (0,8.5) node [above] {$k_2'$};

  \foreach \x in {0,...,8}{
    \foreach \y in {0,...,8}{
      \fill (\x,\y) circle (.1);
    }
  }

  \draw [red, thick] (0,0) -- (8.6,5.73);  
  \draw [red, thick] (0,0) -- (2.87,8.6);  

  \fill[fill=gray, fill opacity=0.2] (0,0) -- (8.6,5.73) -- (8.6,8.6) -- (2.87,8.6) -- cycle;

\fill[red] (1,1) circle (.1);
\end{tikzpicture}};
\end{tikzpicture}}
\end{center}
\caption{\textit{(left)} Shape of the allowed $(k'_1,k'_2)$ region in a situation where $a\geq 2$ and $b \geq 1$ (here $a=3$ and $b=2$). \textit{(right)} Shape of the allowed $(k'_1,k'_2)$ region in a situation where $a=1$ and $b \geq 4$ (here $b=6$). In both cases, the unique minimal non-zero quiver is the red dot.}
\label{fig:plots}
\end{figure}

\begin{table}[]
    \centering
\begin{tabular}{ccccc}
\toprule 
Unframed & Framed & Condition & Geometry & Comments \\ 
\midrule 
U(2) with $g$ adjoints & - & $g \geq 2$ & $D_{g+1}$ & \cite{Hanany:2010qu} \\ 
\raisebox{-.5\height}{\begin{tikzpicture}
\tikzset{node distance = 0.75cm};
        \node (g1) [gauge,label=below:{\footnotesize{$1$}}] {};
        \node (g3) [gauge,left of=g1,label=below:{\footnotesize{$1$}}] {};
        \draw[line width=3pt,gray] (g1)--(g3) node[midway, above, black] {$n , \ell$};
         \draw (-0.5,0.15)--(-0.3,0)--(-0.5,-0.15);
\end{tikzpicture}} & \raisebox{-.5\height}{\begin{tikzpicture}
\tikzset{node distance = 0.75cm};
        \node (g1) [gauge,label=below:{\footnotesize{$1$}}] {};
        \node (g3) [flavour,left of=g1,label=below:{\footnotesize{$n$}}] {};
        \draw (g1)--(g3);
\end{tikzpicture}} & $n \geq 2 , \ell \geq 1$ & $A_{n-1}$ & SQED quiver \\ 
\raisebox{-.5\height}{\begin{tikzpicture}
\tikzset{node distance = 0.75cm};
        \node (g1) [gauge,label=below:{\footnotesize{$1$}}] {};
        \node (g3) [gauge,left of=g1,label=below:{\footnotesize{$2$}}] {};
        \draw[line width=3pt,gray] (g1)--(g3) node[midway, above, black] {$\ell$};
         \draw (-0.5,0.15)--(-0.3,0)--(-0.5,-0.15);
\end{tikzpicture}} & -  & $\ell \geq 4$  & $ \mathcal{Y}(\ell)$ & \cite{bellamy2023new} \cite{zz} \cite{Bourget:2022tmw} \\ 
\raisebox{-.5\height}{\begin{tikzpicture}
        \node (g1) at (0,0) [gauge,label=below:{\footnotesize{$1$}}] {};
        \node (g2) at (1,0) [gauge,label=below:{\footnotesize{$1$}}] {};
        \node (g3) at (0,1) [gauge,label=left:{\footnotesize{$1$}}] {};
        \draw[line width=3pt,gray] (g1)--(g2) node[midway, below, black] {$\ell_2$};
        \draw[line width=3pt,gray] (g1)--(g3) node[midway, left, black] {$\ell_1 $};
        \draw[line width=3pt,gray] (g2)--(g3) node[midway, above right, black] {$\ell_1 \ell_2$};
         \draw (0.3,0.15)--(0.5,0)--(0.3,-0.15);
         \draw (0.15,0.5)--(0,0.3)--(-0.15,0.5);
         \draw (.4,.4)--(.6,.4)--(.6,.6);
\end{tikzpicture}} &  
\raisebox{-.5\height}{\begin{tikzpicture}
        \node (g1) at (0,0) [gauge,label=below:{\footnotesize{$1$}}] {};
        \node (g2) at (1,0) [gauge,label=below:{\footnotesize{$1$}}] {};
        \node (g3) at (0,1) [flavour,label=left:{\footnotesize{$1$}}] {};
        \node (g4) at (1,1) [flavour,label=right:{\footnotesize{$1$}}] {};
        \draw[line width=3pt,gray] (g1)--(g2) node[midway, below, black] {$\ell_2$};
        \draw (g1)--(g3);
        \draw (g2)--(g4);
         \draw (0.3,0.15)--(0.5,0)--(0.3,-0.15);
\end{tikzpicture}}
 & $\ell_1 ,  \ell_2 \geq 1$ & $ \bar{h}_{2 , \ell_2}$ & \cite{Bourget:2021siw} \\ 
\raisebox{-.5\height}{\begin{tikzpicture}
\tikzset{node distance = 0.75cm};
        \node (g1) [gauge,label=below:{\footnotesize{$1$}}] {};
        \node (g2) [gauge,right of=g1,label=below:{\footnotesize{$1$}}] {};
        \node (g3) [gauge,left of=g1,label=below:{\footnotesize{$1$}}] {};
        \draw[line width=3pt,gray] (g1)--(g2) node[midway, above, black] {$\ell_1$};
        \draw[line width=3pt,gray] (g1)--(g3) node[midway, above, black] {$\ell_1 \ell_2$};
         \draw (0.5,0.15)--(0.3,0)--(0.5,-0.15);
         \draw (-0.5,0.15)--(-0.3,0)--(-0.5,-0.15);
\end{tikzpicture}} & 
\raisebox{-.5\height}{\begin{tikzpicture}
\tikzset{node distance = 0.75cm};
        \node (g1) [gauge,label=below:{\footnotesize{$1$}}] {};
        \node (g2) [gauge,right of=g1,label=below:{\footnotesize{$1$}}] {};
        \node (g3) [flavour,left of=g1,label=below:{\footnotesize{$1$}}] {};
        \draw[line width=3pt,gray] (g1)--(g2) node[midway, above, black] {$\ell_1$};
        \draw (g1)--(g3);
         \draw (0.5,0.15)--(0.3,0)--(0.5,-0.15);
\end{tikzpicture}}
& $\ell_1  \geq 2 , \ell_2  \geq 1$  & $ h_{2 , \ell_1}$ & \cite{Bourget:2021siw} \\ 
\raisebox{-.5\height}{\begin{tikzpicture}
\tikzset{node distance = 0.75cm};
        \node (g1) [gauge,label=below:{\footnotesize{$2$}}] {};
        \node (g2) [gauge,right of=g1,label=below:{\footnotesize{$1$}}] {};
        \node (g3) [gauge,left of=g1,label=below:{\footnotesize{$1$}}] {};
        \draw[transform canvas={xshift=0pt,yshift=-2pt}](g1)--(g2);
        \draw[transform canvas={xshift=0pt,yshift=0pt}](g1)--(g2);
        \draw[transform canvas={xshift=0pt,yshift=2pt}](g1)--(g2);
        \draw[line width=3pt,gray] (g1)--(g3) node[midway, above, black] {$\ell$};
         \draw (0.3,0.15)--(0.5,0)--(0.3,-0.15);
         \draw (-0.5,0.15)--(-0.3,0)--(-0.5,-0.15);
\end{tikzpicture}} & \raisebox{-.5\height}{\begin{tikzpicture}
\tikzset{node distance = 0.75cm};
        \node (g1) [gauge,label=below:{\footnotesize{$2$}}] {};
        \node (g2) [gauge,right of=g1,label=below:{\footnotesize{$1$}}] {};
        \node (g3) [flavour,left of=g1,label=below:{\footnotesize{$1$}}] {};
        \draw[transform canvas={xshift=0pt,yshift=-2pt}](g1)--(g2);
        \draw[transform canvas={xshift=0pt,yshift=0pt}](g1)--(g2);
        \draw[transform canvas={xshift=0pt,yshift=2pt}](g1)--(g2);
        \draw (g1)--(g3);
         \draw (0.3,0.15)--(0.5,0)--(0.3,-0.15);
\end{tikzpicture}} & $\ell  \geq  1$  & $g_2$ & Affine DD \\ 
\raisebox{-.5\height}{\begin{tikzpicture}
\tikzset{node distance = 0.75cm};
        \node (g1) [gauge,label=below:{\footnotesize{$2$}}] {};
        \node (g2) [gauge,right of=g1,label=below:{\footnotesize{$1$}}] {};
        \node (g3) [gauge,left of=g1,label=below:{\footnotesize{$1$}}] {};
        \draw[transform canvas={xshift=0pt,yshift=-1pt}](g1)--(g2);
        \draw[transform canvas={xshift=0pt,yshift=1pt}](g1)--(g2);
        \draw[transform canvas={xshift=0pt,yshift=-1pt}](g1)--(g3);
        \draw[transform canvas={xshift=0pt,yshift=1pt}](g1)--(g3);
         \draw (0.3,0.15)--(0.5,0)--(0.3,-0.15);
         \draw (-0.3,0.15)--(-0.5,0)--(-0.3,-0.15);
\end{tikzpicture}} & - & -  & $d_3$ & Twisted affine DD \\ 
\raisebox{-.5\height}{\begin{tikzpicture}
\tikzset{node distance = 0.75cm};
        \node (g1) [gauge,label=below:{\footnotesize{$2$}}] {};
        \node (g2) [gauge,right of=g1,label=below:{\footnotesize{$1$}}] {};
        \node (g3) [gauge,left of=g1,label=below:{\footnotesize{$1$}}] {};
        \draw[transform canvas={xshift=0pt,yshift=-2pt}](g1)--(g2);
        \draw[transform canvas={xshift=0pt,yshift=0pt}](g1)--(g2);
        \draw[transform canvas={xshift=0pt,yshift=2pt}](g1)--(g2);
        \draw[transform canvas={xshift=0pt,yshift=-1pt}](g1)--(g3);
        \draw[transform canvas={xshift=0pt,yshift=1pt}](g1)--(g3);
         \draw (0.3,0.15)--(0.5,0)--(0.3,-0.15);
         \draw (-0.3,0.15)--(-0.5,0)--(-0.3,-0.15);
\end{tikzpicture}} & - &-  & $\mathcal{J}_{2,3}$ & \cite{Bourget:2022tmw} \\ 
\raisebox{-.5\height}{\begin{tikzpicture}
\tikzset{node distance = 0.75cm};
        \node (g1) [gauge,label=below:{\footnotesize{$2$}}] {};
        \node (g2) [gauge,right of=g1,label=below:{\footnotesize{$1$}}] {};
        \node (g3) [gauge,left of=g1,label=below:{\footnotesize{$1$}}] {};
        \draw[transform canvas={xshift=0pt,yshift=-2pt}](g1)--(g2);
        \draw[transform canvas={xshift=0pt,yshift=0pt}](g1)--(g2);
        \draw[transform canvas={xshift=0pt,yshift=2pt}](g1)--(g2);
        \draw[transform canvas={xshift=0pt,yshift=-2pt}](g1)--(g3);
        \draw[transform canvas={xshift=0pt,yshift=0pt}](g1)--(g3);
        \draw[transform canvas={xshift=0pt,yshift=2pt}](g1)--(g3);
         \draw (0.3,0.15)--(0.5,0)--(0.3,-0.15);
         \draw (-0.3,0.15)--(-0.5,0)--(-0.3,-0.15);
\end{tikzpicture}} &- & - & $\mathcal{J}_{3,3}$ & \cite{Bourget:2022tmw} \\ 
\raisebox{-.5\height}{\begin{tikzpicture}
\tikzset{node distance = 0.75cm};
        \node (g1) [gauge,label=below:{\footnotesize{$2$}}] {};
        \node (g2) [gauge,right of=g1,label=below:{\footnotesize{$1$}}] {};
        \node (g3) [gauge,left of=g1,label=below:{\footnotesize{$2$}}] {};
        \draw[transform canvas={xshift=0pt,yshift=-1pt}](g1)--(g2);
        \draw[transform canvas={xshift=0pt,yshift=1pt}](g1)--(g2);
        \draw[transform canvas={xshift=0pt,yshift=-1pt}](g1)--(g3);
        \draw[transform canvas={xshift=0pt,yshift=1pt}](g1)--(g3);
         \draw (0.3,0.15)--(0.5,0)--(0.3,-0.15);
         \draw (-0.5,0.15)--(-0.3,0)--(-0.5,-0.15);
\end{tikzpicture}} &- & - & $a_{4}$ & Twisted affine DD \\ 
\rowcolor{black!5!white!95!} \raisebox{-.5\height}{\begin{tikzpicture}
\tikzset{node distance = 0.75cm};
        \node (g1) [gauge,label=below:{\textcolor{black}{\footnotesize{$2$}}}] {};
        \node (g2) [gauge,right of=g1,label=below:{\textcolor{black}{\footnotesize{$1$}}}] {};
        \node (g3) [gauge,left of=g1,label=below:{\textcolor{black}{\footnotesize{$2$}}}] {};
        \draw[transform canvas={xshift=0pt,yshift=-1pt}](g1)--(g2);
        \draw[transform canvas={xshift=0pt,yshift=1pt}](g1)--(g2);
        \draw[transform canvas={xshift=0pt,yshift=-2pt}](g1)--(g3);
        \draw[transform canvas={xshift=0pt,yshift=0pt}](g1)--(g3);
        \draw[transform canvas={xshift=0pt,yshift=2pt}](g1)--(g3);
         \draw (0.3,0.15)--(0.5,0)--(0.3,-0.15);
         \draw (-0.5,0.15)--(-0.3,0)--(-0.5,-0.15);
\end{tikzpicture}} &- & - & ? & See text.  \\ 
\raisebox{-.5\height}{\begin{tikzpicture}
\tikzset{node distance = 0.75cm};
        \node (g1) [gauge,label=below:{\footnotesize{$2$}}] {};
        \node (g2) [gauge,right of=g1,label=below:{\footnotesize{$1$}}] {};
        \node (g3) [gauge,left of=g1,label=below:{\footnotesize{$3$}}] {};
        \draw[transform canvas={xshift=0pt,yshift=0pt}](g1)--(g2);
        \draw[transform canvas={xshift=0pt,yshift=-2pt}](g1)--(g3);
        \draw[transform canvas={xshift=0pt,yshift=0pt}](g1)--(g3);
        \draw[transform canvas={xshift=0pt,yshift=2pt}](g1)--(g3);
         \draw (-0.5,0.15)--(-0.3,0)--(-0.5,-0.15);
\end{tikzpicture}} &- & - & $d_4$ & Twisted affine DD \\ \bottomrule 
\end{tabular}
    \caption{Complete list of good unitary quivers with 3 nodes or less whose Coulomb branch is an ICSS. The first column gives the raw results from the systematic search; the quivers are unframed and depend on a number of parameters, some of which may be redundant from the geometric perspective. The second column gives the framed version, which is not redundant. The geometry is given in the fourth column, and the way it is identified in the fifth. DD stands for Dynkin Diagram. The shaded line corresponds to a new singularity, discussed in the text.   }
    \label{tab:smallQuivers}
\end{table}

The case $n=1$ is essentially trivial: the matrix $A$ reduces to an even integer $A \geq -2$. If $A = -2$, no good quiver can be constructed. If $A \geq 0$, every quiver with $K \geq 2$ gives a non-trivial singularity, and whenever $K \geq 3$, the quiver can fission or decay. Therefore, the $n=1$ case corresponds to the family of quivers $(A,K) = (2g , 2)$, or in other words, U(2) with $g-1$ adjoint hypermultiplets.

Let us move on to $n=2$. We can restrict to the case where there are no loops, i.e.\ $A_{1,1} = A_{2,2} = -2$. By symmetry, we can parametrize the quivers as 
\begin{equation}
    (A , K) = \left( \left( \begin{array}{cc}
        -2 & a \\
        ab & -2
    \end{array}  \right) , (k_1 , k_2) \right)
    \label{eq:2nodeQuiv}
\end{equation}
with $k_1 , k_2 , a, b \in \mathbb{Z}_{>0}$. The conditions $\mathcal{P}_1$ and $\mathcal{P}_2$ are never satisfied by \eqref{eq:2nodeQuiv}, so the set $\mathcal{V}$ is defined as the set of pairs $(k'_1 , k'_2) \in \mathbb{N}^2$ such that $k'_1 \leq k_1$, $k'_2 \leq k_2$ and 
\begin{equation}
    -2 k'_1 + a k'_2 \geq 0   \, , \quad 
    ab k'_1 -2 k'_2 \geq 0   \, . 
    \label{eq:listInequalities2nodes}
\end{equation}
For there to be a solution, it is necessary that $a^2 b \geq 4$. This leads to two types of solutions, as illustrated on Figure \ref{fig:plots}: 
\begin{itemize}
    \item $a \geq 2$ and $b \geq 1$. In this case, $(k'_1 , k'_2) = (1,1)$ satisfies \eqref{eq:listInequalities2nodes} and is minimal, which gives the quiver 
    \begin{equation}
\begin{tikzpicture}
\tikzset{node distance = 0.75cm};
        \node (g1) [gauge,label=below:{\footnotesize{$1$}}] {};
        \node (g3) [gauge,left of=g1,label=below:{\footnotesize{$1$}}] {};
        \draw[line width=3pt,gray] (g1)--(g3) node[midway, above, black] {$a, b$};
         \draw (-0.5,0.15)--(-0.3,0)--(-0.5,-0.15);
\end{tikzpicture}
\end{equation}
 This corresponds to the $A_{a-1}$ singularity. 
 \item $a = 1$ and $b \geq 4$.  Now $(k'_1 , k'_2) = (1,2)$ satisfies \eqref{eq:listInequalities2nodes} and is minimal, which gives the quiver 
    \begin{equation}
\begin{tikzpicture}
\tikzset{node distance = 0.75cm};
        \node (g1) [gauge,label=below:{\footnotesize{$1$}}] {};
        \node (g3) [gauge,left of=g1,label=below:{\footnotesize{$2$}}] {};
        \draw[line width=3pt,gray] (g1)--(g3) node[midway, above, black] {$b$};
         \draw (-0.5,0.15)--(-0.3,0)--(-0.5,-0.15);
\end{tikzpicture}
\end{equation}
This corresponds to the $\mathcal{Y}(b)$ singularities. 
\end{itemize}

Finally, we can similarly analyze the $n=3$ case, using similar -- but much longer -- arguments. All in all, we finally obtain the list shown in Table \ref{tab:smallQuivers}. It is interesting to note that almost all the quivers that emerge from this study have been studied in the literature, although very recently for some of them, see the comment column in Table \ref{tab:smallQuivers}. However, we find one outlier (the shaded row in the table) for which we believe the geometry has not been studied. It has isometry $\mathfrak{sp}(2)$ and the Coulomb branch Hilbert series reads:
\begin{align}
    \mathrm{HS}(t)&=\frac{1+4 t^2+31 t^4-7 t^6+22 t^8-102 t^{10}+22 t^{12}-7 t^{14}+31 t^{16}+4
   t^{18}+t^{20}}{\left(-1+t^2\right)^8 \left(1+t^2\right)^2 \left(-1+t^4\right)^2}\\
   &=1+10 t^2+80 t^4+359 t^6+1295 t^8+3751 t^{10}+9560 t^{12}+21675 t^{14}+45313
   t^{16}+O\left(t^{18}\right) \notag.
\end{align}
The highest weight generating function reads (in conventions where $\alpha_1$ is the short simple root and $\alpha_2$ the long simple root of $\mathfrak{sp}(2)$): 
\begin{eqnarray}
    \mathrm{PE} \left[ \mu _1^2 t^2 + (1+  \mu _2^2 + \mu _2^3) t^4 + \mu _2^3   t^6 -  \mu _2^6 t^{12}\right] &=& 1+\mu _1^2 t^2+\left(\mu _1^4+\mu _2^3+\mu _2^2+1\right) t^4 \\ & & +\left(\mu _1^6+\mu _2^3 \mu _1^2+\mu _2^2 \mu _1^2+\mu _1^2+\mu _2^3\right)
   t^6+O\left(t^7\right) \, .  \nonumber
\end{eqnarray}

Two natural questions then arise from this study: first, can this new quiver be obtained from a brane construction or geometric engineering in string theory? And second, can one generalize this search to arbitrary number of nodes? We leave these fundamental  challenges for future work, and turn now to applications of our algorithm to various physical situations.

 \section{Decay and fission in action: selected examples}
 \label{sec:examples}
We now apply the decay and fission algorithm on selected examples in space-time dimension $d=3,4,5,6$. This case study illuminates and demonstrates the arising Higgsing pattern between different theories.

\subsection{\texorpdfstring{3d $\Ncal=4$ theories}{3d N=4 theories}}
\label{sec:3d}

\subsubsection{Intuition via mirror symmetry}
In this subsection, we consider a class of 3d $\Ncal=4$ theories for which one can directly use 3d mirror symmetry in order to prove the validity of the decay algorithm (no fission occurs here). In this specific framework, there is a duality between decay and the quiver subtraction of \cite{Bourget:2019aer}: one corresponds to moving on the Higgs branch, while the other corresponds to moving on the Coulomb branch.  

\paragraph{Higgs and Coulomb branch Higgsing.}
Consider a Lagrangian gauge theory, with some scalars transforming in a representation $R$ of the gauge group $G$. The scalars may acquire non-trivial vacuum expectation values (VEVs), which breaks the gauge group to a subgroup $H$ depending. Depending on the type of the residual gauge group, the phases in a given vacuum carry different names. For example, for $R$ the fundamental representation: $H$ is a subgroup of $G$ with reduced rank --- these are called Higgs vacua, in analogy to the Standard Model. On the other hand, for $R$ the adjoint representation: $H$ is a subgroup of $G$ with the same rank (adjoint Higgsing) --- these are called Coulomb vacua, because $H$, in general, contains $\urm(1)$ factors.

A large class of 3d $\Ncal=4$ theories have two maximal branches of the moduli spaces of supersymmetric vacua: the Higgs and Coulomb branch, which are parameterised by VEVs of scalar fields in hypermultiplet and vector multiplets, respectively. Consequently, there are two types of Higgs mechanisms: one triggered by a VEV for a hypermultiplet scalar and the other trigger by a VEV for a vector multiplet scalar --- in short, a Higgs or Coulomb branch Higgs mechanisms, respectively. 

\begin{sidewaysfigure}
    \centering
    \begin{subfigure}[t]{0.45\textwidth}
    \flushleft
  \includegraphics[page=1]{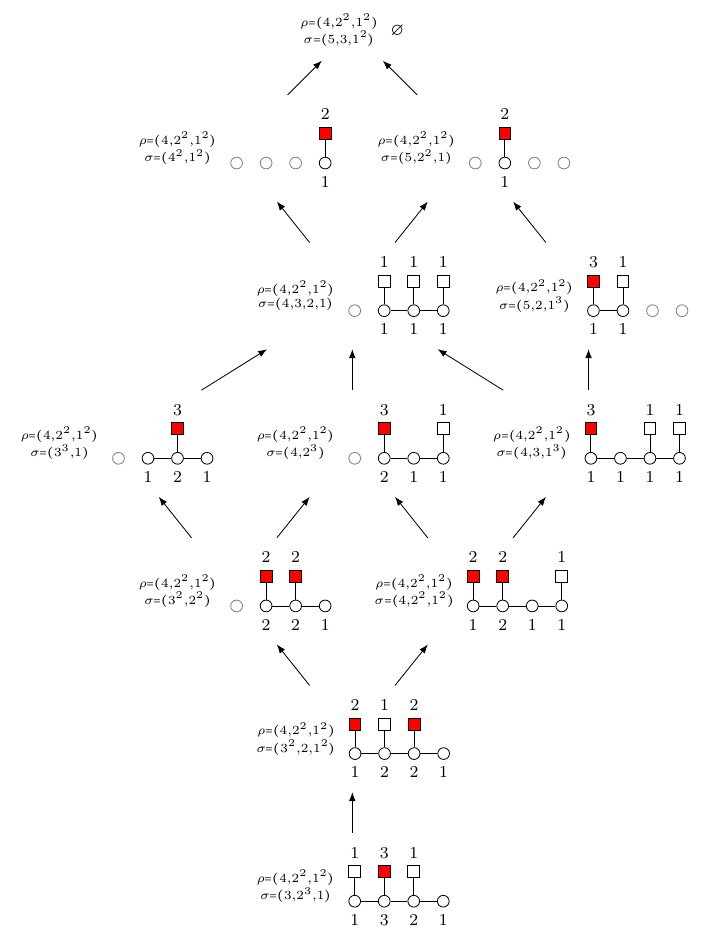}
  \caption{Higgs branch Higgsing of theory $\Tcal$}
  \label{fig:HB_higgs}
    \end{subfigure}
    \hfill
    \begin{subfigure}[t]{0.45\textwidth}
    \flushright
  \includegraphics[page=2]{figures_inverted_sub_3d.pdf}
  \caption{Coulomb branch Higgsing of mirror theory $\Tcal^\vee$}
    \label{fig:CB_higgs}
    \end{subfigure}
    \caption{Partial Higgs mechanism of a theory $\Tcal=T_{\rho}^\sigma[\surm(10)]$ with $\rho=(4,2^2,1^2)$, $\sigma=(3,2^3,1)$ and its mirror $\Tcal^\vee=T^{\rho}_\sigma[\surm(15)]$. The red coloured flavour nodes indicated non-abelian global symmetries that trigger minimal Higgs branch transitions. On the mirror, this translates to the set of balanced nodes indicated by green. At each step, mirror symmetry is manifest via exchanging $\rho\leftrightarrow \sigma$.}
    \label{fig:3d_Higgs_framed}
\end{sidewaysfigure}

To gain intuition consider the linear quiver gauge theories $T_\rho^\sigma[\surm(N)]$.
 Higgs branch Higgsing can be effectively realised on the quiver theory by 
 \begin{compactenum}[({HB}1)]
 \item \label{HB1} Mesons of a single gauge: subtracting the SQED with $N$ flavours quiver $(1)-[N]$ from the gauge node. Here, the single gauge node $\urm(k) \to \urm(k-1)$ is partially broken. 
 \item \label{HB2} Gauge invariant spanning several gauge nodes: subtracting the linear quiver $[1]-(1)-\ldots-(1)-[1]$ between two flavour nodes. Here, a whole sequence of gauge group factors $\urm(k_i)\to\urm(k_i-1)$ is partially broken.
 \end{compactenum}
To illustrate, consider the partial Higgs mechanisms along the Higgs branch in Figure~\ref{fig:HB_higgs}. Note in particular, that the balance of the gauge nodes is preserved during Higgs branch Higgsing, which is evident from the D3-D5-NS5 brane configurations \cite{Hanany:1996ie}. In terms of quiver subtraction, this is known as the re-balancing condition after the subtraction and leads in general to a change of flavour nodes. 
On a related note, there exists an algorithm in the mathematics literature \cite{crawley2001geometry,crawley2001normality,bellamy2021symplectic} that produces these quivers for the Higgs branch Higgsing. Related to decay via 3d mirror dual, this algorithm also shares a remarkable feature that it defines a quiver via the adjacency matrix $A$ and a rank vector $K$, and subsequently generates quivers for each leaf of the Higgs branch. 

One the mirror side (or in terms of magnetic quivers), these Higgs branch transitions are partial Higgs mechanisms along the Coulomb branch. Since the flavour nodes have been the indicator for the Higgs branch transitions (because Higgs branch gauge invariant operators need to start and end at flavour nodes), the topological symmetry (i.e.\ the balance of the gauge nodes) is the smoking gun for Coulomb branch Higgs transitions. Again, this is particularly transparent in the brane realisation. On the level of the quiver theory, Coulomb branch Higgsing has equally straightforward implementation \cite{Gu:2022dac}:
\begin{compactenum}[({CB}1)]
    \item \label{CB1} For a connected set of balanced gauge nodes, Coulomb branch Higgsing breaks all nodes partially: $\urm(k_i)\to \urm(k_i -1)$, but the flavour nodes are not affected. This can be realised by subtracting the finite weighted $A$-type Dynkin quiver $(1)-(1)-\ldots-(1)$.
    \item \label{CB2} For an over-balanced node, not connected to any balanced nodes, Coulomb branch Higgsing only breaks this gauge group factor via $\urm(k)\to\urm(k-1)$. On the quiver, one simply subtracts a $(1)$ node.
\end{compactenum}
This is exemplified in Figure~\ref{fig:CB_higgs}. In contrast to Higgs branch transitions, the partial Higgs mechanisms CB\ref{CB1}--\ref{CB2} do not preserve the balance of the gauge nodes.

The point to appreciate is the following: given a theory $\Tcal$ that admits a Higgs branch transition HB\ref{HB1} or HB\ref{HB2} to $\Tcal_1$, then, assuming that the mirror theory of $\Tcal$ is $\Tcal^\vee$, the mirror of $\Tcal_1$ is simply obtained from $\Tcal^\vee$ by either CB\ref{CB1} or CB\ref{CB2}.

It is a simple exercise to generalise the rule CB\ref{CB1} for Coulomb branch Higgsing to other unitary quiver theories, such as $BCD$-type Dynkin quivers, see \cite{Gu:2022dac}. Using for instance O5 or ON planes, one deduces that also weighted finite Dynkin diagrams of type $BCD$ can be subtracted. 
As a result, (CB\ref{CB1}-\ref{CB2}) motivate the practical implementation (\ref{Rule:1}-\ref{Rule:2}), which in turn can be interpreted in terms of decays. 

\begin{sidewaysfigure}
\centering
    \begin{subfigure}[t]{0.495\textwidth}
    \flushleft
  \includegraphics[page=3]{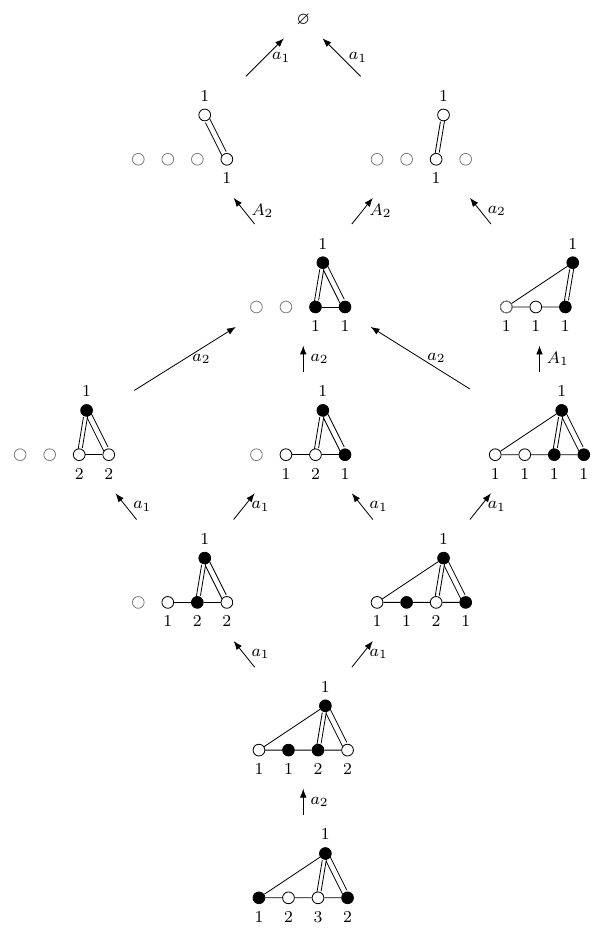}
  \caption{Decay algorithm.}
  \label{subtraction:new}
  \hfill
    \end{subfigure}
    \begin{subfigure}[t]{0.495\textwidth}
    \flushright
  \includegraphics[page=4]{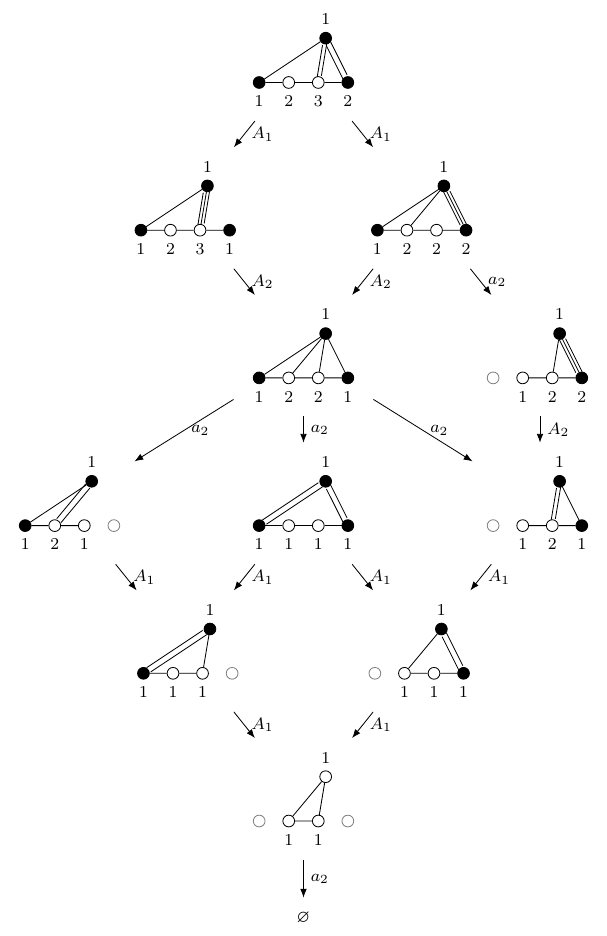}
  \caption{Quiver subtraction.}
   \label{subtraction:traditional}
    \end{subfigure}
    \caption{Comparison of quiver subtraction and the decay algorithm. White gauge nodes $\Circle$ are balanced, black nodes $\CIRCLE$ are over-balanced. Note that for 3d $\Ncal=4$ gauge theories, Coulomb branch Higgsing on a framed quiver $\mathsf{Q}$ is identical to decay (and fission) algorithm on the unframed version of $\mathsf{Q}$. In other words, compare Figures \ref{fig:CB_higgs} and \ref{subtraction:new}.}
    \label{fig:3d_Higgs_unframed}
\end{sidewaysfigure}

\paragraph{Magnetic quivers and decay transitions.}
As an example, consider the theory $\Tcal=T_{\rho}^\sigma[\surm(10)]$ with $\rho=(4,2^2,1^2)$, $\sigma=(3,2^3,1)$. The Higgsings on the Higgs and Coulomb branches are shown in Figure~\ref{fig:3d_Higgs_framed}. The quiver subtraction algorithm generates magnetic quivers $\MQ^{\Lcal_i}$ for the closure of each leaf $\overline{\Lcal}_i$ in the stratification of the Higgs branch of $\Tcal$.
However, \emph{none} of these $\MQ^{\Lcal_i}$ is the magnetic quiver for any of the theories $\Tcal$ can be Higgsed to, i.e.\ the quivers in Figure~\ref{fig:HB_higgs}.
In contrast, directly utilising the \emph{decay algorithm} (Figure~\ref{subtraction:new}) generates a different set of quivers $\MQ_{\Lcal_i}$. 
These \emph{are} the magnetic quivers for the different theories $\Tcal_i$ that can be obtained from $\Tcal$ via partial Higgsing along the Higgs branch (Figure~\ref{fig:HB_higgs}). 
The special feature in 3d $\Ncal=4$ is 
\begin{align}
    \MQ_{\Lcal_i} = \left(\Tcal_i\right)^\vee \;,
\end{align}
meaning that the magnetic quivers $\MQ_{\Lcal_i}$ obtained from the decay and fission algorithm are, in fact, the mirror theories of the $\Tcal_i$ themselves. This is evident on our example from Figure \ref{fig:CB_higgs} and \ref{subtraction:new}. 

\paragraph{Framed vs unframed quivers.}
A magnetic quiver is typically represented as an unframed quiver (without flavour nodes). In cases where all the gauge groups are unitary (which is always the case in this paper), this implies that there is an overall $\urm(1)$ that decouples from the theory. Choosing to decouple this $\urm(1)$ from a $\urm(1)$ gauge group creates flavour nodes, thereby framing the quiver. In instances like the $T^\sigma_\rho(\surm(N))$ family, applying the decay and fission algorithm to either the unframed (Figure \ref{fig:3d_Higgs_unframed}) or framed version (Figure \ref{fig:3d_Higgs_framed}) of the magnetic quivers yields the same result. However, this is not generally the case, and only by considering unframed magnetic quivers can we capture all the Higgsings without missing any\footnote{It is possible to only work with framed quivers, one just needs to consider \emph{all} the different choices of $\urm(1)$ decoupling when performing the subtraction.}.

\subsubsection{Mixed U/SU Quivers}
\begin{sidewaysfigure}
    \centering
    \begin{subfigure}[t]{0.31\textwidth}
    \flushleft
  \includegraphics[page=5]{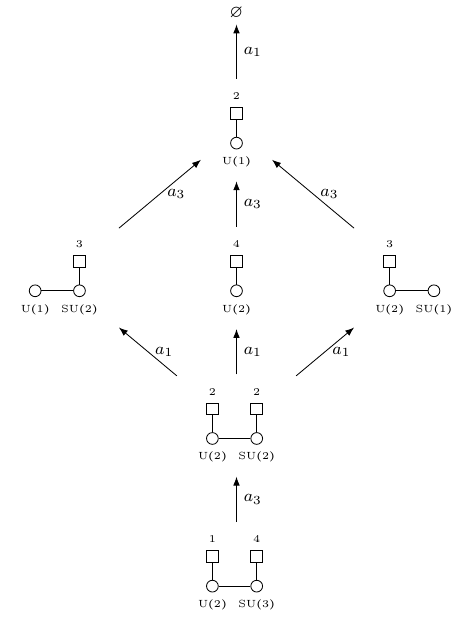}
  \caption{Higgsing}
  \label{fig:SU-U_Higgs}
    \end{subfigure}
    \hfill
    \begin{subfigure}[t]{0.31\textwidth}
    \flushright
  \includegraphics[page=6]{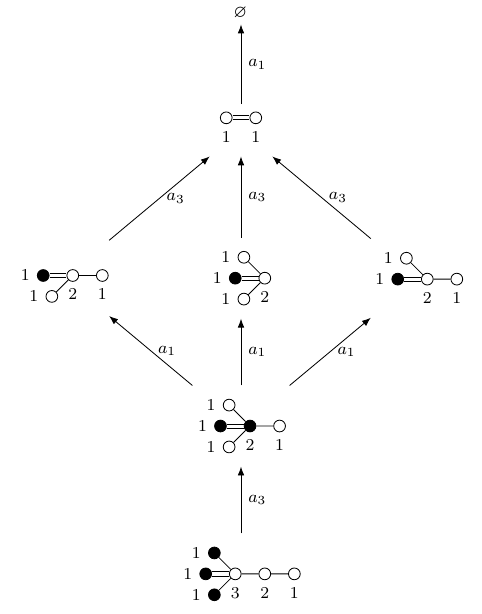}
  \caption{Decay and Fission}
    \label{fig:SU-U_inverted}
    \end{subfigure}
    \hfill
    \begin{subfigure}[t]{0.31\textwidth}
    \flushright
  \includegraphics[page=7]{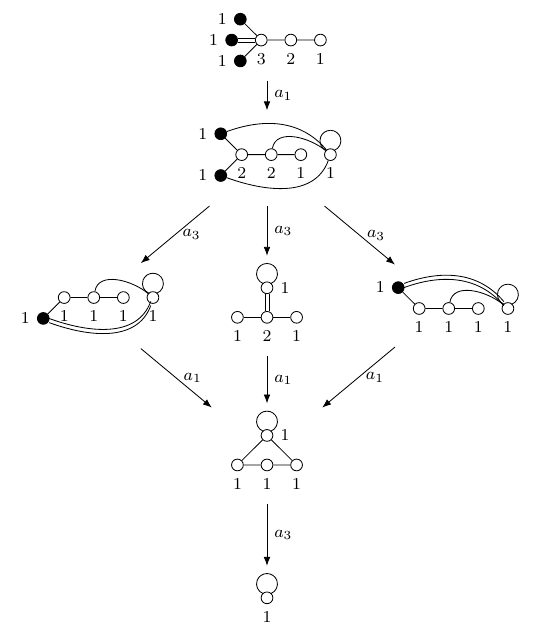}
  \caption{Quiver Subtraction}
    \label{fig:SU-U_standard}
    \end{subfigure}
    \caption{Example for a mixed U/SU quiver. \subref{fig:SU-U_Higgs} displays the partial Higgsing pattern of the (electric) U/SU quiver with $\urm(2)\times \surm(2)$ gauge group. \subref{fig:SU-U_inverted} derives the Higgsing pattern from the IQS starting from the magnetic quiver. \subref{fig:SU-U_standard} provides a comparison with QS.}
    \label{fig:SU-U_example}
\end{sidewaysfigure}

Figures~\ref{fig:CB_higgs} and \ref{subtraction:new} illustrate the application of the decay and fission algorithm on a pair of quivers belonging to the $T^\sigma_\rho(\surm(N))$ family. This family, a broad spectrum of quiver theories, was initially introduced in \cite{Gaiotto} and is characterised by the mirror pairs being linear quivers with unitary gauge groups. The $T^\sigma_\rho(\surm(N))$ family was further generalised in \cite{Bourget:2021jwo} to accommodate linear quivers with an assortment of both unitary and special unitary gauge groups. This expansion significantly broadened the repertoire of recognised 3d mirror pairs. Regarding these U/SU quivers, which we represent as $T^\sigma_\rho(\surm(N))_{U/SU}$, the 3d mirrors continue to consist solely of unitary gauge groups. However, it should be noted that the mirrors no longer conform to the linear quiver structure.
The $T^\sigma_\rho(\surm(N))_{U/SU}$ represents a Lagrangian theory where Higgsings can be conducted utilising group theoretic analysis, as outlined in \cite{Bourget:2019aer}. However, due to the incorporation of special unitary gauge groups, this procedure becomes significantly more complex, as evidenced by the intricate Hasse diagrams detailed in \cite{Bourget:2021jwo}. Despite these complexities, the 3d mirror of the $T^\sigma_\rho(\surm(N))_{U/SU}$ consists entirely of unitary quivers, thus allowing us to once again implement the decay and fission algorithm. The quivers, after decays, can be mapped back to members of the $T^\sigma_\rho(\surm(N))_{U/SU}$ theories, using the algorithms provided in \cite{Bourget:2021jwo}, giving us the Higgsing tree. 
A proof of concept is given in Figure \ref{fig:SU-U_example}.

\subsubsection{Fission}
\begin{figure}[h]
    \centering
    \begin{subfigure}[t]{0.485\textwidth}
    \centering
\includegraphics[page=9]{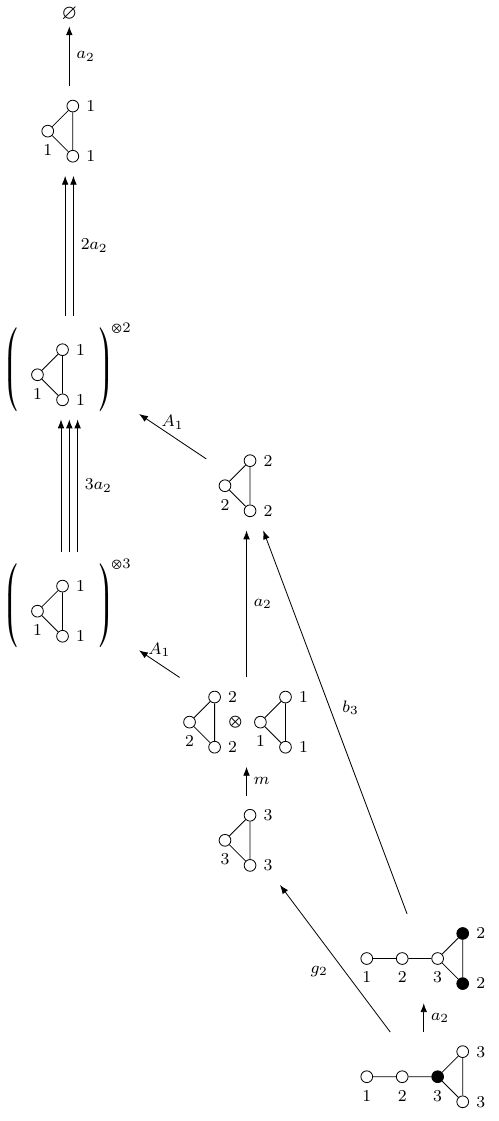}  
        \caption{Decay and fission}
        \label{subfig:instanton_inverted}
    \end{subfigure}
\begin{subfigure}[t]{0.485\textwidth}
\centering
\includegraphics[page=10]{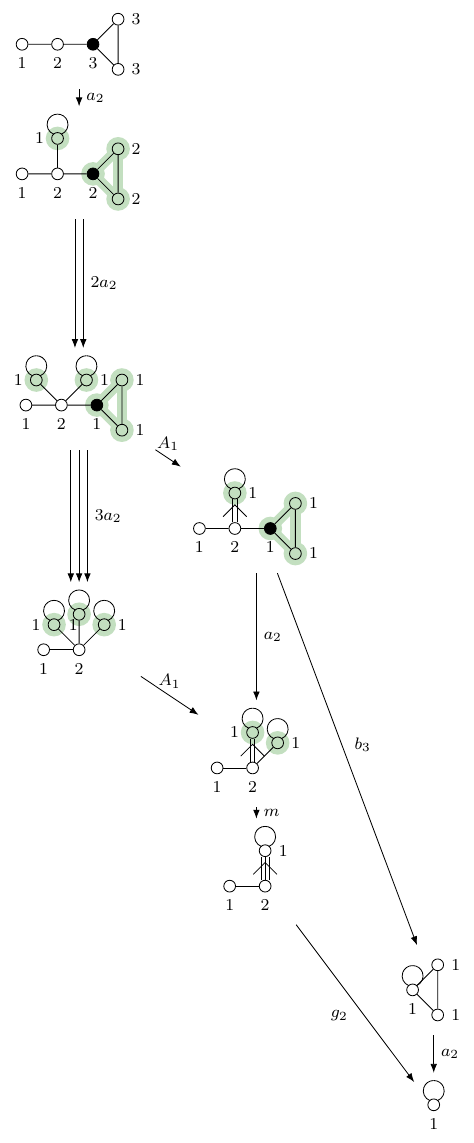}
        \caption{Quiver Subtraction}
        \label{subfig:instanton_standard}
    \end{subfigure}
    \caption{Hasse diagram of the magnetic quiver \eqref{g2quiver}. \subref{subfig:instanton_inverted}: via decay and fission algorithm. \subref{subfig:instanton_standard}: via quiver subtraction, wherein green indicates ``decoration''. Note that fission reproduces all the transverse slices that are unions, e.g.\ $3 a_2 = a_2 \cup a_2 \cup a_2$.}
    \label{fig:diagramInstantons}
\end{figure}

To illustrate all aspects of decay and fission, and to motivate the introduction of the condition $\mathcal{P}_2$ in \eqref{eq:decay_fission_products}, consider the following quiver:
\begin{align}
\raisebox{-.5\height}{
\includegraphics[page=5]{figures_inverted_sub_affine.pdf}
}
\label{g2quiver}
\end{align}
for which the decay and fission algorithm produces the Hasse diagram shown in Figure \ref{subfig:instanton_inverted}. For comparison, the Hasse diagram obtained via quiver subtraction is compiled in Figure \ref{subfig:instanton_standard}. A visual inspection confirms two facts: firstly, the transverse slice geometries associated to minimal transitions are identical in both algorithms. Secondly, the magnetic quivers obtain in both algorithms are substantially different. Let us now comment the phenomena in more detail.

In the decay algorithm, the general idea is to list all quivers of the same shape, but with smaller rank vector. However, in certain specific cases, such a quiver can still contain free hypermultiplets. The most well-known example is a $\urm(1)$ attached to an $n$-stacked affine Dynkin diagram of algebra $\mathfrak{g}$, such as \eqref{g2quiver}. The resulting moduli space is the $n-\mathfrak{g}$ instanton moduli space, plus a free hypermultiplet. However, rather than decaying to a theory with a decoupled free hypermultiplet, we expect the free hypermultiplet to enhance (or possibly fibrate) the transverse slice in between instead. This is the essence of condition $\mathcal{P}_2$ in \eqref{eq:decay_fission_products}. 

Returning to \eqref{g2quiver} and recalling the practical implementation (\ref{Rule:1}--\ref{Rule:3}), the two decays are the $a_2$ coming from the left two balanced nodes and the $a_2$ from the right two balanced nodes. However, the naive $a_2$ decay in the left nodes would lead to the quiver of the three-$\surm(3)$ instantons on $\mathbb{C}^2$. This contains a free $\mathbb{H}$ factor and is excluded in the decay algorithm, as discussed above. Thus, the complete algorithm step is simply:
\begin{align}
\raisebox{-.5\height}{
\includegraphics[page=6]{figures_inverted_sub_affine.pdf}
}
\end{align}
and the transverse slice is obtained as described in Section \ref{sec:formal}. One finds
\begin{align}
\raisebox{-.5\height}{
\includegraphics[page=7]{figures_inverted_sub_affine.pdf}
}
\end{align}
which yields a $g_2$ slice due to the non-simply laced rebalancing with $\gcd=3$.

\FloatBarrier

\subsection{\texorpdfstring{4d $\Ncal=2$ theories}{4d theories}}
The landscape of 4d $\Ncal=2$ theories accommodates a host of non-Lagrangian theories; see \cite{Seiberg:1994aj,Seiberg:1994rs,Argyres:1995jj,Argyres:1995xn,Argyres:1996eh,Gaiotto:2009we,Gaiotto:2009hg,Argyres:2007cn} and subsequent works. For those, the partial Higgs mechanisms (or Higgs branch RG-flows) are far from obvious and we demonstrate that the \emph{decay and fission} algorithm is a powerful technique to trace out the entire Higgs branch Hasse diagram.

\subsubsection{\texorpdfstring{Class $\Scal$ theories }{Class S theories}}
\begin{figure}[h]
    \centering
    \begin{subfigure}[t]{1\textwidth}
    \centering
\includegraphics[page=6]{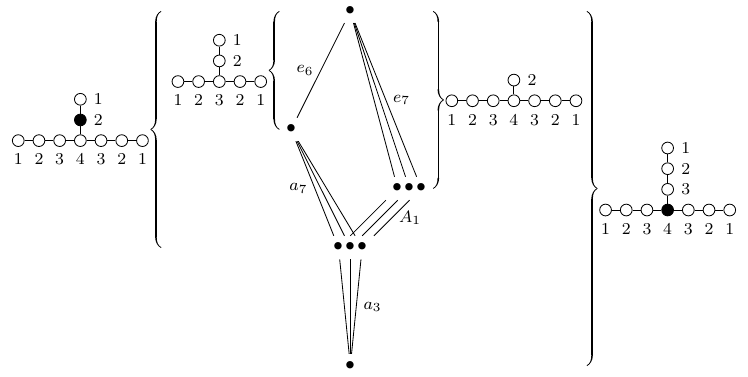}  
        \caption{Decay algorithm}
        \label{subfig:T4_inverted_sub}
    \end{subfigure}
\begin{subfigure}[t]{1\textwidth}
\centering
\includegraphics[page=7]{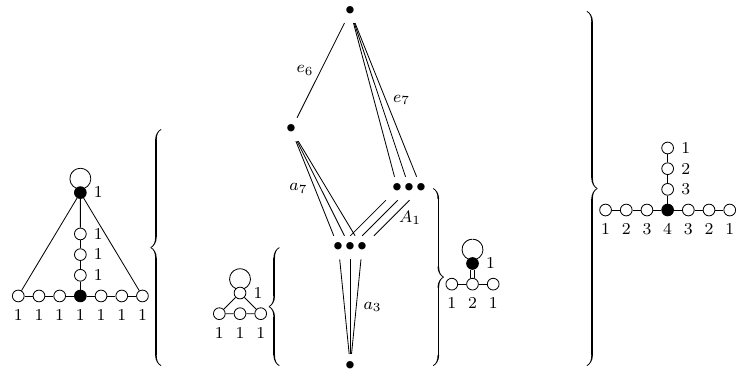} 
        \caption{Quiver subtraction}
        \label{subfig:T4_standard_sub}
    \end{subfigure}
    \caption{The quiver subtraction and decay algorithms for the $T_4$ magnetic quiver generate two sets of quivers. The Coulomb branches of these quivers are equivalent to the transverse spaces between the symplectic leaves (dots) indicated by the brackets in the Hasse diagram.  }
    \label{fig:4d_inverted_vs_standard}
\end{figure}

A large class of 4d $\Ncal=2$ SCFTs is generated using the class $\mathcal{S}$ framework. 
In this case, the decay and fission algorithm reproduces the Higgsing from one theory to the next by partially closing punctures. 
Let us consider as a simple example the $T_4$ theory. It can be constructed as a class $\mathcal{S}$ theory labelled by a punctured Riemann surface \cite{Gaiotto} with three maximal punctures. Its magnetic quiver reads \cite{Benini:2010uu}
\begin{align}
    \raisebox{-.5\height}{
\includegraphics[page=1]{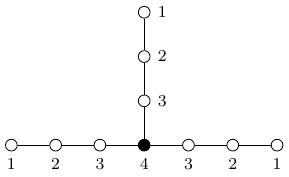}
}
\end{align}
and applying the decay and fission algorithm results in the Hasse diagram shown in Figure \ref{subfig:T4_inverted_sub}. For comparison, the quiver subtraction result is show in Figure \ref{subfig:T4_standard_sub}.

\begin{figure}[h]
    \centering
\includegraphics[page=5]{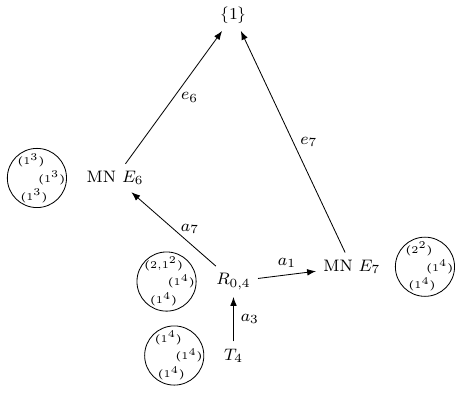}
    \caption{Higgsing pattern relating 4d $\Ncal=2$ theories starting from $T_4$. $E_6$, $E_7$ denote the Minahan-Nemeschansky (MN) theories. Note that, in contrast to Figure \ref{fig:4d_inverted_vs_standard}, the equivalent transitions are not displayed, due to symmetry. }
    \label{fig:4d_Higgsing}
\end{figure}

As advertised in the Introduction, the algorithm indeed generalises the closing of punctures in the class $\mathcal{S}$ language; to see this, it is useful to translate Figure \ref{fig:4d_inverted_vs_standard} into a Higgsing pattern of 4d SCFTs, shown in Figure \ref{fig:4d_Higgsing}. 
The $a_3$ decay transition partially closes a regular puncture: $(1^4) \to (2,1^2)$ and realises the Higgsing of $T_4\to R_{0,4}$. Similarly, the subsequent $a_1$ decay is the further closure of the puncture $(2,1^2) \to (2^2)$ and realises the $R_{0,4}\to \mathrm{MN} E_7$ Higgsing. However, the Higgsing $R_{0,4} \to \mathrm{MN} E_6$ involves changing the rank of class $\mathcal{S}$ theory from $A_3$ to $A_2$ which is more involved than just partially closing a puncture\footnote{This Higgsing can be done as follows. First, partially closing one of the puncture leads to a magnetic quiver where one of the gauge nodes has balance $-1$. This means the quiver contains some free hypermultiplets that needs to be removed. One then needs to do a set of Seiberg dualities to ensure all the nodes have positive balance. The resulting quiver is the magnetic quiver of the Higgsed theory which can then be mapped back to the $\mathrm{MN} E_6$. In general, this step of performing Seiberg dualities on all the nodes make the Higgsing procedure much more inefficient and can often lead to a non-minimal Higgsing.}. 
Furthermore, fission is not covered by closing partial punctures; this is particularly evident when the quiver contains stacks of affine Dynkin subquivers.
We conclude that even for class $\mathcal{S}$ theories, the closing of partial puncture does not realise all possible Higgsings.

\subsubsection{Argyres-Douglas theories} 
One can also consider Argyres-Douglas (AD) theories \cite{Argyres:1995jj}, which are constructed in class $\mathcal{S}$ making use of \emph{irregular} punctures \cite{DistlerA}. 
The magnetic quivers of these theories are known \cite{Xie:2012hs, Giacomelli:2020ryy,Xie:2021ewm} and the irregular puncture contributes a \emph{complete graph} to the magnetic quiver. We see below that the Higgsing pattern associated to irregular punctures is encapsulated in fission transitions. 

\begin{figure}[h]
 \includegraphics[width=\textwidth,page=9]{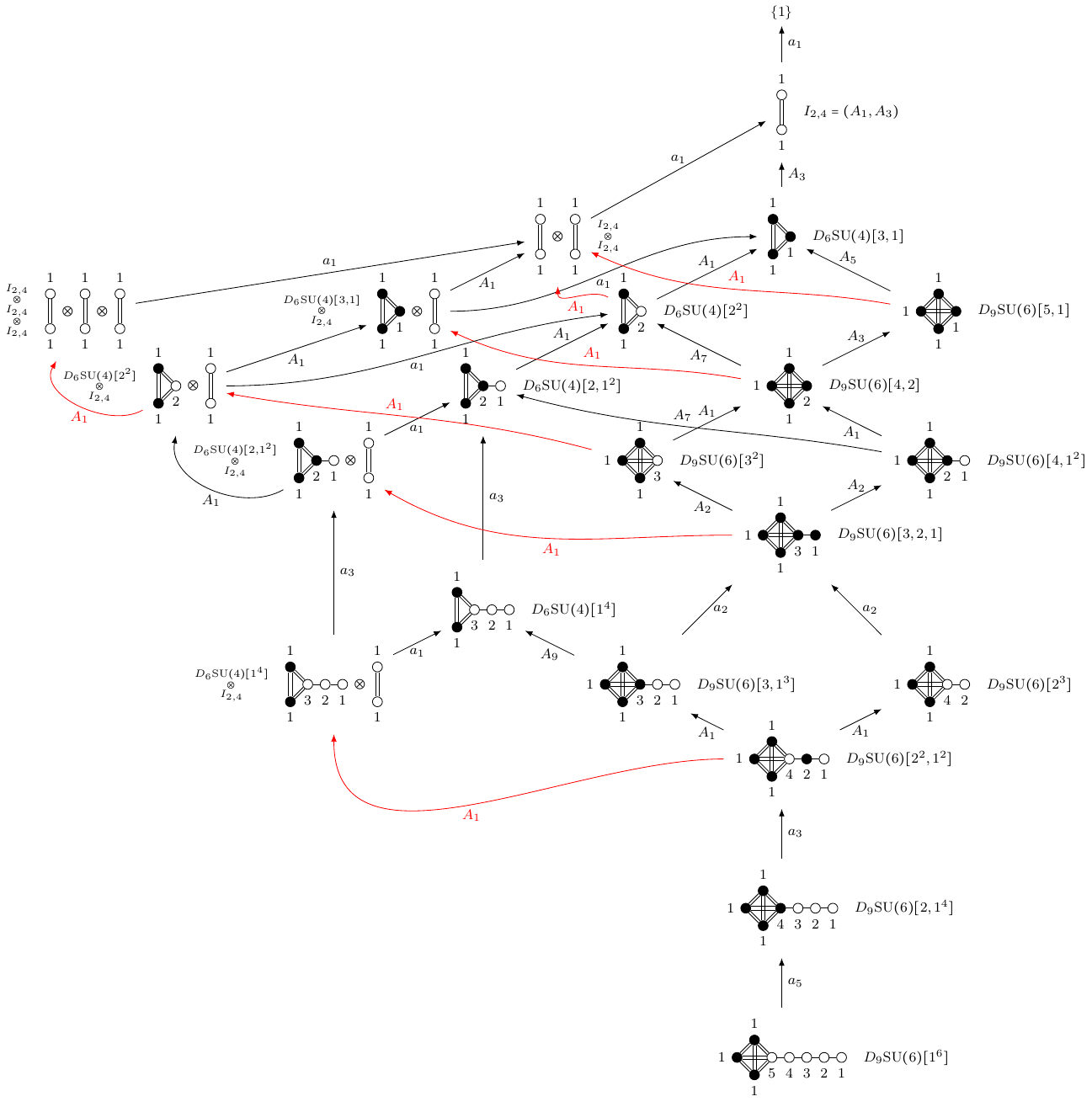}  
    \caption{Higgsings of the $D_9\surm(6)[1^6]$ AD theory. For readability, equivalent transitions are not depicted and fissions are highlighted in red. 
    The $\{1\}$ here refers to the Higgs branch of the AD theory is trivial as it is possible that the SCFT is not completely Higgsable.}
    \label{fig:Higgs_D9SU6}
\end{figure}

Let us, for example, consider the $(A_2,A_5)[1^6]$ AD theory and its magnetic quiver, which was studied in \cite{Giacomelli:2020ryy} and denoted as $D_9\surm(6)[1^6]$ therein. Here, the label $(A_2,A_5)$ denotes the nature of the irregular puncture, whereas $[1^6]$ is the partition that defines the regular puncture.  The AD theory in class $\mathcal{S}$ notation and its magnetic quiver are:
\begin{align}
\raisebox{-.5\height}{
 \includegraphics[page=8]{figures_inverted_sub_4d.pdf}
 }
\end{align}
The pattern of partial Higgsing of the $D_9\surm(6)[1^6]$ theory obtained from decay and fission algorithm is summarised in Figure \ref{fig:Higgs_D9SU6}.

To begin with, let us analyse the decays.
The Higgsings for $(A_2,A_5)[\rho]$, with $\rho$ a partition of $6$, can be understood as closing the regular puncture, which changes the partitions and in turn shortens the tail of the magnetic quiver. However, the transitions like the $A_7$ or $A_9$ decay are achieved by removing one of the overbalanced $\urm(1)$ gauge nodes in the complete graph\footnote{
Naturally, there are three equivalent $A_7$ or $A_9$ decays and we only wrote down one of them to prevent cluttering the diagram.}. This decay can be interpreted as Higgsing the irregular puncture which changes its nature from $(A_2,A_5)$ to $(A_1,A_3)$. The resulting AD theories are $(A_1,A_3)[1^4]$ and $(A_1,A_3)[2^2]$, respectively. Thus, we see from the decay point of view, that the Higgsing of the irregular puncture is naturally captured as well.

Crucially, the irregular puncture also triggers fissions, due to the presence of highly overbalanced nodes in the complete graph. These transitions are depicted in red in Figure~\ref{fig:Higgs_D9SU6}. 

For AD theories with more than one regular puncture, e.g.\ \cite{Xie:2012hs}, the same Higgsing process can be done as long as the magnetic quiver is made of only unitary gauge groups. 
For some 4d SCFTs it is well known that they may not be completely Higgsable. Therefore, in such cases, the top of the Higgsing diagram where we denote the theory with $\{1\}$ really refers to an SCFT with a trivial Higgs branch rather than a trivial theory.

\subsubsection{SCFTs with non-simply laced magnetic quivers}
\begin{figure}[h]
    \centering
    \begin{subfigure}[t]{0.45\textwidth}
    \centering
\includegraphics[page=2]{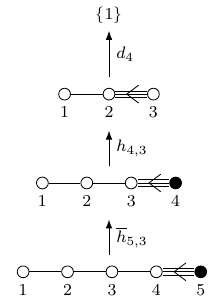}  
        \caption{Decay and fission}
        \label{subfig:2A3_tree}
    \end{subfigure}
\begin{subfigure}[t]{0.45\textwidth}
\centering
\includegraphics[page=3]{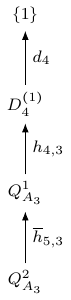}  
        \caption{Higgsing}
        \label{subfig:2A3_Higgs}
    \end{subfigure}
    \caption{\subref{subfig:2A3_tree}: Magnetic quivers from the decay algorithm. \subref{subfig:2A3_Higgs}: Higgsing pattern of the $Q_{A_3}^2$ SCFT. Here, $Q_{A_3}^{1}$ labels an $\Ncal=2$ SCFT with Higgs branch global symmetry $A_3$ \cite{Bourget:2020asf} and $D^{(1)}_4$ is a 4d SCFT with Higgs branch the one-$\sorm(8)$ instanton moduli space.
}
    \label{fig:non-simply_laced_new}
\end{figure}

As shown, for example, in the exhaustive lists of \cite{Bourget:2020asf,Bourget:2021csg}, magnetic quivers for 4d $\Ncal=2$ SCFTs are often non-simply laced, and the decay and fission algorithm applies in this case as well. 
For instance, consider the following quiver:
\begin{align}
\raisebox{-.5\height}{
\includegraphics[page=1]{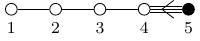}
}
\label{eq:ex_non-simply}
\end{align}
which is the magnetic quiver \cite{Bourget:2020asf} of a rank 2 $4d$ $\Ncal=2$ SCFT introduced in \cite{Zafrir:2016wkk}, labelled as $Q_{A_3}^{2}$ as it is part of the  $Q_{A_3}^{n}$ family of SCFTs, or labelled $\mathfrak{su}(5)_{16}$ in \cite{Martone:2021ixp,Bourget:2021csg}.  
Using the decay and fission algorithm, the arising Higgsing pattern between 4d $\Ncal=2$ SCFTs can be summarised as in Figure \ref{fig:non-simply_laced_new}.

\begin{figure}[h]
    \centering
    \begin{subfigure}[t]{0.45\textwidth}
    \centering
\includegraphics[page=8]{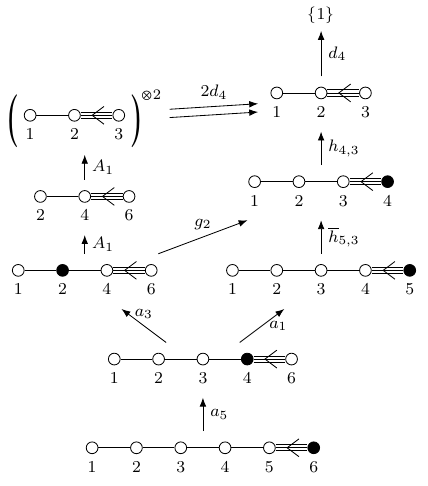}  
        \caption{}
        \label{subfig:nonsimplylacedtree}
    \end{subfigure}
\begin{subfigure}[t]{0.45\textwidth}
\centering
\includegraphics[page=9]{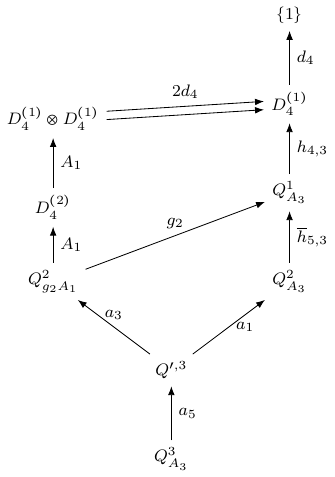}  
        \caption{}
        \label{subfig:nonsimplylacedHiggs}
    \end{subfigure}
    \caption{\subref{subfig:nonsimplylacedtree}: Decay and fission algorithm on the magnetic quiver of $Q_{A_3}^3$. 
    \subref{subfig:nonsimplylacedHiggs}: The Higgsings between 4d SCFTs. Here $Q^{2}_{g_2 A_1}$ is a rank 2 SCFT introduced in \cite{Argyres:2022lah}; its Higgs branch global symmetry is $\mathfrak{g}_2\times \mathfrak{su}(2)$.
    The transition $Q_{g_2 A_1}\to D_4^{(1)}\otimes D_4^{(1)}$ was conjectured in \cite{Bourget:2020mez} to be $k_{3}$. However, we conjecture that $k_3$ is \emph{not} an elementary transition, but rather composed of two consecutive $A_1$ transitions. The fission $D_4^{(2)}\to D_4^{(1)} \otimes D_4^{(1)}$ is analogous to the example in Section~\ref{sec:informal}; where $D_4^{(2)}$ is the rank 2 SCFT with Higgs branch the two $\sorm(8)$ instanton moduli space.
    The second magnetic quiver from the bottom in \ref{subfig:nonsimplylacedtree} has not yet been matched with a 4d theory. This is likely because the theory is a rank 3 4d theory, which are not yet classified. The candidate 4d theory has $\surmL(4)\times \surmL(2)$ global symmetry and is labelled as $Q^{\prime,3}$. }
    \label{fig:non-simply_laced}
\end{figure}

Applying the decay and fission algorithm to the rank 3 SCFT $Q_{A_3}^{3}$ in the same family, one obtains Figure \ref{subfig:nonsimplylacedtree}; which translates into a Higgsing pattern between 4d SCFTs as shown in Figure \ref{subfig:nonsimplylacedHiggs}.

\subsubsection{\texorpdfstring{An $\Scal$-fold theory -- Multiple affine Dynkin diagrams}{An S-fold theory -- Multiple affine Dynkin diagrams}}
\label{sec:stacksofaffine}

Consider the theory $\mathring{\Tcal}^{r=3}_{A_2,2}$ introduced in \cite{Bourget:2020mez}. This theory is an orbifold of $\Scal$-fold theories with the following magnetic quiver
\begin{align}
\raisebox{-.5\height}{
\includegraphics[page=2]{figures_inverted_sub_affine.pdf}
} \, . 
\label{fig:S-fold_example}
\end{align}
One observes a stack of three $a_2$ affine Dynkin diagrams. This signals the possibility of fission. The full diagram can be obtained by running the algorithm; we do not present it here, but focus on the bottom part, which shows \emph{four} possible Higgsing transitions: 
\begin{align}
\raisebox{-.5\height}{
\includegraphics[page=3]{figures_inverted_sub_affine.pdf}
}
\end{align}
This analysis shows that the magnetic quiver \eqref{fig:S-fold_example} admits two decay and two fission transitions; hence, the $\Scal$-fold theory $\mathring{\Tcal}^{r=3}_{A_2,2}$ should admit 4 distinct Higgs branch RG-flows:
\begin{align}
\raisebox{-.5\height}{
\includegraphics[page=4]{figures_inverted_sub_affine.pdf}
}
\end{align}
which completes the preliminary results of \cite[Fig.\ 12]{Bourget:2020mez}. The $\mathring{\mathcal{S}}^{r}_{G,\ell}$ theories are another type of $\Scal$-fold theories, and $I_G^{r}$ denotes the $r$ $G$-instanton theories.
Working out the entire Hasse diagram produces 21 symplectic leaves in total. This is straightforward, and they are not detailed here.

\subsection{\texorpdfstring{5d $\Ncal=1$ theories}{5d N=1 theories}} 

In this section we consider partial Higgsing of 5d SCFTs which can be deformed to SQCD theories with $\surm(N_c)_{|k|}$ with $N_f$ fundamental flavours, $N_{\Lambda^2}$ antisymmetrics and Chern-Simons levels $k$. The magnetic quivers of these theories have been studied in \cite{Cabrera:2018jxt,vanBeest:2020kou,VanBeest:2020kxw}. 
\subsubsection{Union of two cones}
A common feature of $5d$ $\Ncal=1$ SCFTs is that their Higgs branch can be the union of several hyper-K\"ahler cones. For example, consider $\surm(6)_{2}$ with $N_f=8$ fundamental flavors and CS-level $2$. The SCFT Higgs branch is composed of two cones, each associated with a magnetic quiver \cite[Tab.\ 7]{Cabrera:2018jxt}
\begin{align}
\raisebox{-.5\height}{
\includegraphics[page=1]{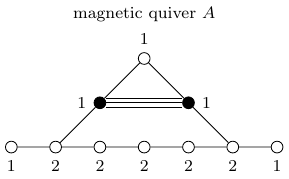}
}
\qquad 
\raisebox{-.5\height}{
\includegraphics[page=2]{figures_inverted_sub_5d.pdf}
}
\end{align}
The set of balanced nodes indicate two possible Higgsings: an $a_7$ and $a_1$ transition. Since the two cones intersects non-trivially, one should also inspect the magnetic quiver for the intersection between the two cones:
\begin{align}
    \raisebox{-.5\height}{
\includegraphics[page=3]{figures_inverted_sub_5d.pdf}
}
\end{align}
From the intersection quiver, one observes that the $a_7$ and $a_1$ transitions are indeed common to both quivers. This means that the $a_7$ and $a_1\cong A_1$ transition are part of the intersection of the two cones.

Performing the $a_7$ Higgsing via the decay algorithm results in
\begin{align}
    \raisebox{-.5\height}{
\includegraphics[page=4]{figures_inverted_sub_5d.pdf}
}
\end{align}
and the 5d $\Ncal=1$ SCFT whose Higgs branch is given by this set of magnetic quivers is $\surm(5)_2$ with $N_f=6$ fundamental hypermultiplets, see \cite[Tab.\ 7]{Cabrera:2018jxt}. 

\begin{figure}[t]
    \centering
    \includegraphics[page=7]{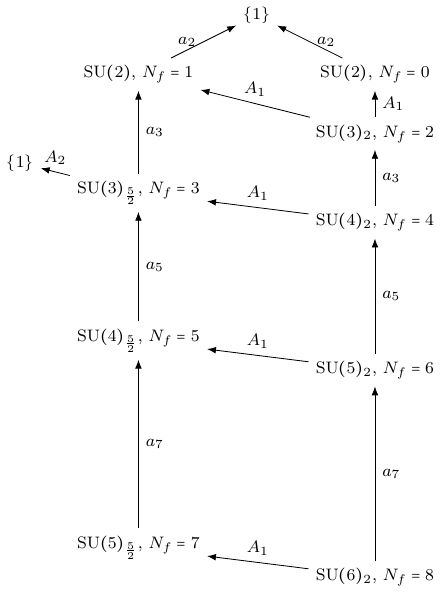}
    \caption{Higgsing between 5d SCFTs starting from 5d $\Ncal=1$ $\surm(6)_2$ with 8 fundamental hypermultiplets. 
    Furthermore, there could be discrete `fat points' in the moduli space as in \cite{Cremonesi:2015lsa} which are not sensitive in the magnetic quiver and hence the procedure here. 
   }
    \label{fig:5d_Higgsing}
\end{figure}

Alternatively, the $a_1$ transition leads to
\begin{align}
    \raisebox{-.5\height}{
\includegraphics[page=5]{figures_inverted_sub_5d.pdf}
}
\end{align}
which corresponds to the conformal fixed point of 5d $\Ncal=1$ $\surm(5)_{\frac{5}{2}}$ with $N_f=7$ fundamental flavours, as follows from comparing the magnetic quivers \cite[Tab.\ 2]{Cabrera:2018jxt}. 
Repeating the decay and fission algorithm reveals the partial Higgsing pattern between 5d SCFTs as depicted in Figure \ref{fig:5d_Higgsing}.

Besides the transitions that are at the intersection of both cones, there exist, in general, a transition $\mathfrak{g}$ that is allowed in one of the magnetic quivers, but not in the other. This is observed in the upper part of Figure \ref{fig:5d_Higgsing} both at $\surm(3)_{\frac{5}{2}}, N_f=3$ and at $\surm(3)_{2}, N_f=2$. For instance, with $\surm(3)_{\frac{5}{2}}, N_f=3$, the magnetic quivers are the following:
\begin{align}
    \raisebox{-.5\height}{
\includegraphics[page=6]{figures_inverted_sub_5d.pdf}
}
\end{align}
Here, the $A_2$ transition is only applicable to the left magnetic quiver. Naively, after performing this Higgsing, one of the cones gets Higgsed into a smaller cone whilst the other remains the same. We argue against such process. One argument is that the remaining magnetic quiver on the right is not a known magnetic quiver of 5d SCFT, which is fully classified for lower rank theories\footnote{An even more concrete example is to take the $E_3$ theory of $\surm(2)$ with 2 flavors in 5d. This theory's Higgs branch is a union of two cones that intersect trivially. One of the cone is $\mathbb{C}_2/\mathbb{Z}_2$ whereas the other is the one-$\mathfrak{su}(3)$ instanton moduli space. If doing the $A_1$ transition, thus Higgsing away one cone, does not affect the other cone, means that the remaining 5d SCFT is a rank 1 or rank 0 theory with  one-$\mathfrak{su}(3)$ instanton moduli space as its Higgs branch. Such a theory does not exist in the literature and we do not expect it to. This is also observed in the 5-brane web, Higgsing in on direction implies that no longer two maximal decompositions exist. Hence, there are no more two cones}. Therefore, what we expect that happens is that once such an asymmetric Higgsing occurs, the cone not involved in the Higgsing disappears entirely. In other words, the hypermultiplets that generate the other cone become free fields. 

\subsubsection{Higgsing between 5d SCFTs}
\begin{sidewaysfigure}
\centering
\resizebox{\textheight}{!}{
\includegraphics[page=1,width=1.2\textwidth]{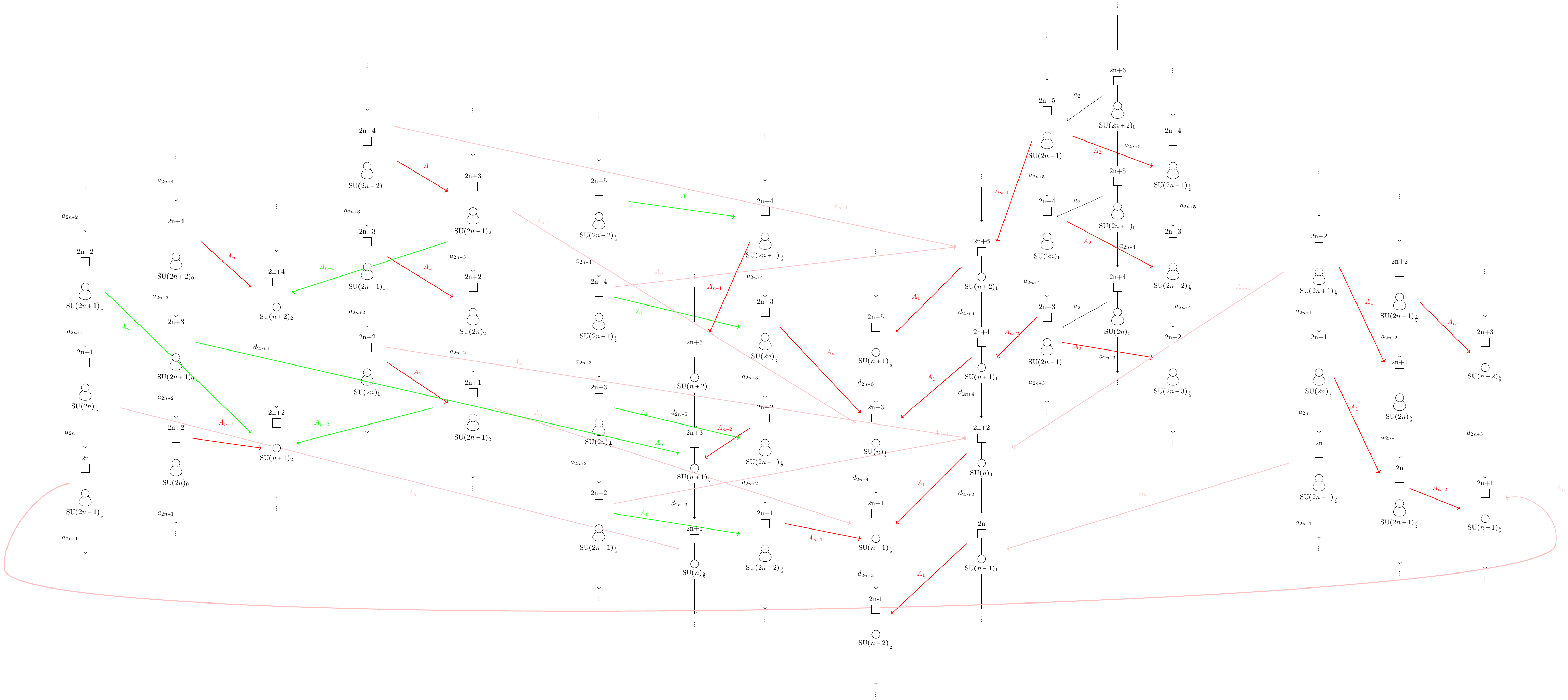}}
\caption{Pattern of minimal Higgs transitions between 5d $\Ncal=1$ SCFTs as predicted by the decay and fission algorithm. The 5d gauge theory descriptions are single $\surm(N)$ gauge groups, $N_f$ fundamental hypermultiplets (the square box), $N_{\Lambda^2}=1$ anti-symmetric hypermultiplets (the loops), and a CS-level.  The algorithm utilises magnetic quivers found in \cite{Cabrera:2018jxt,VanBeest:2020kxw}. The black transitions are non-Abelian, whereas the pink, green, and red transitions are all Kleinian-$A_k=\mathbb{C}^2/\mathbb{Z}_{k+1}$-type. The reason for using green, red, and pink coloured transitions is to improve readability. Despite the drawing, the Higgsing pattern is more apparent when more members of each family is drawn. }
\label{hugeeeetable}
\end{sidewaysfigure}

The Higgsings of $5d$ $\Ncal=1$ SCFTs is a fairly non-trivial process since the UV theories do not admit a Lagrangian description. Furthermore, massless instantons contribute to the Higgs branch as well, so apart from Higgsings where VEVs are given to hypermultiplets, there are also Higgsings where VEVs are given to the massless instantons. In Figure \ref{hugeeeetable}, we \emph{predict via decay and fission} a single Higgsing tree where many families of 5d SCFTs, whose magnetic quivers are detailed in \cite{Cabrera:2018jxt,VanBeest:2020kxw,vanBeest:2020kou}, are shown to be Higgsed into each other.

\subsection{\texorpdfstring{6d $\Ncal=(1,0)$ theories}{6d N=(1,0) theories}}
The existence of 6d supersymmetric theories (e.g.\ superconformal and little string theories) have revolutionised the understanding of quantum field theories. Again, such theories are inherently strongly coupled and the systematic analysis of the Higgs branch RG-flows is challenging, but of utmost importance. Some earlier works include \cite{Heckman:2015ola,Heckman:2015axa,Heckman:2016ssk,Mekareeya:2016yal,Hassler:2019eso,Baume:2021qho,Giacomelli:2022drw,Fazzi:2022hal,Fazzi:2022yca}. 

\subsubsection{\texorpdfstring{$\surm(6)$ with $N_f=14$ fundamentals and one antisymmetric}{SU6 with 14 fundamentals and 1 antisymmetric}}
\begin{figure}[h]
    \centering
    \begin{subfigure}[t]{0.485\textwidth}
        \flushleft
        \includegraphics[page=8]{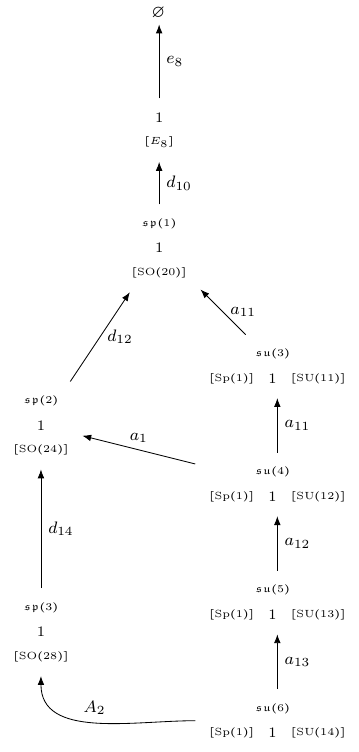}
        \caption{Higgsing}
         \label{subfig:6d_ex_Higgs}
    \end{subfigure}
      \begin{subfigure}[t]{0.485\textwidth}
        \flushright
        \includegraphics[page=5]{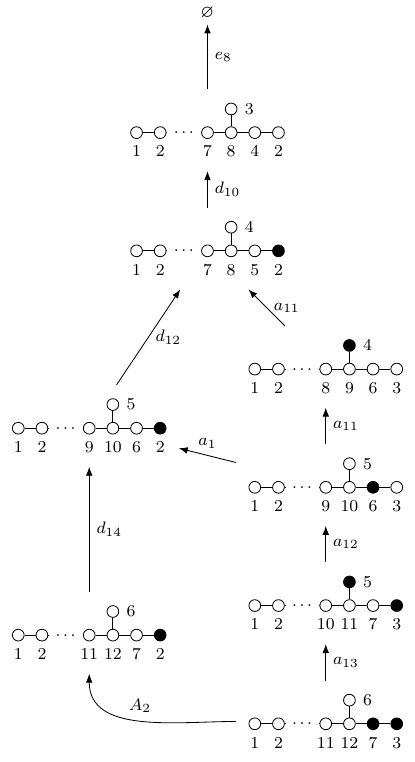}
        \caption{Decay algorithm}
         \label{subfig:6d_ex_inverted}
    \end{subfigure}
    \caption{The Higgsing pattern between 6d SCFTs starting from $\surm(6)$ with 14 fundamentals hypermultiplets and one anti-symmetric hypermultiplet. \subref{subfig:6d_ex_Higgs}: The SCFTs are labelled by their 6d generalised quiver description $\underset{[G_f]}{\overset{\mathfrak{g}}{n}}$, with gauge algebra $\mathfrak{g}$ and flavour symmetry $G_f$. Eg.\ the theory $\underset{[E_8]}{1}$, i.e.\ the E-string theory, gives rise to the one-$E_8$ instanton moduli space in the tensionless limit. \subref{subfig:6d_ex_inverted}: The Higgsing pattern derived via the decay algorithm (no fission occurs here).}
    \label{fig:6d_Higgsing}
\end{figure}

We consider an example of 6d $\Ncal=(1,0)$ SCFTs
\begin{equation}
    [\sprm(1)]\overset{\surmL(6)}{1}[\surm(14)] \, , 
\end{equation}
with a deformation to $\surm(6)$ gauge theory with $N_f=14$ fundamentals and one antisymmetric. 
For this theory, there is a known superconformal fixed point at the origin of the tensor branch and the magnetic quiver is given by \cite{Mekareeya:2017jgc,Cabrera:2019izd}
\begin{equation}
\raisebox{-.5\height}{
\includegraphics[page=1]{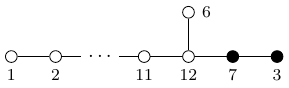}
}
\label{eq:MagQuiv:SU14_6d}
\end{equation}
Applying the decay and fission algorithm leads to the Higgsing pattern between 6d $\Ncal=(1,0)$ SCFTs  depicted in Figure \ref{fig:6d_Higgsing}. 
This diagram neatly ties together two features: 
Firstly, the partial Higgsing pattern of the two families 6d SCFTs defined on a single $-1$ curves \cite[Fig.\ 5]{DelZotto:2018tcj}, which have known magnetic quivers.
Secondly, the geometric data reproduces the conjectured Hasse diagram of \cite[Fig.\ 18]{Bourget:2019aer}. 

\subsubsection{Orbi-instanton and higher-rank E-string}
\begin{sidewaysfigure}
    \centering
    \resizebox{\textheight}{!}{\includegraphics[page=11]{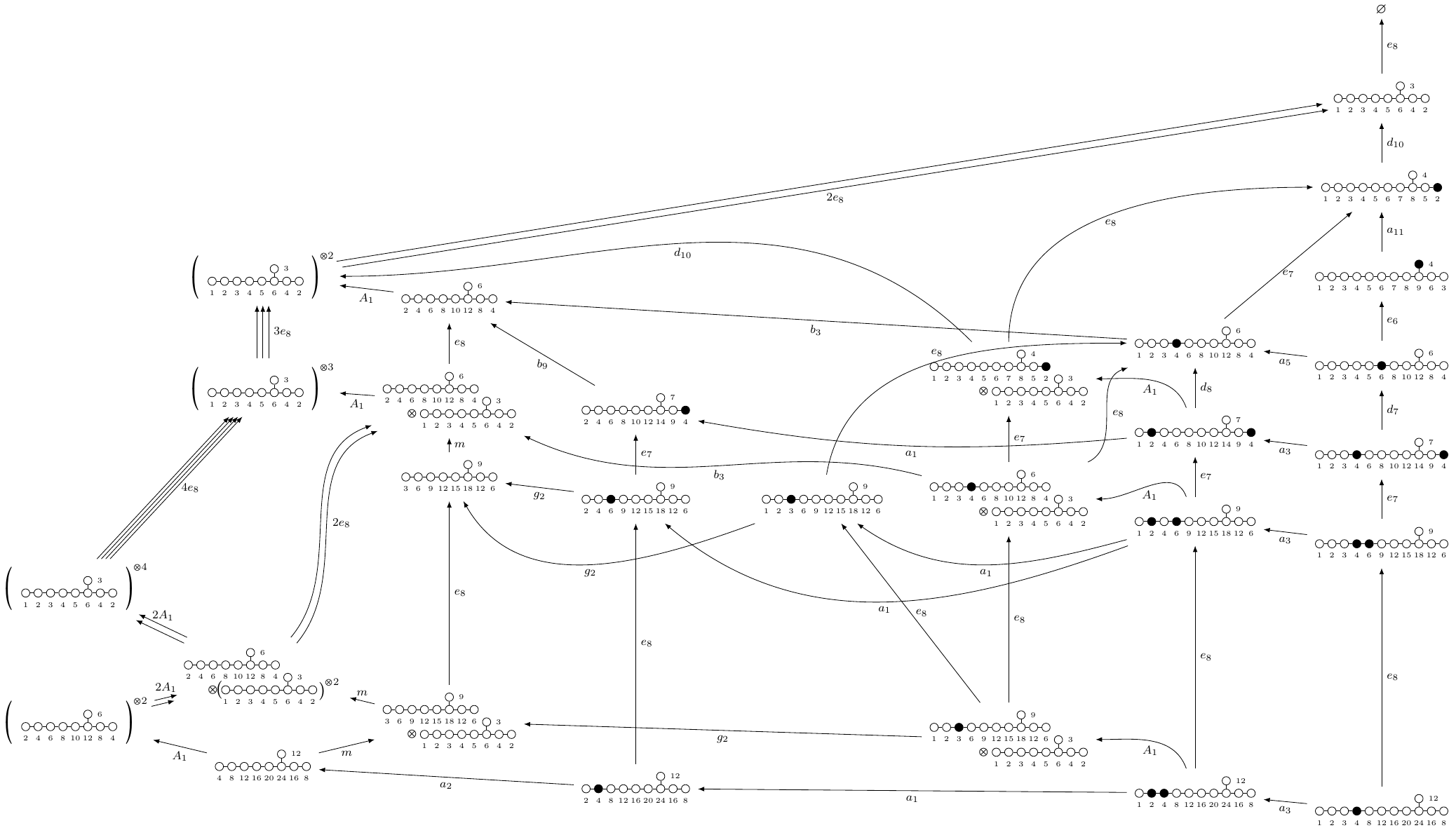}}
    \caption{Decay and fission algorithm applied to the magnetic quiver of the orbi-instanton theory \eqref{eq:orbi-instanton}. The right-hand column focuses on the decays that encode the change of homomorphisms $\ZZ_k \hookrightarrow E_8$ for the M9 wall. The left-hand parts detail the physics of $E_8$-instantons.}
    \label{fig:6d_ex_inverted}
\end{sidewaysfigure}

\begin{sidewaysfigure}
    \centering
    \resizebox{\textheight}{!}{\includegraphics[page=12]{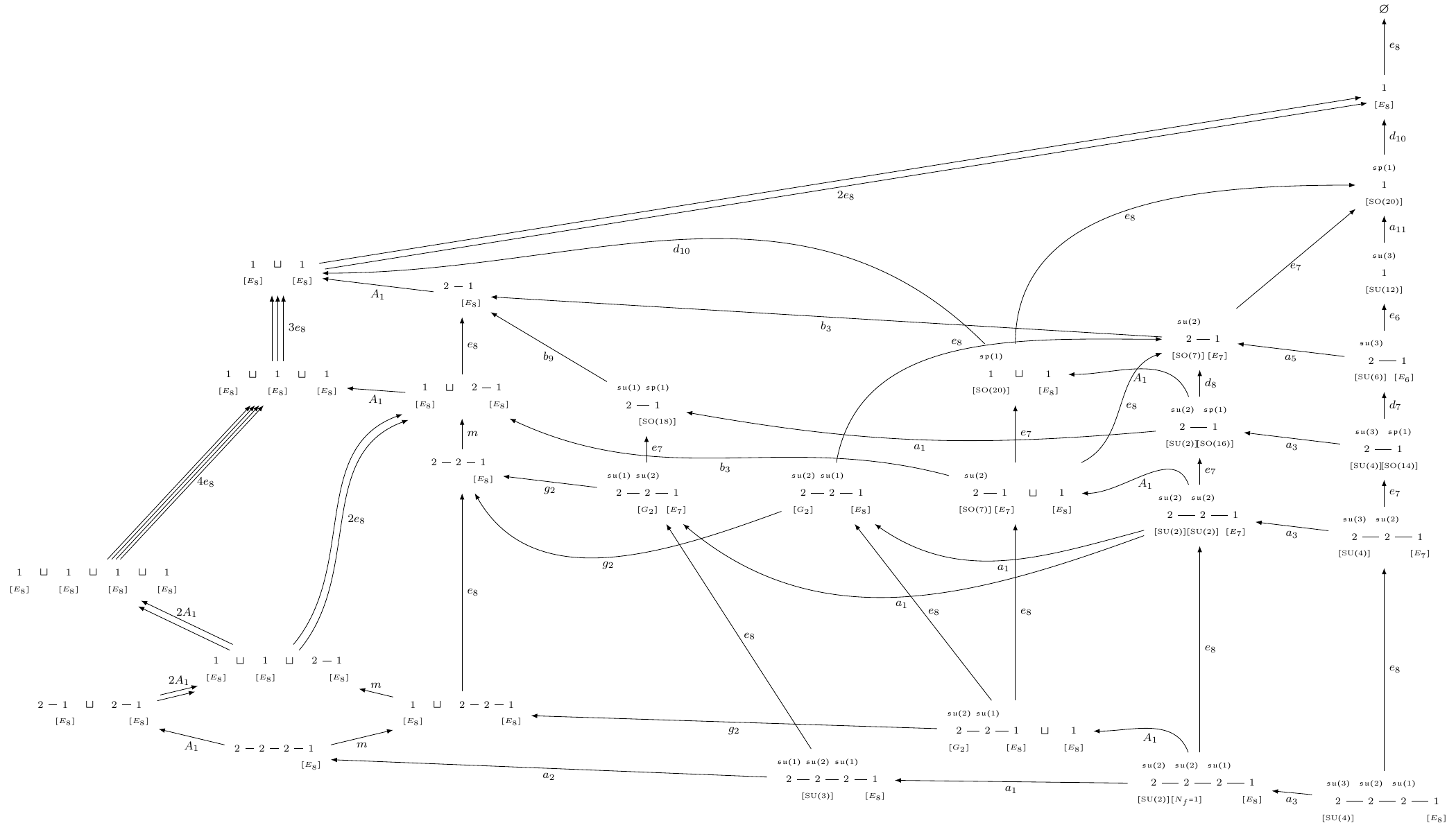}}
    \caption{Higgs branch RG-flows between between the descendants of the orbi-instanton theory \eqref{eq:orbi-instanton}. The right-hand column shows flows that change of homomorphisms $\ZZ_k \hookrightarrow E_8$. The left-hand parts detail the physics of $E_8$-instantons: 4 M5 branes in an M9 wall.}
    \label{fig:6d_ex_Higgs}
\end{sidewaysfigure}

Consider the 6d $\Ncal=(1,0)$ theories originating from M5 branes on an A-type ALE space near an M9 plane. Specifically, consider 4 M5 branes, the $\C^2 \slash \ZZ_3$ ALE space, and choose the trivial embedding $\ZZ_3 \hookrightarrow E_8$
\begin{align}
[\surm(3)]\underset{[N_f=1]}{\overset{\surmL(3)}{2}} -   \overset{\surmL(2)}{2}-\overset{\surmL(1)}{2}-\underset{[E_8]}{1}
\qquad \cong \qquad
    \underset{[\surm(4)]}{\overset{\surmL(3)}{2}} -   \overset{\surmL(2)}{2}-\overset{\surmL(1)}{2}-\underset{[E_8]}{1} 
    \label{eq:orbi-instanton}
\end{align}
wherein the last description keeps the flavour symmetries manifest. The magnetic quiver is readily available \cite{Mekareeya:2017jgc,Cabrera:2019izd} and serves as starting point for the decay and fission algorithm. The result is the Higgsing pattern of Figure~\ref{fig:6d_ex_inverted}, which can be translated to the Higgs mechanism of the 6d theory shown in Figure~\ref{fig:6d_ex_Higgs}.

A few comments are in order.
Firstly, the 6d theory 
\begin{align}
\overset{\surmL(1)}{2}-\underset{[\surm(3)]}{\overset{\surmL(2)}{2}} -   \overset{\surmL(1)}{2}-\underset{[E_8]}{1} 
\end{align}
admits an RG-flow to the rank-4 E-string theory
\begin{align}
2-2 - 2-\underset{[E_8]}{1} 
\end{align}
i.e.\ 4 M5 branes near an M9 plane. The transition type is deduced as follows:
\begin{align}
\raisebox{-.5\height}{
\includegraphics[page=13]{figures_inverted_sub_6d.pdf}
}
\end{align}
which is in fact the twisted affine $a_2$ Dynkin quiver.

Next, in Figure \ref{fig:6d_ex_Higgs} there are two 6d theories with a $G_2$ global symmetry factor that can flow to the rank-2 E-string theory. To begin with, consider 
\begin{align}
    \underset{[G_2]}{\overset{\surmL(2)}{2}} -   \overset{\surmL(1)}{2}-\underset{[E_8]}{1} 
\end{align}
and the $g_2$ Higgs branch flows is deduced via
\begin{align}
\raisebox{-.5\height}{
\includegraphics[page=14]{figures_inverted_sub_6d.pdf}
}
\end{align}
which produced the affine $G_2$ Dynkin diagram due to $\gcd=3$ of the decay product.
The other 6d  theory is
\begin{align}
   \overset{\surmL(1)}{2}- \underset{[G_2]}{\overset{\surmL(2)}{2}} -   \underset{[E_7]}{1} 
\end{align}
and the transverse geometry of the RG-flow is seen via
\begin{align}
\raisebox{-.5\height}{
\includegraphics[page=15]{figures_inverted_sub_6d.pdf}
}
\end{align}
which again results in the affine $G_2$ Dynkin quiver.

Via similar arguments, one deduces that the transition geometries for the RG-flows  
\begin{alignat}{3}
    &\underset{[\sorm(7)]}{\overset{\surmL(1)}{2}} -
    \underset{[\sorm(16)]}{\overset{\sprmL(1)}{1}} 
  & \qquad  &\xrightarrow[]{\quad b_9 \quad }    \qquad&
   &2 -\underset{[E_8]}{1}  \\
   &\underset{[\sorm(7)]}{\overset{\surmL(2)}{2}} -\underset{[E_7]}{1} 
  & \qquad  &\xrightarrow[]{\quad b_3 \quad }    \qquad &
   &2 -\underset{[E_8]}{1} 
\end{alignat}
ending in the rank-2 E-string theory.

Besides the decays there are also fissions. A natural fission occurs for the orbi-instanton theory of $N$ M5 branes on $\C^2\slash \mathbb{Z}_k$ with trivial homomorphism $[1^k]: \mathbb{Z}_k \hookrightarrow E_8$ (assuming $N-1 \geq k$)  
\begin{align}
\label{eq:fission_orbi-instanton}
\begin{aligned}
\overbrace{\underset{[\surm(k)]}{\overset{\surmL(k)}{2}}
- 
\overset{\surmL(k)}{2}
-
\cdots
-
\overset{\surmL(k)}{2}
-
\underset{[N_f=1]}{\overset{\surmL(k)}{2}} 
-
\overset{\surmL(k-1)}{2}
-
\cdots
-
\overset{\surmL(2)}{2}
-
\overset{\surmL(1)}{2}}^{\text{$N-1$ curves of self-intersection $-2$ }} -   \underset{[E_8]}{1} \\
\longrightarrow \quad 
\overbrace{\underset{[\surm(k)]}{\overset{\surmL(k)}{2}}
-
\cdots
-
\overset{\surmL(k)}{2}
-
\underset{[N_f=1]}{\overset{\surmL(k)}{2}} 
-
\overset{\surmL(k-1)}{2}
-
\cdots
-
\overset{\surmL(2)}{2}
-
\overset{\surmL(1)}{2}}^{\text{$N-2$ curves of self-intersection $-2$ }} -   \underset{[E_8]}{1}
\quad \sqcup \quad 
  \underset{[E_8]}{1}
\end{aligned}
\end{align}
into the orbi-instanton theory with $N-1$ M5 branes and a single E-string theory, provided $N-2\geq k$. In the limiting case $N-1 =k$, no fission occurs, as removing an M5 renders it impossible to realise the trivial boundary conditions $[1^k]$ in a supersymmetry preserving way. That is the reason why the initial theory \eqref{eq:orbi-instanton} cannot fission, while its daughter theory 
\begin{align}
    \underset{[\surm(2)]}{\overset{\surmL(2)}{2}} 
    - 
    \underset{[N_f=1]}{\overset{\surmL(2)}{2}}
    -
    \overset{\surmL(1)}{2}
    -
    \underset{[E_8]}{1} 
  \quad   \longrightarrow \quad
     \underset{[G_2]}{\overset{\surmL(2)}{2}} 
    - 
   \overset{\surmL(1)}{2}
    -
    \underset{[E_8]}{1}
    \quad \sqcup \quad
    \underset{[E_8]}{1}
\end{align}
can. Analogous arguments hold for fissions of orbi-instanton theories with different homomorphisms $\mathbb{Z}_k \hookrightarrow E_8$, as exemplified in Figure~\ref{fig:6d_ex_Higgs}.

From the magnetic quiver perspective, the starting point for the fission \eqref{eq:fission_orbi-instanton} is 
\begin{align}
\raisebox{-.5\height}{
\includegraphics[page=16]{figures_inverted_sub_6d.pdf}
}
\end{align}
which shows that for $N-1>k$ both the $\urm(k)$ and $\urm(N)$ nodes are over-balanced. Therefore, the removal of an M5, given by $N\to N-1$, leads to another good magnetic quiver for the orbi-instanton fission product. For $N-1=k$, however, the $\urm(k)$ nodes is balanced, which implies that the hypothetical fission step $N\to N - 1$ yields a bad magnetic quiver.

Lastly, the fission of stacks of affine $E_8$ Dynkin diagrams has a natural 6d manifestation: a stack of $k$ affine $E_8$ Dynkin diagrams describes the rank $k$ E-string theory\footnote{The Higgs branch of the rank $k$ E-string theory is the moduli space of $k$ $E_8$ instantons \cite{Cordova:2015fha}.}. The splitting $k\to \ell+ (k-\ell)$ is then the fission of the rank $k$ E-string into the rank $\ell$ E-string theory and the rank $(k-\ell)$ E-string theory, which is indicated by the $\sqcup$. This is a known Higgs branch RG-flow \cite{Heckman:2015ola,Cordova:2015fha}; in M-theory language: the stack of $k$ M5 branes is separated into a stack of $\ell$ M5s and a stack of $(k-\ell)$ M5 branes \emph{along} a direction parallel to the M9 plane. 
As a consequence, the subset of RG-flows belonging to the E-string theories display the physics of $k$ M5 branes within an M9 --- giving rise to $\mathrm{Sym}^n(\C^2)$. Specifically, the splitting of the rank-4 E-string into to product theory of rank-$n_i$ E-string theories with $\{n_i\}$ an integer partition of 4 gives rise to the Hasse diagram of $\mathrm{Sym}^4(\C^2)$, see \cite[eq.\ (B.21)]{Bourget:2022ehw} and \cite[Fig.\ 2]{Bourget:2022tmw} for a corrected version. The complete Hasse diagram of the RG-flows of the higher-rank E-string theories is then a special case of an instanton moduli space \cite{Bourget:2022ehw}, as recently reviewed in \cite{Lawrie:2023uiu}. It is, however, crucial to remark that the decay and fission algorithm captures all these subtleties (e.g.\ transitions that are unions, like $2 A_1 = A_1 \cup A_1$) without prior input and solely from the initial quiver.

\subsubsection{Higgs branches of little string theories}
\label{sec:LST}
\begin{figure}[h]
    \centering
    \includegraphics[page=3]{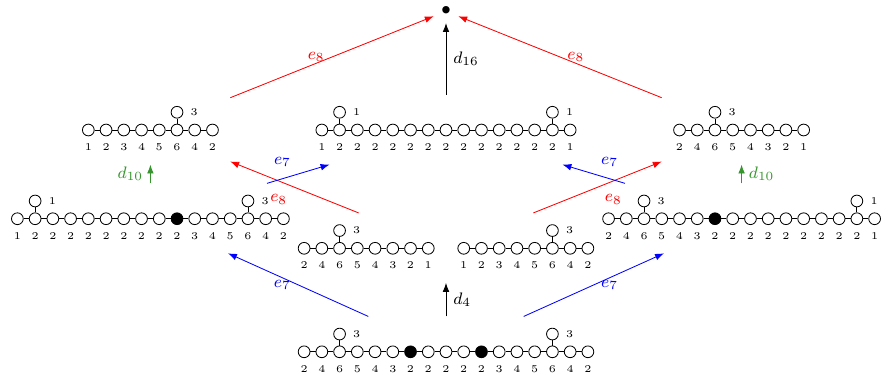}
    \caption{Exemplary Hasse diagram for the LST \eqref{eq:LST_E7} the decay algorithm.}
    \label{fig:Hasse_LST_E7}
\end{figure}

The decay and fission algorithm can also be applied to magnetic quivers of little string theories (LSTs), which have recently been proposed \cite{DelZotto:2023nrb,Lawrie:2023uiu,Mansi:2023faa}. Here, a proof of concept is provided by focusing on simple examples \cite{DelZotto:2023nrb}: the $(e^\prime)$ LST of type A with $G=\surm(2)$ and two choices $([\rho_L],[\rho_R])$ of $\mathbb{Z}_2 \to E_8$ embeddings. 
\paragraph{A first example.}
The simplest case preserves $(E_7\times \surm(2))\times (\surm(2)\times E_7)$, i.e.\ embedding $([2],[2])$, \cite{DelZotto:2022ohj}
\begin{align}
\raisebox{-.5\height}{
    \includegraphics[page=1]{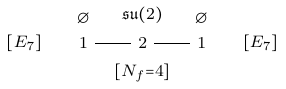}
    }
    \label{eq:LST_E7}
\end{align}
and already provides an intricate pattern of Higgs branch RG-flows predicted by the decay and fission algorithm. The starting point is the magnetic quiver \cite{DelZotto:2023nrb} 
\begin{align}
\raisebox{-.5\height}{
    \includegraphics[page=2]{figures_inverted_sub_LST.pdf}
    }
\label{eq:MQ_LST_E7}
\end{align}
for which the decay process traces out the Hasse diagram shown in Figure \ref{fig:Hasse_LST_E7}. 

Here, a new phenomenon appears: the magnetic quiver after the $d_4$ transitions is \emph{reducible}, see Section \ref{sec:formal}. Therefore, deriving the geometry of the slice requires a rebalancing node for each \emph{irreducible} component. In detail:
\begin{align}
\raisebox{-.5\height}{
    \includegraphics[page=5]{figures_inverted_sub_LST.pdf}
    }
    \label{eq:LST_E7_enhanced}
\end{align}
so that the transition is identified as $d_4$. The reducibility can be understood from the physical system: one starts from \eqref{eq:LST_E7} and ends with a curve configuration that does not support any gauge algebras. The M-theory picture is that of two M9 walls on a finite interval with one M5 brane inside each $E_8$ wall. Each of them yields an  affine $E_8$ Dynkin quiver for its magnetic degrees of freedom (i.e.\ Higgs branch moduli).

\begin{figure}[h]
    \centering
    \includegraphics[page=4]{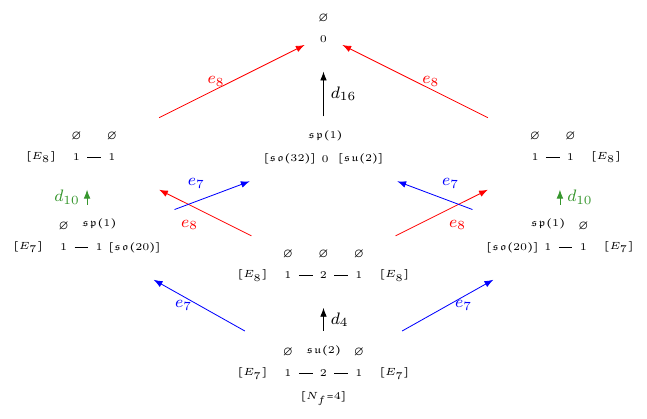}
    \caption{Higgsing pattern of LST \eqref{eq:LST_E7}.}
    \label{fig:Higgs_LST_E7}
\end{figure}

The magnetic quivers in Figure \ref{fig:Hasse_LST_E7} can be identified with descendants of the little string theory \eqref{eq:LST_E7} and the predicted Higgs branch RG-flow pattern is summarised in Figure \ref{fig:Higgs_LST_E7}.

Note that the initial LST curve configuration $1-2-1$ can reach the LST defined on a curve $0$ of vanishing self-intersection by collapsing each of the initial $-1$ curves. The Higgs branch RG-flows of these $0$ curve models have been analysed in \cite{Mansi:2023faa}.

\begin{sidewaysfigure}
    \centering
    \resizebox{\textheight}{!}{
    \includegraphics[page=8]{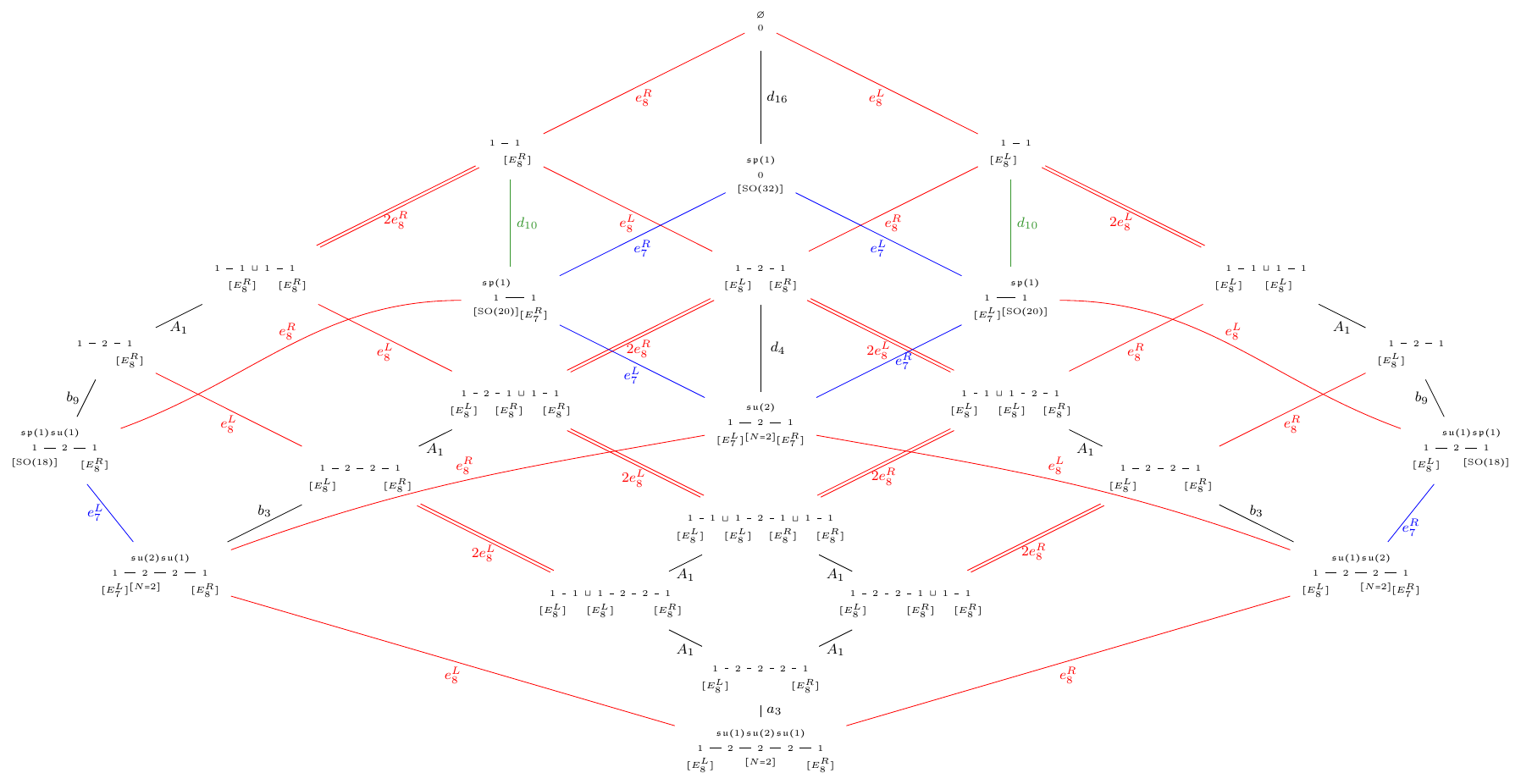}}
    \caption{Exemplary Hasse diagram for the LST \eqref{eq:LST_E8}. Besides decays, there are fission present. These originate from the fact that the system has M-theory description of two M9 wall at the ends of the interval $S^1\slash \mathbb{Z}_2$ with two M5 branes on each wall. Then, the stacks of M5 can undergo fission.
    Note also that the Hasse diagram of Figure \ref{fig:Hasse_LST_E7} is clearly contained as subdiagram.}
    \label{fig:Hasse_LST_E8}
\end{sidewaysfigure}

\paragraph{A second example.}
A more elaborate example is given by the choice of $([1^2],[1^2])$ embeddings \cite{DelZotto:2022ohj}
\begin{align}
\raisebox{-.5\height}{
\includegraphics[page=6]{figures_inverted_sub_LST.pdf} 
}
\label{eq:LST_E8}
\end{align}
which preserves the full $E_8 \times E_8$ symmetry. Starting again from the magnetic 
quiver \cite{DelZotto:2023nrb}
\begin{align}
\raisebox{-.5\height}{
\includegraphics[page=7]{figures_inverted_sub_LST.pdf} 
}
\label{eq:MQ_LST_E8}
\end{align}
one can apply decay and fission algorithm to traces out the Hasse diagram, shown in Figure \ref{fig:Hasse_LST_E8}.

This already contains some more new features that deserved to be commented on. 
Firstly, the bottom $a_3$ transition again leads to a \emph{reducible} quiver
\begin{align}
\raisebox{-.5\height}{
\includegraphics[page=9]{figures_inverted_sub_LST.pdf} 
}
\end{align}
and the geometry of the slice is deduced again by a rebalancing node for each reducible component. In addition, each reducible component has $\gcd =2$, such that non-simply laced edges are required. As a result, the slice is read off to be $a_3$. Physically, this transitions takes the curve configuration \eqref{eq:LST_E8} with non-trivial gauge algebras to the same curve configuration but with trivial gauge algebras. The reason why this yields two reducible magnetic quivers is understood from the M-theory picture: two M9 walls on a finite interval with two M5 branes inside each M9. This setting also illuminates the subsequent diamond of $A_1$ fissions: each stack of two M5s can fission independently. The M5 that is moved off along the, here right-hand side, M9 leads to the LST $1- 1 [E_8]$.

Similarly, the $b_3$ transition in Figure \ref{fig:Hasse_LST_E8} is the result of a reducible quiver after a decay transition: 
\begin{align}
\raisebox{-.5\height}{
\includegraphics[page=10]{figures_inverted_sub_LST.pdf} 
}
\end{align}
such that the slice is read off by introducing two rebalancing nodes. Here, the two irreducible quivers have $\gcd=2$ and $1$, respectively. The geometry is identified as $b_3$. Again, the transition starts from the curve configuration $1-2-2-1$ with non-trivial gauge algebras and ends up with trivial gauge algebras. Thus, the reducible magnetic quiver is clear from the M-theory picture: one M9 wall hosts 2 M5 branes, while the other holds only one. As a result, there exists a subsequent $A_1$ fission of the stack of 2 M5s.

Likewise, the $b_9$ transition in Figure \ref{fig:Hasse_LST_E8} stems from a non-trivial $\gcd$ of the decay product: 
\begin{align}
\raisebox{-.5\height}{
\includegraphics[page=12]{figures_inverted_sub_LST.pdf} 
}
\end{align}
such that the magnetic quiver for the slice has a non-simply laced edge. As with the other two examples, the starting point is a $1-2-1$ curve configuration with gauge algebras that become trivial during the decay. The decay product originates from the M-theory setting of two M9 walls on a finite interval and one wall contains a stack of two M5 branes. Again, this opens up the possibility of a subsequent fission into $1-1[E_8] \sqcup 1-1[E_8]$, i.e.\ two LSTs.

\FloatBarrier

\section{Discussion, conclusions, and open questions}
\label{sec:conclusion}
Magnetic quivers have proven to be immensely useful for both providing insights in Higgs branches of strongly coupled supersymmetric theories and revealing geometric features. Particularly important are algorithms of quiver operations. 
In this work, we introduced the \emph{decay and fission} algorithm, which at its core only relies on linear algebra. 
Beyond the simplicity, this algorithm allows us to determine precisely the phase diagram of the theory with the Higgs branch of the Higgsed theories identified along with the transverse slices that underlines the geometric structure of the vacuum. The former allowed us to find the Higgsing diagram of SCFTs and gauge theories with 8 supercharges in dimensions $3,4,5,6$ of which a plethora of examples are shown in the text. 

\subsection{Comparison with standard quiver subtraction}
We now see how the decay and fission algorithm distinguishes from another similar but fundamentally different quiver operation: quiver subtraction \cite{Bourget:2019aer}.
It is insightful to compare the two methods and analyse the different quivers that arise. In several examples, both techniques have been applied and presented: e.g.\ Figures \ref{fig:splitting_ex}, \ref{fig:3d_Higgs_unframed}, \ref{fig:SU-U_example}, \ref{fig:diagramInstantons}, and \ref{fig:4d_inverted_vs_standard}. 
An immediate observation is that the decay and fission algorithm keeps the shape of the quiver similar after subtraction, whereas quiver subtraction drastically changes the shape due to rebalancing gauge nodes. Here, the aim is to discuss the geometric meaning behind the two different sets of magnetic quivers.

\begin{figure}[ht]
    \centering
    \includegraphics[page=1]{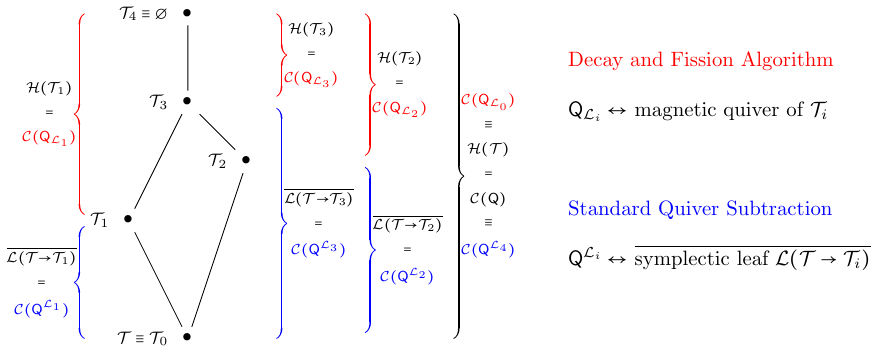}
    \caption{Higgs branch Hasse diagram of a theory $\Tcal$ with 8 supercharges in $d=3,4,5,6$. A theory with this Higgsing pattern is, for example, the 6d $\Ncal=(1,0)$ $\surm(4)$ theory with $N_f=12$  and $N_{\Lambda^2}=1$, see \cite[Fig.\ 6--9]{Bourget:2019aer}.   
    Both, decay and fission as well as quiver subtraction, produce magnetic quivers that capture certain transverse slices. Denote $X=\Higgs(\Tcal) = \Coulomb(\MQ)$ the entire symplectic singularity. Also, label the leaves $\Lcal_i$ with the theories $\Tcal_i$ at the end of the Higgs branch RG-flow: i.e.\ $\Lcal_i \equiv \Lcal(\Tcal \to \Tcal_i)$. Then the slice
    $\Scal(\Lcal_i,X)=\mathcal{H}(\mathcal{T}_i)$ is the Higgs branch of the Higgsed theory $\mathcal{T}_i$ and equals the Coulomb branch $\mathcal{C}(\MQ_{\Lcal_i})$ of the quiver $\MQ_{\Lcal_i}$. These are results obtained from the \textcolor{red}{decay and fission algorithm}, introduced in this work. In contrast, the slices $\Scal(\Lcal_0,\Lcal_i) = \overline{\Lcal}_i = \Coulomb(\MQ^{\Lcal_i})$ are described by the quivers $\MQ^{\Lcal_i}$ obtained from the \textcolor{blue}{standard quiver subtraction}. 
    }
    \label{fig:compare_Hasse}
\end{figure}

Upon Higgsing a theory $\Tcal$ to a theory $\Tcal'$, the Higgs branch geometry changes: a coarse signature of this change is the fact that $\dim \, \Higgs(\Tcal') < \dim \, \Higgs(\Tcal)$. A finer relation is the fact that $\Higgs(\Tcal)$ is a conical symplectic singularity with a (finite) stratification into symplectic leaves, and $\Higgs(\Tcal')$ appears as a \emph{transverse slice} in $\Higgs(\Tcal)$ to a certain leaf. Due to these restrictions, it is often possible to turn the logic around, and to identify, for each transverse slice in $\Higgs(\Tcal)$, which theory $\Tcal'$ possesses this slice as its Higgs branch. Therefore, \emph{understanding the geometry of transverse slices in $\Higgs(\Tcal)$ is often sufficient to identify the possible Higgsings of a given theory $\Tcal$}. 

The decay and fission algorithm introduced in this work achieves precisely this goal, in cases where $\Higgs(\Tcal)$ admits a \emph{unitary magnetic quiver}, i.e.\ when there is a quiver $\MQ$ with unitary gauge nodes such that $\Coulomb^{\mathrm{3d} \, \Ncal=4} (\MQ) = \Higgs(\Tcal)$. Specifically, the \underline{\emph{decay and fission}} algorithm produces 
\begin{compactenum}[($i$)]
    \item The poset of symplectic leaves and elementary degenerations between adjacent leaves.
    \item For each leaf $\mathcal{L}$ in $\Higgs(\Tcal)$ a magnetic quiver $\MQ_\mathcal{L} (\Tcal)$ for the transverse slice to this leaf. 
\end{compactenum}
It is then often possible to identify, using techniques developed elsewhere, theories $\Tcal_{\mathcal{L}}$ with Higgs branch admitting magnetic quivers $\MQ_\mathcal{L} (\Tcal)$. Combined with output $(i)$, this then produces the complete Higgsing graph for $\Tcal$. Multiple examples have been detailed in Section \ref{sec:examples}. Note that a primitive version of the decay and fission algorithm was applied to height 2 nilpotent orbits in \cite{Cabrera:2018ann}. There are also quiver algorithms which achieves this to some level of success in \cite{Giacomelli:2022drw,Gu:2022dac, Fazzi:2022yca}.

Before this, it is important to mention that $(i)$ can also be obtained from the previously introduced \underline{\emph{quiver subtraction}}  algorithm \cite{Bourget:2019aer}, which produces 
\begin{compactenum}
    \item[$(i)$] The poset of symplectic leaves and elementary degenerations between adjacent leaves.
    \item[$(ii')$] For each leaf $\mathcal{L}$ in $\Higgs(\Tcal)$ a magnetic quiver $\MQ^\mathcal{L} (\Tcal)$ for the closure of this leaf. 
\end{compactenum}
Here the poset is obtained inductively beginning from the higher dimensional leaves.

We note that both algorithms output the poset $(i)$: this should be seen as an internal consistency check of the validity of the whole procedure. But the magnetic quivers are definitely not the same, and reflect distinct physical phenomena, see Figures \ref{fig:splitting_ex}, \ref{fig:3d_Higgs_unframed}, \ref{fig:SU-U_example}, \ref{fig:diagramInstantons}, and \ref{fig:4d_inverted_vs_standard}. In terms of Higgsing $\mathcal{T}$ this means the quiver subtraction algorithm can at best provide the Higgs branch dimension of $\mathcal{T}'$ which is unlikely to be enough information to determine the theory $\mathcal{T}'$. On the other hand, the decay and fission tells everything about the Higgs branch of $\mathcal{T}'$, a substantial amount of information that allows an easy identification of $\mathcal{T}'$. 
 
Lastly, the two algorithms are conceptually different: while quiver subtraction is an iterative process relying in an input list of minimal degenerations; in contrast, decay and fission is ``holistic'' in nature. More concretely, the latter takes the magnetic quiver and outputs the entire Hasse diagram solely from the shape the quiver data $(A,K)$, see Section \ref{sec:algorithm}. The appearing minimal degenerations are then byproducts of the first step. For instance, more complicated quivers such as non-simply laced quivers with lengths greater than two or with gauged groups having more than one adjoint hypermultiplet the list of minimal degenerations are likely incomplete. This makes it challenging for the quiver subtraction algorithm, but the decay and fission algorithm, which only relies on the balance of the gauge groups and their relative position in the quiver, generates the phase diagram without difficulty.

The standard quiver subtraction algorithm \cite{Bourget:2019aer} had undergone several improvements \cite{Bourget:2020mez,Bourget:2022ehw,Bourget:2022tmw} to handle quivers with more unique features (such as gauge nodes with adjoint hypermultiplets, non-simply laced edges, and stacks of affine Dynkin quivers). The decay and fission algorithm now incorporates all these features as well. However, in the future, it is possible that both algorithms require further refinements as we explore the quiver landscape further; see below for open questions.

\subsection{Further applications and future directions}
The decay and fission algorithm has other interesting applications beyond just Higgsing theories. 

\paragraph{Identifying any transverse slice.}
The standard quiver subtraction introduced in \cite{Bourget:2019aer} and the decay and fission algorithm, introduced in this paper, can be combined to generate $3d$ $\Ncal=4$ quivers whose Coulomb branch describe any transverse slices between two symplectic leaves in a Hasse diagram. This is clear since combining $(ii)$ in decay and fission and $(ii')$ in quiver subtraction allows one to obtain magnetic quivers $\MQ_{\mathcal{L}}^{\mathcal{L}'}$ for the transverse slice to $\mathcal{L}$ into $\overline{{\mathcal{L}}'}$ for any ${\mathcal{L}} < {\mathcal{L}}'$. 

However, before enthusiastically using this method to find quivers whose Coulomb branch corresponds to more exotic slices such as the 1-dimensional non-normal slice denoted as $m$, it is important to note that the procedure \emph{may not} apply to quivers with ``decorations''. These appear during standard subtraction on quivers containing multiple affine Dynkin diagrams or quivers containing gauge nodes with adjoint hypermultiplets \cite{Bourget:2022ehw}. We leave such possibilities for the future when decorated quivers are better understood. 

A final remark concerns appearing product theories after fissions. Recalling the introductory cartoon in Figure \ref{fig:intro_cartoon} and its detailed version in Figure \ref{fig:splitting_ex}, one observes leaves with product theories, but the Hasse diagram does not contain the product of Hasse diagram as sub-graph. This is a manifestation of the known fact \cite[Fig.\ 3]{Bourget:2022tmw} that the Hasse diagram of a transverse slice does not have to be a subdiagram of the Hasse diagram of the full theory.

\paragraph{Identifying elementary slices.} 
Classifying all elementary slices (to be precise, these are isolated symplectic singularities \cite{beauville2000symplectic,bellamy2023new}) has long been an intriguing goal in the mathematics community. Recently, \cite{Bourget:2021siw,Bourget:2022tmw}, many new quivers that correspond to elementary slices have been discovered and added to the dictionary. An interesting application of the decay and fission algorithm is that it can assist in this search, as explained in Section \ref{sec:classification}. An obvious extension of the results obtained in that section, in which we restricted ourselves to quivers with 3 nodes or less, is the classification of isolated singularities corresponding to unitary quivers with arbitrarily many nodes.

\paragraph{Finding new SCFTs.}
The classification of SCFTs with 8 supercharges in various dimensions has long been an interesting goal of the community. The decay and fission algorithm acting as a Higgsing algorithm can help find missing SCFTs. If, for example, a complete classification up to rank $r$ $4d$ $\mathcal{N}=2$ SCFT is made, then given a rank $r$ SCFT $\mathcal{T}$, all the theories $\mathcal{T}'$ it can Higgs to as indicated by our algorithm \emph{must} appear in the classification as well. 

\paragraph{Orthosymplectic quivers.}
The decay and fission algorithm applies to a large, but still restricted set of (magnetic) quivers. Given the wealth of known magnetic quivers, the decay and fission algorithm needs to be extended to accommodated further quiver theories; for instance, orthosymplectic quivers. For those theories, even standard quiver subtraction is still under development.

Moreover, for all examples considered here, the practical implementation (\ref{Rule:1}--\ref{Rule:3}) is equivalent to the decay algorithm, provided no fission can appear. It might however be that some new minimal transitions are found. These would not be covered by (\ref{Rule:1}--\ref{Rule:3}), but the decay and fission algorithm is fully capable to detect them. This is because decay and fission algorithm \emph{does not} rely on a list of known minimal transition (which is necessarily incomplete).

In the case of orthosymplectic quivers, one thing that prevents us from constructing the standard quiver subtraction is that we do not know all orthosymplectic quivers that correspond to minimal transitions. The number of inequivalent orthosymplectic quivers that leads to the same moduli space seems more diverse than unitary quivers (see e.g.\ orthosymplectic quivers of one-$E_6$ instanton moduli space in \cite{Bourget:2020gzi,Bourget:2020xdz,Bourget:2021xex}). On the other hand, the decay and fission algorithm, as mentioned above, does not require obtaining a list of minimal transitions. So finding all possible daughter/grand daughter orthosymplectic quivers that it can decay to should be a much simpler task.  

\paragraph{Higgs branch RG-flows.}
One application of Higgs branch RG-flows between SCFTs already seen in the literature is to prove certain maximization theorems. For instance, the a-theorem that $a$ anomaly decreases under RG-flows in 6d SCFTs in \cite{Fazzi:2023ulb}, the c-theorem for $c$-anomaly in \cite{Heckman:2015axa}. Also, other quantities such as the F-theorem for $5d$ $\mathcal{N}=1$ SCFTs that the free energy decreases under Higgs branch RG-flows in \cite{Fluder:2020pym}. With the decay and fission algorithm, we now have all possible minimal Higgsings for these SCFTs, which can be easily missed in the literature, and can give a more complete prove of these theorems.  

\paragraph{Acknowledgements.}
We thank Christopher Beem, Simone Giacomelli, Julius Grimminger, Jie Gu, Amihay Hanany, Patrick Jefferson, Yunfeng Jiang, Monica Kang, Hee-Cheol Kim, Sung-Soo Kim, Craig Lawrie, Lorenzo Mansi, Carlos Nunez, Matteo Sacchi, Sakura Sch\"afer-Nameki for many discussions on this topic.
The results of this work have been presented prior publication (sometimes under the preliminary name ``inverted quiver subtraction'') by the authors at the following events: AB at String Theory Seminar of DESY/University Hamburg on Jun 26, 2023. MS at workshop ``Symplectic Singularities and Supersymmetric QFT'' in Amiens on Jul 20, 2023. ZZ at Oxford's seminar on Jan 22, 2023, KIAS String Seminars on Feb 13, 2023, Swansea University  on May 31, 2023, Seoul National University CTP seminars Oct 17, 2023 and IPMU string-math seminar on Dec 22, 2023. We are grateful for the support during these events and the stimulating discussions. AB is partly supported by the ERC Consolidator Grant 772408-Stringlandscape. This research was supported in part by grant NSF PHY-2309135 to the Kavli Institute for Theoretical Physics (KITP).
MS is supported Austrian Science Fund (FWF), START project ``Phases of quantum field theories: symmetries and vacua'' STA 73-N. MS also acknowledges support from the Faculty of Physics, University of Vienna. MS was previously supported by Shing-Tung Yau Center, Southeast University. ZZ is supported by the ERC Consolidator Grant \# 864828 ``Algebraic Foundations of Supersymmetric Quantum Field Theory'' (SCFTAlg). ZZ is grateful of the hospitality of New College, Oxford where the final stages of this work is completed.

\bibliographystyle{JHEP}
\bibliography{bibli.bib}

\providecommand{\href}[2]{#2}\begingroup\raggedright\begin{thebibliography}{10}

\bibitem{englert1964broken}
F.~Englert and R.~Brout, \emph{{Broken Symmetry and the Mass of Gauge Vector
  Mesons}}, \href{https://doi.org/10.1103/PhysRevLett.13.321}{\emph{Phys. Rev.
  Lett.} {\bfseries 13} (1964) 321}.

\bibitem{higgs1964broken}
P.W.~Higgs, \emph{{Broken Symmetries and the Masses of Gauge Bosons}},
  \href{https://doi.org/10.1103/PhysRevLett.13.508}{\emph{Phys. Rev. Lett.}
  {\bfseries 13} (1964) 508}.

\bibitem{guralnik1964global}
G.S.~Guralnik, C.R.~Hagen and T.W.B.~Kibble, \emph{{Global Conservation Laws
  and Massless Particles}},
  \href{https://doi.org/10.1103/PhysRevLett.13.585}{\emph{Phys. Rev. Lett.}
  {\bfseries 13} (1964) 585}.

\bibitem{kibble1967symmetry}
T.W.B.~Kibble, \emph{{Symmetry breaking in nonAbelian gauge theories}},
  \href{https://doi.org/10.1103/PhysRev.155.1554}{\emph{Phys. Rev.} {\bfseries
  155} (1967) 1554}.

\bibitem{Bourget:2019aer}
A.~Bourget, S.~Cabrera, J.F.~Grimminger, A.~Hanany, M.~Sperling, A.~Zajac
  et~al., \emph{{The Higgs mechanism --- Hasse diagrams for symplectic
  singularities}}, \href{https://doi.org/10.1007/JHEP01(2020)157}{\emph{JHEP}
  {\bfseries 01} (2020) 157}
  [\href{https://arxiv.org/abs/1908.04245}{{\ttfamily 1908.04245}}].

\bibitem{Cabrera:2018jxt}
S.~Cabrera, A.~Hanany and F.~Yagi, \emph{{Tropical Geometry and Five
  Dimensional Higgs Branches at Infinite Coupling}},
  \href{https://doi.org/10.1007/JHEP01(2019)068}{\emph{JHEP} {\bfseries 01}
  (2019) 068} [\href{https://arxiv.org/abs/1810.01379}{{\ttfamily
  1810.01379}}].

\bibitem{Cabrera:2019izd}
S.~Cabrera, A.~Hanany and M.~Sperling, \emph{{Magnetic quivers, Higgs branches,
  and 6d $N$=(1,0) theories}}, \href{https://doi.org/10.1007/JHEP07(2019)137,
  10.1007/JHEP06(2019)071}{\emph{JHEP} {\bfseries 06} (2019) 071}
  [\href{https://arxiv.org/abs/1904.12293}{{\ttfamily 1904.12293}}].

\bibitem{Bourget:2019rtl}
A.~Bourget, S.~Cabrera, J.F.~Grimminger, A.~Hanany and Z.~Zhong, \emph{{Brane
  Webs and Magnetic Quivers for SQCD}},
  \href{https://doi.org/10.1007/JHEP03(2020)176}{\emph{JHEP} {\bfseries 03}
  (2020) 176} [\href{https://arxiv.org/abs/1909.00667}{{\ttfamily
  1909.00667}}].

\bibitem{Cabrera:2019dob}
S.~Cabrera, A.~Hanany and M.~Sperling, \emph{{Magnetic Quivers, Higgs Branches,
  and 6d N=(1,0) Theories -- Orthogonal and Symplectic Gauge Groups}},
  \href{https://doi.org/10.1007/JHEP02(2020)184}{\emph{JHEP} {\bfseries 02}
  (2020) 184} [\href{https://arxiv.org/abs/1912.02773}{{\ttfamily
  1912.02773}}].

\bibitem{Bourget:2020gzi}
A.~Bourget, J.F.~Grimminger, A.~Hanany, M.~Sperling and Z.~Zhong,
  \emph{{Magnetic Quivers from Brane Webs with O5 Planes}},
  \href{https://doi.org/10.1007/JHEP07(2020)204}{\emph{JHEP} {\bfseries 07}
  (2020) 204} [\href{https://arxiv.org/abs/2004.04082}{{\ttfamily
  2004.04082}}].

\bibitem{Bourget:2020asf}
A.~Bourget, J.F.~Grimminger, A.~Hanany, M.~Sperling, G.~Zafrir and Z.~Zhong,
  \emph{{Magnetic quivers for rank 1 theories}},
  \href{https://doi.org/10.1007/JHEP09(2020)189}{\emph{JHEP} {\bfseries 09}
  (2020) 189} [\href{https://arxiv.org/abs/2006.16994}{{\ttfamily
  2006.16994}}].

\bibitem{Bourget:2020xdz}
A.~Bourget, J.F.~Grimminger, A.~Hanany, R.~Kalveks, M.~Sperling and Z.~Zhong,
  \emph{{Magnetic Lattices for Orthosymplectic Quivers}},
  \href{https://doi.org/10.1007/JHEP12(2020)092}{\emph{JHEP} {\bfseries 12}
  (2020) 092} [\href{https://arxiv.org/abs/2007.04667}{{\ttfamily
  2007.04667}}].

\bibitem{Closset:2020scj}
C.~Closset, S.~Schafer-Nameki and Y.-N.~Wang, \emph{{Coulomb and Higgs Branches
  from Canonical Singularities: Part 0}},
  \href{https://doi.org/10.1007/JHEP02(2021)003}{\emph{JHEP} {\bfseries 02}
  (2021) 003} [\href{https://arxiv.org/abs/2007.15600}{{\ttfamily
  2007.15600}}].

\bibitem{Akhond:2020vhc}
M.~Akhond, F.~Carta, S.~Dwivedi, H.~Hayashi, S.-S.~Kim and F.~Yagi,
  \emph{{Five-brane webs, Higgs branches and unitary/orthosymplectic magnetic
  quivers}}, \href{https://doi.org/10.1007/JHEP12(2020)164}{\emph{JHEP}
  {\bfseries 12} (2020) 164}
  [\href{https://arxiv.org/abs/2008.01027}{{\ttfamily 2008.01027}}].

\bibitem{vanBeest:2020kou}
M.~van Beest, A.~Bourget, J.~Eckhard and S.~Schafer-Nameki, \emph{{(Symplectic)
  Leaves and (5d Higgs) Branches in the Poly(go)nesian Tropical Rain Forest}},
  \href{https://doi.org/10.1007/JHEP11(2020)124}{\emph{JHEP} {\bfseries 11}
  (2020) 124} [\href{https://arxiv.org/abs/2008.05577}{{\ttfamily
  2008.05577}}].

\bibitem{Giacomelli:2020gee}
S.~Giacomelli, M.~Martone, Y.~Tachikawa and G.~Zafrir, \emph{{More on
  $\mathcal{N} =2$ S-folds}},
  \href{https://doi.org/10.1007/JHEP01(2021)054}{\emph{JHEP} {\bfseries 01}
  (2021) 054} [\href{https://arxiv.org/abs/2010.03943}{{\ttfamily
  2010.03943}}].

\bibitem{Bourget:2020mez}
A.~Bourget, S.~Giacomelli, J.F.~Grimminger, A.~Hanany, M.~Sperling and
  Z.~Zhong, \emph{{S-fold magnetic quivers}},
  \href{https://doi.org/10.1007/JHEP02(2021)054}{\emph{JHEP} {\bfseries 02}
  (2021) 054} [\href{https://arxiv.org/abs/2010.05889}{{\ttfamily
  2010.05889}}].

\bibitem{VanBeest:2020kxw}
M.~Van~Beest, A.~Bourget, J.~Eckhard and S.~Sch\"afer-Nameki, \emph{{(5d
  RG-flow) Trees in the Tropical Rain Forest}},
  \href{https://doi.org/10.1007/JHEP03(2021)241}{\emph{JHEP} {\bfseries 03}
  (2021) 241} [\href{https://arxiv.org/abs/2011.07033}{{\ttfamily
  2011.07033}}].

\bibitem{Closset:2020afy}
C.~Closset, S.~Giacomelli, S.~Schafer-Nameki and Y.-N.~Wang, \emph{{5d and 4d
  SCFTs: Canonical Singularities, Trinions and S-Dualities}},
  \href{https://doi.org/10.1007/JHEP05(2021)274}{\emph{JHEP} {\bfseries 05}
  (2021) 274} [\href{https://arxiv.org/abs/2012.12827}{{\ttfamily
  2012.12827}}].

\bibitem{Akhond:2021knl}
M.~Akhond, F.~Carta, S.~Dwivedi, H.~Hayashi, S.-S.~Kim and F.~Yagi,
  \emph{{Factorised 3d $\mathcal{N}=4$ orthosymplectic quivers}},
  \href{https://doi.org/10.1007/JHEP05(2021)269}{\emph{JHEP} {\bfseries 05}
  (2021) 269} [\href{https://arxiv.org/abs/2101.12235}{{\ttfamily
  2101.12235}}].

\bibitem{Bourget:2021csg}
A.~Bourget, J.F.~Grimminger, M.~Martone and G.~Zafrir, \emph{{Magnetic quivers
  for rank 2 theories}},
  \href{https://doi.org/10.1007/JHEP03(2022)208}{\emph{JHEP} {\bfseries 03}
  (2022) 208} [\href{https://arxiv.org/abs/2110.11365}{{\ttfamily
  2110.11365}}].

\bibitem{vanBeest:2021xyt}
M.~van Beest and S.~Giacomelli, \emph{{Connecting 5d Higgs branches via
  Fayet-Iliopoulos deformations}},
  \href{https://doi.org/10.1007/JHEP12(2021)202}{\emph{JHEP} {\bfseries 12}
  (2021) 202} [\href{https://arxiv.org/abs/2110.02872}{{\ttfamily
  2110.02872}}].

\bibitem{Sperling:2021fcf}
M.~Sperling and Z.~Zhong, \emph{{Balanced B and D-type orthosymplectic quivers
  \textemdash{} magnetic quivers for product theories}},
  \href{https://doi.org/10.1007/JHEP04(2022)145}{\emph{JHEP} {\bfseries 04}
  (2022) 145} [\href{https://arxiv.org/abs/2111.00026}{{\ttfamily
  2111.00026}}].

\bibitem{Nawata:2021nse}
S.~Nawata, M.~Sperling, H.E.~Wang and Z.~Zhong, \emph{{Magnetic quivers and
  line defects \textemdash{} On a duality between 3d $ \mathcal{N} $ = 4
  unitary and orthosymplectic quivers}},
  \href{https://doi.org/10.1007/JHEP02(2022)174}{\emph{JHEP} {\bfseries 02}
  (2022) 174} [\href{https://arxiv.org/abs/2111.02831}{{\ttfamily
  2111.02831}}].

\bibitem{Akhond:2022jts}
M.~Akhond, F.~Carta, S.~Dwivedi, H.~Hayashi, S.-S.~Kim and F.~Yagi,
  \emph{{Exploring the orthosymplectic zoo}},
  \href{https://doi.org/10.1007/JHEP05(2022)054}{\emph{JHEP} {\bfseries 05}
  (2022) 054} [\href{https://arxiv.org/abs/2203.01951}{{\ttfamily
  2203.01951}}].

\bibitem{Giacomelli:2022drw}
S.~Giacomelli, M.~Moleti and R.~Savelli, \emph{{Probing 7-branes on
  orbifolds}}, \href{https://doi.org/10.1007/JHEP08(2022)163}{\emph{JHEP}
  {\bfseries 08} (2022) 163}
  [\href{https://arxiv.org/abs/2205.08578}{{\ttfamily 2205.08578}}].

\bibitem{Hanany:2022itc}
A.~Hanany and M.~Sperling, \emph{{Magnetic quivers and negatively charged
  branes}}, \href{https://doi.org/10.1007/JHEP11(2022)010}{\emph{JHEP}
  {\bfseries 11} (2022) 010}
  [\href{https://arxiv.org/abs/2208.07270}{{\ttfamily 2208.07270}}].

\bibitem{Fazzi:2022hal}
M.~Fazzi and S.~Giri, \emph{{Hierarchy of RG flows in 6d (1, 0)
  orbi-instantons}}, \href{https://doi.org/10.1007/JHEP12(2022)076}{\emph{JHEP}
  {\bfseries 12} (2022) 076}
  [\href{https://arxiv.org/abs/2208.11703}{{\ttfamily 2208.11703}}].

\bibitem{Bourget:2022tmw}
A.~Bourget and J.F.~Grimminger, \emph{{Fibrations and Hasse diagrams for 6d
  SCFTs}}, \href{https://doi.org/10.1007/JHEP12(2022)159}{\emph{JHEP}
  {\bfseries 12} (2022) 159}
  [\href{https://arxiv.org/abs/2209.15016}{{\ttfamily 2209.15016}}].

\bibitem{Fazzi:2022yca}
M.~Fazzi, S.~Giacomelli and S.~Giri, \emph{{Hierarchies of RG flows in 6d (1,
  0) massive E-strings}},
  \href{https://doi.org/10.1007/JHEP03(2023)089}{\emph{JHEP} {\bfseries 03}
  (2023) 089} [\href{https://arxiv.org/abs/2212.14027}{{\ttfamily
  2212.14027}}].

\bibitem{Nawata:2023rdx}
S.~Nawata, M.~Sperling, H.E.~Wang and Z.~Zhong, \emph{{3d $\mathcal{N}=4$
  mirror symmetry with 1-form symmetry}},
  \href{https://doi.org/10.21468/SciPostPhys.15.1.033}{\emph{SciPost Phys.}
  {\bfseries 15} (2023) 033}
  [\href{https://arxiv.org/abs/2301.02409}{{\ttfamily 2301.02409}}].

\bibitem{Bourget:2023cgs}
A.~Bourget, J.F.~Grimminger, A.~Hanany, R.~Kalveks, M.~Sperling and Z.~Zhong,
  \emph{{A Tale of N Cones}},
  \href{https://arxiv.org/abs/2303.16939}{{\ttfamily 2303.16939}}.

\bibitem{DelZotto:2023nrb}
M.~Del~Zotto, M.~Fazzi and S.~Giri, \emph{{The Higgs branch of heterotic ALE
  instantons}},  \href{https://arxiv.org/abs/2307.11087}{{\ttfamily
  2307.11087}}.

\bibitem{Lawrie:2023uiu}
C.~Lawrie and L.~Mansi, \emph{{The Higgs Branch of Heterotic LSTs: Hasse
  Diagrams and Generalized Symmetries}},
  \href{https://arxiv.org/abs/2312.05306}{{\ttfamily 2312.05306}}.

\bibitem{Mansi:2023faa}
L.~Mansi and M.~Sperling, \emph{{Unravelling T-Duality: Magnetic Quivers in
  Rank-zero Little String Theories}},
  \href{https://arxiv.org/abs/2312.12510}{{\ttfamily 2312.12510}}.

\bibitem{Fazzi:2023ulb}
M.~Fazzi, S.~Giri and P.~Levy, \emph{{Proving the 6d a-theorem with the double
  affine Grassmannian}},  \href{https://arxiv.org/abs/2312.17178}{{\ttfamily
  2312.17178}}.

\bibitem{Ferlito:2016grh}
G.~Ferlito and A.~Hanany, \emph{{A tale of two cones: the Higgs Branch of Sp(n)
  theories with 2n flavours}},
  \href{https://arxiv.org/abs/1609.06724}{{\ttfamily 1609.06724}}.

\bibitem{Cremonesi:2013lqa}
S.~Cremonesi, A.~Hanany and A.~Zaffaroni, \emph{{Monopole operators and Hilbert
  series of Coulomb branches of $3d$ $\mathcal{N} = 4$ gauge theories}},
  \href{https://doi.org/10.1007/JHEP01(2014)005}{\emph{JHEP} {\bfseries 01}
  (2014) 005} [\href{https://arxiv.org/abs/1309.2657}{{\ttfamily 1309.2657}}].

\bibitem{Cremonesi:2014xha}
S.~Cremonesi, G.~Ferlito, A.~Hanany and N.~Mekareeya, \emph{{Coulomb Branch and
  The Moduli Space of Instantons}},
  \href{https://doi.org/10.1007/JHEP12(2014)103}{\emph{JHEP} {\bfseries 12}
  (2014) 103} [\href{https://arxiv.org/abs/1408.6835}{{\ttfamily 1408.6835}}].

\bibitem{Cabrera:2018ann}
S.~Cabrera and A.~Hanany, \emph{{Quiver Subtractions}},
  \href{https://doi.org/10.1007/JHEP09(2018)008}{\emph{JHEP} {\bfseries 09}
  (2018) 008} [\href{https://arxiv.org/abs/1803.11205}{{\ttfamily
  1803.11205}}].

\bibitem{Eckhard:2020jyr}
J.~Eckhard, S.~Sch\"afer-Nameki and Y.-N.~Wang, \emph{{Trifectas for T$_{N}$ in
  5d}}, \href{https://doi.org/10.1007/JHEP07(2020)199}{\emph{JHEP} {\bfseries
  07} (2020) 199} [\href{https://arxiv.org/abs/2004.15007}{{\ttfamily
  2004.15007}}].

\bibitem{Martone:2021ixp}
M.~Martone, \emph{{Testing our understanding of SCFTs: a catalogue of rank-2 $
  \mathcal{N} $ = 2 theories in four dimensions}},
  \href{https://doi.org/10.1007/JHEP07(2022)123}{\emph{JHEP} {\bfseries 07}
  (2022) 123} [\href{https://arxiv.org/abs/2102.02443}{{\ttfamily
  2102.02443}}].

\bibitem{Closset:2021lwy}
C.~Closset, S.~Sch\"afer-Nameki and Y.-N.~Wang, \emph{{Coulomb and Higgs
  branches from canonical singularities. Part I. Hypersurfaces with smooth
  Calabi-Yau resolutions}},
  \href{https://doi.org/10.1007/JHEP04(2022)061}{\emph{JHEP} {\bfseries 04}
  (2022) 061} [\href{https://arxiv.org/abs/2111.13564}{{\ttfamily
  2111.13564}}].

\bibitem{Santilli:2021rlf}
L.~Santilli and M.~Tierz, \emph{{Crystal bases and three-dimensional
  \ensuremath{\mathcal{N}} = 4 Coulomb branches}},
  \href{https://doi.org/10.1007/JHEP03(2022)073}{\emph{JHEP} {\bfseries 03}
  (2022) 073} [\href{https://arxiv.org/abs/2111.05206}{{\ttfamily
  2111.05206}}].

\bibitem{Arias-Tamargo:2021ppf}
G.~Arias-Tamargo, A.~Bourget and A.~Pini, \emph{{Discrete gauging and Hasse
  diagrams}},
  \href{https://doi.org/10.21468/SciPostPhys.11.2.026}{\emph{SciPost Phys.}
  {\bfseries 11} (2021) 026}
  [\href{https://arxiv.org/abs/2105.08755}{{\ttfamily 2105.08755}}].

\bibitem{Mu:2023uws}
J.~Mu, Y.-N.~Wang and H.N.~Zhang, \emph{{5d SCFTs from Isolated Complete
  Intersection Singularities}},
  \href{https://arxiv.org/abs/2311.05441}{{\ttfamily 2311.05441}}.

\bibitem{Kraft1980}
H.~Kraft and C.~Procesi, \emph{Minimal singularities in ${GL}_n$},
  {\emph{Inventiones mathematicae} {\bfseries 62} (1980) 503}.

\bibitem{Kraft1982}
H.~Kraft and C.~Procesi, \emph{On the geometry of conjugacy classes in
  classical groupes.}, {\emph{Commentarii mathematici Helvetici} {\bfseries 57}
  (1982) 539}.

\bibitem{fu2017generic}
B.~Fu, D.~Juteau, P.~Levy and E.~Sommers, \emph{Generic singularities of
  nilpotent orbit closures}, {\emph{Advances in Mathematics} {\bfseries 305}
  (2017) 1} [\href{https://arxiv.org/abs/1502.05770}{{\ttfamily 1502.05770}}].

\bibitem{bellamy2023new}
G.~Bellamy, C.~Bonnaf{\'e}, B.~Fu, D.~Juteau, P.~Levy and E.~Sommers, \emph{A
  new family of isolated symplectic singularities with trivial local
  fundamental group}, {\emph{Proceedings of the London Mathematical Society}
  {\bfseries 126} (2023) 1496}
  [\href{https://arxiv.org/abs/2112.15494}{{\ttfamily 2112.15494}}].

\bibitem{Bourget:2021xex}
A.~Bourget, J.F.~Grimminger, A.~Hanany, R.~Kalveks, M.~Sperling and Z.~Zhong,
  \emph{{Folding orthosymplectic quivers}},
  \href{https://doi.org/10.1007/JHEP12(2021)070}{\emph{JHEP} {\bfseries 12}
  (2021) 070} [\href{https://arxiv.org/abs/2107.00754}{{\ttfamily
  2107.00754}}].

\bibitem{Bourget:2022ehw}
A.~Bourget, J.F.~Grimminger, A.~Hanany and Z.~Zhong, \emph{{The Hasse diagram
  of the moduli space of instantons}},
  \href{https://doi.org/10.1007/JHEP08(2022)283}{\emph{JHEP} {\bfseries 08}
  (2022) 283} [\href{https://arxiv.org/abs/2202.01218}{{\ttfamily
  2202.01218}}].

\bibitem{Bourget:2023dkj}
A.~Bourget, M.~Sperling and Z.~Zhong, \emph{{Decay and Fission of Magnetic
  Quivers}}, \href{https://doi.org/10.1103/PhysRevLett.132.221603}{\emph{Phys.
  Rev. Lett.} {\bfseries 132} (2024) 221603}
  [\href{https://arxiv.org/abs/2312.05304}{{\ttfamily 2312.05304}}].

\bibitem{Chacaltana:2012zy}
O.~Chacaltana, J.~Distler and Y.~Tachikawa, \emph{{Nilpotent orbits and
  codimension-two defects of 6d N=(2,0) theories}},
  \href{https://doi.org/10.1142/S0217751X1340006X}{\emph{Int. J. Mod. Phys. A}
  {\bfseries 28} (2013) 1340006}
  [\href{https://arxiv.org/abs/1203.2930}{{\ttfamily 1203.2930}}].

\bibitem{vinberg1972class}
{\`E}.B.~Vinberg and V.L.~Popov, \emph{On a class of quasihomogeneous affine
  varieties}, {\emph{Mathematics of the USSR-Izvestiya} {\bfseries 6} (1972)
  743}.

\bibitem{Nekrasov:2012xe}
N.~Nekrasov and V.~Pestun, \emph{{Seiberg-Witten geometry of four dimensional
  N=2 quiver gauge theories}},
  \href{https://arxiv.org/abs/1211.2240}{{\ttfamily 1211.2240}}.

\bibitem{Hanany:2010qu}
A.~Hanany and N.~Mekareeya, \emph{{Tri-vertices and SU(2)'s}},
  \href{https://doi.org/10.1007/JHEP02(2011)069}{\emph{JHEP} {\bfseries 02}
  (2011) 069} [\href{https://arxiv.org/abs/1012.2119}{{\ttfamily 1012.2119}}].

\bibitem{zz}
Z.~Zhong, \emph{Private communication}, .

\bibitem{Bourget:2021siw}
A.~Bourget, J.F.~Grimminger, A.~Hanany, M.~Sperling and Z.~Zhong,
  \emph{{Branes, Quivers, and the Affine Grassmannian}},
  \href{https://doi.org/10.2969/aspm/08810331}{\emph{Adv. Stud. Pure Math.}
  {\bfseries 88} (2023) 331}
  [\href{https://arxiv.org/abs/2102.06190}{{\ttfamily 2102.06190}}].

\bibitem{Hanany:1996ie}
A.~Hanany and E.~Witten, \emph{{Type IIB superstrings, BPS monopoles, and
  three-dimensional gauge dynamics}},
  \href{https://doi.org/10.1016/S0550-3213(97)00157-0,
  10.1016/S0550-3213(97)80030-2}{\emph{Nucl. Phys.} {\bfseries B492} (1997)
  152} [\href{https://arxiv.org/abs/hep-th/9611230}{{\ttfamily
  hep-th/9611230}}].

\bibitem{crawley2001geometry}
W.~Crawley-Boevey, \emph{Geometry of the moment map for representations of
  quivers}, {\emph{Compositio Mathematica} {\bfseries 126} (2001) 257}.

\bibitem{crawley2001normality}
W.~Crawley-Boevey, \emph{Normality of marsden-weinstein reductions for
  representations of quivers},
  \href{https://arxiv.org/abs/math/0105247}{{\ttfamily math/0105247}}.

\bibitem{bellamy2021symplectic}
G.~Bellamy and T.~Schedler, \emph{Symplectic resolutions of quiver varieties},
  {\emph{Selecta Mathematica} {\bfseries 27} (2021) 36}
  [\href{https://arxiv.org/abs/1602.00164}{{\ttfamily 1602.00164}}].

\bibitem{Gu:2022dac}
J.~Gu, Y.~Jiang and M.~Sperling, \emph{{Rational $Q$-systems, Higgsing and
  Mirror Symmetry}},
  \href{https://doi.org/10.21468/SciPostPhys.14.3.034}{\emph{SciPost Phys.}
  {\bfseries 14} (2023) 034}
  [\href{https://arxiv.org/abs/2208.10047}{{\ttfamily 2208.10047}}].

\bibitem{Gaiotto}
D.~Gaiotto, \emph{{N=2 dualities}},
  \href{https://doi.org/10.1007/JHEP08(2012)034}{\emph{JHEP} {\bfseries 08}
  (2012) 034} [\href{https://arxiv.org/abs/0904.2715}{{\ttfamily 0904.2715}}].

\bibitem{Bourget:2021jwo}
A.~Bourget, J.F.~Grimminger, A.~Hanany, R.~Kalveks and Z.~Zhong, \emph{{Higgs
  branches of U/SU quivers via brane locking}},
  \href{https://doi.org/10.1007/JHEP08(2022)061}{\emph{JHEP} {\bfseries 08}
  (2022) 061} [\href{https://arxiv.org/abs/2111.04745}{{\ttfamily
  2111.04745}}].

\bibitem{Seiberg:1994aj}
N.~Seiberg and E.~Witten, \emph{{Monopoles, duality and chiral symmetry
  breaking in N=2 supersymmetric QCD}},
  \href{https://doi.org/10.1016/0550-3213(94)90214-3}{\emph{Nucl. Phys. B}
  {\bfseries 431} (1994) 484}
  [\href{https://arxiv.org/abs/hep-th/9408099}{{\ttfamily hep-th/9408099}}].

\bibitem{Seiberg:1994rs}
N.~Seiberg and E.~Witten, \emph{{Electric - magnetic duality, monopole
  condensation, and confinement in N=2 supersymmetric Yang-Mills theory}},
  \href{https://doi.org/10.1016/0550-3213(94)90124-4}{\emph{Nucl. Phys. B}
  {\bfseries 426} (1994) 19}
  [\href{https://arxiv.org/abs/hep-th/9407087}{{\ttfamily hep-th/9407087}}].

\bibitem{Argyres:1995jj}
P.C.~Argyres and M.R.~Douglas, \emph{{New phenomena in SU(3) supersymmetric
  gauge theory}},
  \href{https://doi.org/10.1016/0550-3213(95)00281-V}{\emph{Nucl. Phys. B}
  {\bfseries 448} (1995) 93}
  [\href{https://arxiv.org/abs/hep-th/9505062}{{\ttfamily hep-th/9505062}}].

\bibitem{Argyres:1995xn}
P.C.~Argyres, M.R.~Plesser, N.~Seiberg and E.~Witten, \emph{{New N=2
  superconformal field theories in four-dimensions}},
  \href{https://doi.org/10.1016/0550-3213(95)00671-0}{\emph{Nucl. Phys. B}
  {\bfseries 461} (1996) 71}
  [\href{https://arxiv.org/abs/hep-th/9511154}{{\ttfamily hep-th/9511154}}].

\bibitem{Argyres:1996eh}
P.C.~Argyres, M.R.~Plesser and N.~Seiberg, \emph{{The Moduli space of vacua of
  N=2 SUSY QCD and duality in N=1 SUSY QCD}},
  \href{https://doi.org/10.1016/0550-3213(96)00210-6}{\emph{Nucl. Phys.}
  {\bfseries B471} (1996) 159}
  [\href{https://arxiv.org/abs/hep-th/9603042}{{\ttfamily hep-th/9603042}}].

\bibitem{Gaiotto:2009we}
D.~Gaiotto, \emph{{N=2 dualities}},
  \href{https://doi.org/10.1007/JHEP08(2012)034}{\emph{JHEP} {\bfseries 08}
  (2012) 034} [\href{https://arxiv.org/abs/0904.2715}{{\ttfamily 0904.2715}}].

\bibitem{Gaiotto:2009hg}
D.~Gaiotto, G.W.~Moore and A.~Neitzke, \emph{{Wall-crossing, Hitchin systems,
  and the WKB approximation}},
  \href{https://doi.org/10.1016/j.aim.2012.09.027}{\emph{Adv. Math.} {\bfseries
  234} (2013) 239} [\href{https://arxiv.org/abs/0907.3987}{{\ttfamily
  0907.3987}}].

\bibitem{Argyres:2007cn}
P.C.~Argyres and N.~Seiberg, \emph{{S-duality in N=2 supersymmetric gauge
  theories}}, \href{https://doi.org/10.1088/1126-6708/2007/12/088}{\emph{JHEP}
  {\bfseries 12} (2007) 088} [\href{https://arxiv.org/abs/0711.0054}{{\ttfamily
  0711.0054}}].

\bibitem{Benini:2010uu}
F.~Benini, Y.~Tachikawa and D.~Xie, \emph{{Mirrors of 3d Sicilian theories}},
  \href{https://doi.org/10.1007/JHEP09(2010)063}{\emph{JHEP} {\bfseries 09}
  (2010) 063} [\href{https://arxiv.org/abs/1007.0992}{{\ttfamily 1007.0992}}].

\bibitem{DistlerA}
O.~Chacaltana and J.~Distler, \emph{{Tinkertoys for Gaiotto Duality}},
  \href{https://doi.org/10.1007/JHEP11(2010)099}{\emph{JHEP} {\bfseries 11}
  (2010) 099} [\href{https://arxiv.org/abs/1008.5203}{{\ttfamily 1008.5203}}].

\bibitem{Xie:2012hs}
D.~Xie, \emph{{General Argyres-Douglas Theory}},
  \href{https://doi.org/10.1007/JHEP01(2013)100}{\emph{JHEP} {\bfseries 01}
  (2013) 100} [\href{https://arxiv.org/abs/1204.2270}{{\ttfamily 1204.2270}}].

\bibitem{Giacomelli:2020ryy}
S.~Giacomelli, N.~Mekareeya and M.~Sacchi, \emph{{New aspects of
  Argyres--Douglas theories and their dimensional reduction}},
  \href{https://doi.org/10.1007/JHEP03(2021)242}{\emph{JHEP} {\bfseries 03}
  (2021) 242} [\href{https://arxiv.org/abs/2012.12852}{{\ttfamily
  2012.12852}}].

\bibitem{Xie:2021ewm}
D.~Xie, \emph{{3d mirror for Argyres-Douglas theories}},
  \href{https://arxiv.org/abs/2107.05258}{{\ttfamily 2107.05258}}.

\bibitem{Zafrir:2016wkk}
G.~Zafrir, \emph{{Compactifications of 5d SCFTs with a twist}},
  \href{https://doi.org/10.1007/JHEP01(2017)097}{\emph{JHEP} {\bfseries 01}
  (2017) 097} [\href{https://arxiv.org/abs/1605.08337}{{\ttfamily
  1605.08337}}].

\bibitem{Argyres:2022lah}
P.C.~Argyres and M.~Martone, \emph{{The rank 2 classification problem I: scale
  invariant geometries}},  \href{https://arxiv.org/abs/2209.09248}{{\ttfamily
  2209.09248}}.

\bibitem{Cremonesi:2015lsa}
S.~Cremonesi, G.~Ferlito, A.~Hanany and N.~Mekareeya, \emph{{Instanton
  Operators and the Higgs Branch at Infinite Coupling}},
  \href{https://doi.org/10.1007/JHEP04(2017)042}{\emph{JHEP} {\bfseries 04}
  (2017) 042} [\href{https://arxiv.org/abs/1505.06302}{{\ttfamily
  1505.06302}}].

\bibitem{Heckman:2015ola}
J.J.~Heckman, D.R.~Morrison, T.~Rudelius and C.~Vafa, \emph{{Geometry of 6D RG
  Flows}}, \href{https://doi.org/10.1007/JHEP09(2015)052}{\emph{JHEP}
  {\bfseries 09} (2015) 052}
  [\href{https://arxiv.org/abs/1505.00009}{{\ttfamily 1505.00009}}].

\bibitem{Heckman:2015axa}
J.J.~Heckman and T.~Rudelius, \emph{{Evidence for C-theorems in 6D SCFTs}},
  \href{https://doi.org/10.1007/JHEP09(2015)218}{\emph{JHEP} {\bfseries 09}
  (2015) 218} [\href{https://arxiv.org/abs/1506.06753}{{\ttfamily
  1506.06753}}].

\bibitem{Heckman:2016ssk}
J.J.~Heckman, T.~Rudelius and A.~Tomasiello, \emph{{6D RG Flows and Nilpotent
  Hierarchies}}, \href{https://doi.org/10.1007/JHEP07(2016)082}{\emph{JHEP}
  {\bfseries 07} (2016) 082}
  [\href{https://arxiv.org/abs/1601.04078}{{\ttfamily 1601.04078}}].

\bibitem{Mekareeya:2016yal}
N.~Mekareeya, T.~Rudelius and A.~Tomasiello, \emph{{T-branes, Anomalies and
  Moduli Spaces in 6D SCFTs}},
  \href{https://doi.org/10.1007/JHEP10(2017)158}{\emph{JHEP} {\bfseries 10}
  (2017) 158} [\href{https://arxiv.org/abs/1612.06399}{{\ttfamily
  1612.06399}}].

\bibitem{Hassler:2019eso}
F.~Hassler, J.J.~Heckman, T.B.~Rochais, T.~Rudelius and H.Y.~Zhang,
  \emph{{T-Branes, String Junctions, and 6D SCFTs}},
  \href{https://doi.org/10.1103/PhysRevD.101.086018}{\emph{Phys. Rev. D}
  {\bfseries 101} (2020) 086018}
  [\href{https://arxiv.org/abs/1907.11230}{{\ttfamily 1907.11230}}].

\bibitem{Baume:2021qho}
F.~Baume, M.J.~Kang and C.~Lawrie, \emph{{Two 6D origins of 4D SCFTs: Class S
  and 6D (1,\,0) on a torus}},
  \href{https://doi.org/10.1103/PhysRevD.106.086003}{\emph{Phys. Rev. D}
  {\bfseries 106} (2022) 086003}
  [\href{https://arxiv.org/abs/2106.11990}{{\ttfamily 2106.11990}}].

\bibitem{Mekareeya:2017jgc}
N.~Mekareeya, K.~Ohmori, Y.~Tachikawa and G.~Zafrir, \emph{{E$_{8}$ instantons
  on type-A ALE spaces and supersymmetric field theories}},
  \href{https://doi.org/10.1007/JHEP09(2017)144}{\emph{JHEP} {\bfseries 09}
  (2017) 144} [\href{https://arxiv.org/abs/1707.04370}{{\ttfamily
  1707.04370}}].

\bibitem{DelZotto:2018tcj}
M.~Del~Zotto and G.~Lockhart, \emph{{Universal Features of BPS Strings in
  Six-dimensional SCFTs}},
  \href{https://doi.org/10.1007/JHEP08(2018)173}{\emph{JHEP} {\bfseries 08}
  (2018) 173} [\href{https://arxiv.org/abs/1804.09694}{{\ttfamily
  1804.09694}}].

\bibitem{Cordova:2015fha}
C.~Cordova, T.T.~Dumitrescu and K.~Intriligator, \emph{{Anomalies,
  renormalization group flows, and the a-theorem in six-dimensional (1, 0)
  theories}}, \href{https://doi.org/10.1007/JHEP10(2016)080}{\emph{JHEP}
  {\bfseries 10} (2016) 080}
  [\href{https://arxiv.org/abs/1506.03807}{{\ttfamily 1506.03807}}].

\bibitem{DelZotto:2022ohj}
M.~Del~Zotto, M.~Liu and P.-K.~Oehlmann, \emph{{Back to heterotic strings on
  ALE spaces. Part I. Instantons, 2-groups and T-duality}},
  \href{https://doi.org/10.1007/JHEP01(2023)176}{\emph{JHEP} {\bfseries 01}
  (2023) 176} [\href{https://arxiv.org/abs/2209.10551}{{\ttfamily
  2209.10551}}].

\bibitem{beauville2000symplectic}
A.~Beauville, \emph{Symplectic singularities}, {\emph{Invent. Math.} {\bfseries
  139} (2000) 541} [\href{https://arxiv.org/abs/math/9903070}{{\ttfamily
  math/9903070}}].

\bibitem{Fluder:2020pym}
M.~Fluder and C.F.~Uhlemann, \emph{{Evidence for a 5d F-theorem}},
  \href{https://doi.org/10.1007/JHEP02(2021)192}{\emph{JHEP} {\bfseries 02}
  (2021) 192} [\href{https://arxiv.org/abs/2011.00006}{{\ttfamily
  2011.00006}}].

\end{thebibliography}\endgroup
\end{document}